\documentstyle[12pt,puthesis,palatino,graphics,epsf]{report}

\begin{document}
\renewenvironment{description}{\list{}{ \itemsep 1mm \parsep 0mm
    \itemindent -10mm \leftmargin 10mm }} {\endlist}
\newcommand{\bgm}[1]{\mbox{\boldmath $#1$}}
\newcommand{\bgms}[1]{\mbox{{\scriptsize \boldmath $#1$}}}
\newcommand{\ul}[1]{\underline{#1}} \newcommand{\bgt}[1]{{\boldmath
    $#1$}}
\newcommand{\Beta}{{\rm B}}

\title{Information Theory and Learning:\\ A Physical Approach}
\author{Ilya Mark Nemenman}

\dept{Physics}
\principaladviser{William Bialek}
\firstreader{Curtis Callan}
\secondreader{William Bialek}

\submitdate{November 2000}

\figurespagetrue
\signaturefalse

\beforepreface

\prefacesection{Abstract}

We try to establish a unified information theoretic approach to
learning and to explore some of its applications.  First, we define
{\em predictive information} as the mutual information between the
past and the future of a time series, discuss its behavior as a
function of the length of the series, and explain how other quantities
of interest studied previously in learning theory---as well as in
dynamical systems and statistical mechanics---emerge from this
universally definable concept. We then prove that predictive
information provides the {\em unique measure for the complexity} of
dynamics underlying the time series and show that there are classes of
models characterized by {\em power--law growth of the predictive
  information} that are qualitatively more complex than any of the
systems that have been investigated before. Further, we investigate
numerically the learning of a nonparametric probability density, which
is an example of a problem with power--law complexity, and show that
the proper Bayesian formulation of this problem provides for the
`Occam' factors that punish overly complex models and thus allow one
{\em to learn not only a solution within a specific model class, but
  also the class itself} using the data only and with very few a
priori assumptions.  We study a possible {\em information theoretic
  method} that regularizes the learning of an undersampled discrete
variable, and show that learning in such a setup goes through stages
of very different complexities.  Finally, we discuss how all of these
ideas may be useful in various problems in physics, statistics, and,
most importantly, biology.


\prefacesection{Acknowledgements}

Most importantly, I thank my family and my dearest friends; for if not
for their wisdom, knowledge, love, and support I would never be who I
am now.  And I thank Bill Bialek, who is not just a perfect advisor,
but a friend to me.

This thesis would not be what it is now if not for many, too many to
name them all, people who mentored life and physics to me.  So, in
order of their appearance in my life I am grateful to Leonid
Demikhovsky, Mikhail Polozov, Albert Minkevich, Valentin Rusak, people
of the Department of Theoretical Physics at Belarusian State
University, Betty Young, people of Santa Clara University, Gerald
Fisher, Ronald Adler, people of the Physics and Astronomy Department
at San Francisco State University, Anatoly Spitkovsky, Alexander
Silbergleit, Gravity Probe B theory group, Curtis Callan, Vipul
Periwal, the late Howard Stone, Ale\-xander Polyakov, Olexei
Motrunich, Akakii Melikidze, Stanislav Boldyrev, Sergei Gukov, Andrei
Mikhailov, Timur Shutenko, people of the Department of Physics at
Princeton University, Naftali Tishby, Gonzalo Garcia de Polavieja
Embid, Methods in Computational Neuroscience course at Marine
Biological Laboratory, Rob de Ruyter van Steveninck, Adrienne
Fairhall, Jonathan Miller, Dmitry Rinberg, people of NEC Research
Institute, and many others.

\vspace{\fill}

\centerline{\Large Thank you all! }

\vspace{\fill}


\prefacesection{Collaborators} This thesis is based on the work done
in collaboration with William Bialek, Naftali Tishby, Adrienne
Fairhall, and Jonathan Miller. In particular, Chapters
\ref{predictive} and \ref{numeric} largely follow the papers by
Bialek, Nemenman, and Tishby (2000), and Bialek and Nemenman (2000)
respectively, and Chapter \ref{it_reg} is a part of the work in
progress by Bialek, Fairhall, Miller, and Nemenman.

\afterpreface

\newpage

\noindent
\begin{flushleft}
  ``All of the books in the world contain no more information than is
  broadcast as video in a single large American city in a single
  year.   Not all bits have equal value.''\\
  \hspace{2cm} Carl Sagan\footnote{All quotations shown on this page
    can be found at the electronic archive\\
    http://www.starlingtech.com/quotes/} \vspace{\fill} \noindent
  \\
  ``My interest is in the future because I am going to spend the rest
  of my life there.''\\
  \hspace{2cm}Charles F.~Kettering \vspace{\fill} \noindent
  \\
  ``That is what learning is. You suddenly understand something
  you've understood all your life, but in a new way.''\\
  \hspace{2cm}Doris Lessing \vspace{\fill} \noindent
  \\
  ``Learning is not compulsory$\dots$ Neither is survival.''\\
  \hspace{2cm}W.~Edwards Deming \vspace{\fill} \noindent
  \\
  ``Where is  the knowledge we've lost in information?''\\
  \hspace{2cm}T.~S.~Eliot \vspace{\fill} \noindent
  \\
  ``What most experimenters take for granted before they begin their
  experiments is infinitely more interesting than any results to which
  their experiments lead.''\\
  \hspace{2cm}Norbert Wiener
\end{flushleft}


\chapter{Introduction: what do we know?}
\label{introduction}

We hope that while reading this work our readers will unsurprisingly
realize that they actually are learning something. However, what may
come as a surprise is that they learn a lot more than they think:
while reading this very sentence the photoreceptors in the eyes
estimate the mean intensity of the ambient light and adapt to it; the
auditory cortex monitors the surroundings and warns if a visitor
knocks on the door. The reader skips the endings of some long,
complicated words because he has already guessed what is coming; he
then notices peculiarities in the stylistics of the text and soon
learns to distinguish sentences written late at night.  And then,
finally, there is the ``true'' learning of the thoughts that the
authors try to convey in their writing.

Learning is everywhere around and inside us, and it is absolutely
essential for our second--to--second survival. In fact, because of its
utmost importance and omnipresence each one of us has a well developed
personal, unique intuition on what ``learning'' means, and how it
works. One might think that such enormous experience would come in
handy when studying learning from a scientific perspective, but the
situation is quite the opposite: it is extremely difficult to build a
theory that unites the enormous spectrum of possible learning
problems. Intuition built up for the case of learning to play a
musical instrument may be totally useless (and even destructive) for
studying, for example, how we learn our first language, or master
mathematical concepts. A multitude of ideas and approaches, each
treating its specific problem and having only a slight relation to
another, is indeed what we see in learning science now.

In fact, there even is no such thing as the ``learning theory.'' There
is statistical learning theory, which builds probabilistic bounds on
our ability to estimate the parameters of models that describe some
observations, and its formalism seems completely disjoint from the
designs of psychological and physiological experiments that study
learning in humans and animals.  Then there is the Minimal Description
Length paradigm, which states that the shorter is the code for a set
of samples, the better is the knowledge of the structure inside the
samples; it is not clear how to connect these ideas to numerous
learning curves defined in specific contexts of neural networks.  Then
there are ideas that since the speed or (conversely) the difficulty of
learning is related intuitively to the complexity of the studied
problem, learning and complexity should be studied together; this
opens the Pandora box of different approaches to complexity (later in
this work we list over a dozen of definitions of this quantity!) and
does not even come close to quantifying learning and complexity of,
say, some simple geometric concept. We can continue this list, but the
point is clear. We believe that specific learning scenarios, however
interesting and practical they may be, are not going to bring any more
insight to our current understanding of learning (and, for that
matter, complexity).  What we need at this stage is not another
example---there are too many of them to comprehend already---but a
unifying, generalizing theory.

What do we expect from such a theory? We want it to be physical in its
spirit. That is, it must explain and unify all accumulated knowledge
of the subject (and thus necessarily have an element of a review), but
this explanation should bring a new level of understanding to the old
problems, a level from which all the problems appear as different
realizations of one general phenomenon. However, explaining old data is
just a half of a good theory. Using new tools we must also be able to
ask and answer meaningful new questions, thus the theory should be
constructive enough to serve as a kernel for development.

We build our presentation to address all of these questions. In
Chapter~\ref{predictive} we introduce a version of the theory of
learning and complexity which is built on information theory and the
notion of predictability. After finishing the construction, we
extensively analyze the literature to show that most of prior
knowledge of the subject is subsumed in our more general approach.
Then we try to show that the ideas do not only explain the old results
but can be used to study new problems as well. For this, we discuss a
broad spectrum of possible applications to physics, to computer
science, and to biology, and then single out two examples for a
detailed analysis.  In Chapter~\ref{numeric} we study applications of
our ideas to the learning of nonparametric continuous probability
densities, and we show how complexity penalizing Occam factors work in
this case.  Then in Chapter~\ref{it_reg} we turn to the seemingly
easier problem of learning a probability distribution of a discrete
variable, and we study how regularization based only on information
theory makes learning possible in the undersampled regime.

One may argue that the examples we discuss are not enough to claim for
certain that our theory indeed is constructive. We hope to resolve
these fears in the nearest future by studying other possible
applications that we mention throughout our work.  However, we want to
stress here explicitly that we believe that the theory itself is
complete, the definitions that we make are sensible and unique, and
the conclusions are general and universal.


\chapter{Predictability, Complexity, and Learning}

\label{predictive}

\section{Why study predictability?}

There is an obvious interest in having practical algorithms for
predicting the future, and there is a correspondingly large literature
on the problem of time series extrapolation.\footnote{The classic
  papers are by Kolmogoroff (1939, 1941) and Wiener (1949), who
  essentially solved all the extrapolation problems that could be
  solved by linear methods. Our understanding of predictability was
  changed by developments in dynamical systems, which showed that
  apparently random (chaotic) time series could arise from simple
  deterministic rules, and this led to vigorous exploration of
  nonlinear extrapolation algorithms (Abarbanel et al. 1993).  For a
  review comparing different approaches, see the conference
  proceedings edited by Weigend and Gershenfeld (1994).}  But
prediction is both more and less than extrapolation: we might be able
to predict, for example, the chance of rain in the coming week even if
we cannot extrapolate the trajectory of temperature fluctuations. In
the spirit of its thermodynamic origins, information theory (Shannon
1948) characterizes the potentialities and limitations of all possible
prediction algorithms, as well as unifying the analysis of
extrapolation with the more general notion of predictability.
Specifically, we can define a quantity---the {\em predictive
  information}---that measures how much our observations of the past
can tell us about the future.  The predictive information
characterizes the world we are observing, and we shall see that this
characterization is close to our intuition about the complexity of the
underlying dynamics.

Prediction is one of the fundamental problems in neural computation.
Much of what we admire in expert human performance is predictive in
character---the point guard who passes the basketball to a place where
his teammate will arrive in a split second, the chess master who knows
how moves made now will influence the end game two hours hence, the
investor who buys a stock in anticipation that it will grow in the
year to come.  More generally, we gather sensory information not for
its own sake but in the hope that this information will guide our
actions (including our verbal actions). But acting takes time, and
sense data can guide us only to the extent that those data inform us
about the state of the world at the time of our actions, so the only
components of the incoming data that have a chance of being useful are
those that are predictive. Put bluntly, {\em nonpredictive information
  is useless to the organism}, and it therefore makes sense to isolate
the predictive information. It will turn out that most of the
information we collect over a long period of time is nonpredictive, so
that isolating the predictive information must go a long way toward
separating out those features of the sensory world that are relevant
for behavior.

One of the most important examples of prediction is the phenomenon of
generalization in learning.  Learning is formalized as finding a model
that explains or describes a set of observations, but again this is
useful precisely (and only) because we expect this model will continue
to be valid: in the language of learning theory [see, for example,
Vapnik (1998)] an animal can gain selective advantage not from its
performance on the training data but only from its performance at
generalization. Generalizing---and not ``overfitting'' the training
data---is precisely the problem of isolating those features of the
data that have predictive value (see also Bialek and Tishby, in
preparation).  Further, we know that the success of generalization
hinges on controlling the complexity of the models that we are willing
to consider as possibilities. Finally, learning a model to describe a
data set can be seen as an encoding of those data, as emphasized by
Rissanen (1989), and the quality of this encoding can be measured
using the ideas of information theory.  Thus the {\em exploration of
  learning problems should provide us with explicit links among the
  concepts of entropy, predictability, and complexity}.

The notion of complexity arises not only in learning theory, but also
in several other contexts.  Some physical systems exhibit more complex
dynamics than others (turbulent vs. laminar flows in fluids), and some
systems evolve toward more complex states than others (spin glasses
vs.~ferromagnets).  The problem of characterizing complexity in
physical systems has a substantial literature of its own [for an
overview see Bennett (1990)].  In this context several authors have
considered complexity measures based on entropy or mutual information,
although as far as we know no clear connections have been drawn among
the measures of complexity that arise in learning theory and those
that arise in dynamical systems and statistical mechanics.

An essential difficulty in quantifying complexity is to distinguish
complexity from randomness.  A true random string cannot be compressed
and hence requires a long description; it thus is complex in the sense
defined by Kolmogorov (1965, Li and Vit{\'a}nyi 1993, Vit{\'a}nyi and
Li 2000), yet the physical process that generates this string may have
a very simple description.  Both in statistical mechanics and in
learning theory our intuitive notions of complexity correspond to the
statements about complexity of the underlying process, and not
directly to the description length or Kolmogorov complexity.

Our central result is that {\em the predictive information provides a
  general measure of complexity} which includes as special cases some
relevant concepts from learning theory and from dynamical systems.
While the work on the complexity of models in learning theory rests
specifically on the idea that one is trying to infer a model from
data, the predictive information is a property of the data (or, more
precisely, of an ensemble of data) itself without reference to a
specific class of underlying models.  If the data are generated by a
process in a known class but with unknown parameters, then we can
calculate the predictive information explicitly and show that this
{\em information diverges logarithmically with the size of the data
  set we have observed}; the coefficient of this divergence counts the
number of parameters in the model, or more precisely the effective
dimension of the model class, and this provides a link to known
results of Rissanen and others.  But our approach also allows us to
quantify the complexity of processes that fall outside the finite
dimensional models of conventional learning theory, and we show that
these {\em more complex processes are characterized by a power--law
  rather than a logarithmic divergence of the predictive information}.

By analogy with the analysis of critical phenomena in statistical
physics, the separation of logarithmic from power--law divergences,
together with the measurement of coefficients and exponents for these
divergences, allows us to define ``universality classes'' for the
complexity of data streams. The power--law or nonparametric class of
processes may be crucial in real world learning tasks, where the
effective number of parameters becomes so large that asymptotic
results for finitely parameterizable models are inaccessible in
practice.  There is empirical evidence that simple physical systems
can generate dynamics in this complexity class, and there are hints
that language also may fall in this class.

Finally, we argue that {\em the divergent components of the predictive
  information provide a unique measure of complexity} that is
consistent with certain simple requirements.  This argument is in the
spirit of Shannon's original derivation of entropy as the unique
measure of available information. We believe that this uniqueness
argument provides a conclusive answer to the question of how one
should quantify the complexity of a process generating a time series.

With the evident cost of lengthening our discussion, we have tried to
give a self--contained presentation that develops our point of view,
uses simple examples to connect with known results, and then
generalizes and goes beyond these results.\footnote{Some of the basic
  ideas presented here, together with some connections to earlier
  work, can be found in brief preliminary reports (Bialek 1995; Bialek
  and Tishby 1999). The central results of the present work, however,
  were at best conjectures in these preliminary accounts.}  Even in
cases where at least the qualitative form of our results is known from
previous work, we believe that our point of view elucidates some
issues that may have been less the focus of earlier studies. Last but
not least, we explore the possibilities for connecting our theoretical
discussion with the experimental characterization of learning and
complexity in neural systems.

\section{A curious observation}

\begin{figure}[t]
  \centerline{\epsfxsize=0.8\hsize\epsffile{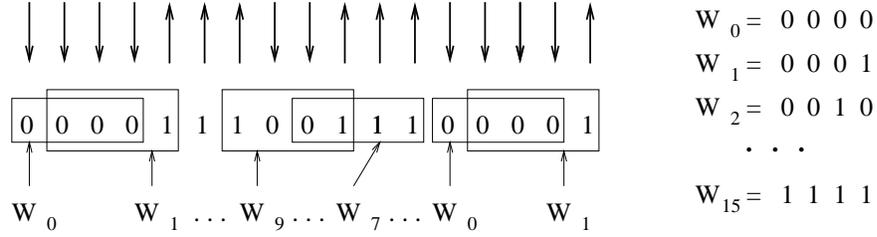}}
\caption[Calculating entropy of spin words.]{Calculating entropy of
  words of length $4$ in a chain of $17$ spins. For this chain,
  $n(W_0)=n(W_1)=n(W_3)=n(W_7)=n(W_{12})=n(W_{14})=2$, $n(W_8)=n(W_9)=1$,
  and all other frequencies are zero. Thus, $S(4)\approx2.95\; {\rm bits}$.} 
\label{spins}
\end{figure}

Before starting the systematic analysis of the problem, we want to
motivate our discussion further by presenting results of some simple
numerical experiments. Suppose we have a 1-dimensional chain of Ising
spins with the Hamiltonian given by
\begin{equation}
H=-\sum_{\rm i,j} J_{\rm ij} \sigma_{\rm i} \sigma_{\rm j},
\end{equation}
where the matrix $J_{\rm ij}$ is not necessarily tridiagonal (that is,
long range interactions are also allowed). One may identify spins
pointing upwards with $1$ and downwards with $0$, and then a spin
chain is equivalent to some sequence of binary digits. This sequence
consists of (overlapping) words of $N$ digits each, $W_{\rm k}$,
$k=0,1 \cdots 2^N-1$.  Even though there are $2^N$ such words total,
they appear with very different frequencies $n(W_k)$ in the spin chain
[see Fig.~(\ref{spins}) for details].
If the number of spins is large, then counting these frequencies
provides a good empirical estimate to $P_N(W_k)$, the probability
distribution of different words of length $N$.  Then one can calculate
the entropy $S(N)$ of this probability distribution by the usual
formula
\begin{equation}
S(N)=-\sum_{k=0}^{2^N-1} P_N(W_k) \log_2 P_N(W_k)\;\;\;\;\;\;{\rm (bits)}.
\end{equation}
Since entropy is an extensive property, $S(N)$ is asymptotically
proportional to $N$ for any spin chain. Choosing a different set of
couplings $J_{\rm ij}$ may change the coefficient of proportionality
(and finding this coefficient is usually the goal of statistical
mechanics) but the linearity is never challenged.

\begin{figure}[t]
  \centerline{\epsfxsize=0.7\hsize\epsffile{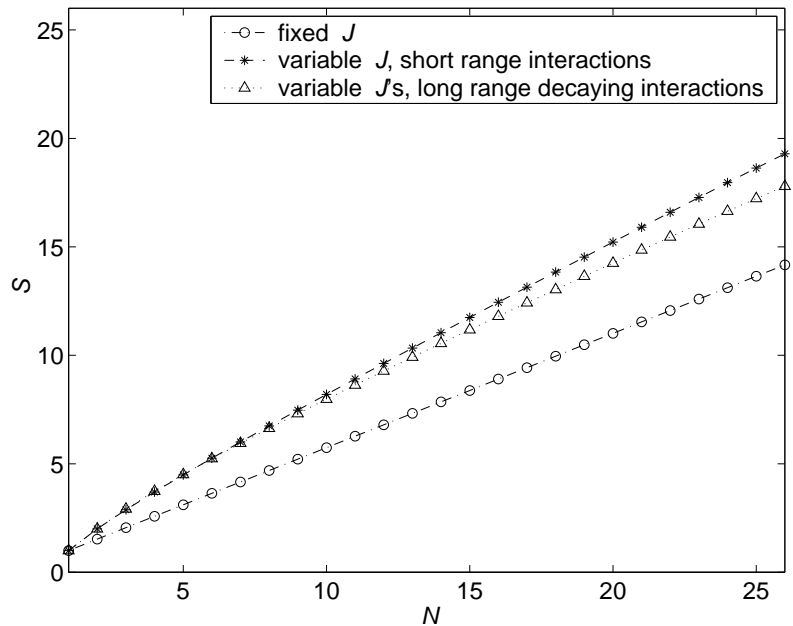}}
\caption[Entropy as a function of the word length]{Entropy as a
  function of the word length for spin chains with different
  interactions. Notice that all lines start from $S(N)=\log_2 2=1$
  since at the values of the coupling we investigated the correlation
  length is much smaller than the chain length ($1\cdot 10^9$ spins).}
\label{entr_lin}
\end{figure}
We investigated this in three different spin chains of one billion
spins each (the temperature is always $k_BT=1$). For the first chain,
only $J_{\rm i, i+1}=1$ was nonzero, and its value was the same for
all ${\rm i}$'s. The second chain was also generated using the nearest
neighbor interactions, but the value of the coupling was reinitialized
every 400,000 spins by taking a random number from a Gaussian
distribution with a zero mean and a unit variance. In the third case,
we again reinitialized at the same frequency, but now interactions
were long--ranged, and the variance of coupling constants decreased
with the distance between the spins as $\langle J_{\rm ij}^2 \rangle
=1/({\rm i}-{\rm j})^2$. We plotted $S(N)$ for all these cases in
Fig.~(\ref{entr_lin}), and, of course, the asymptotically linear
behavior seems to be evident---the extensive entropy shows no
qualitative distinction between the three cases we consider.

However, the situation changes drastically if we remove the asymptotic
linear contribution and plot only the sublinear component $S_1(N)$ of
the entropy.
\begin{figure}[t]
  \centerline{\epsfxsize=0.7\hsize\epsffile{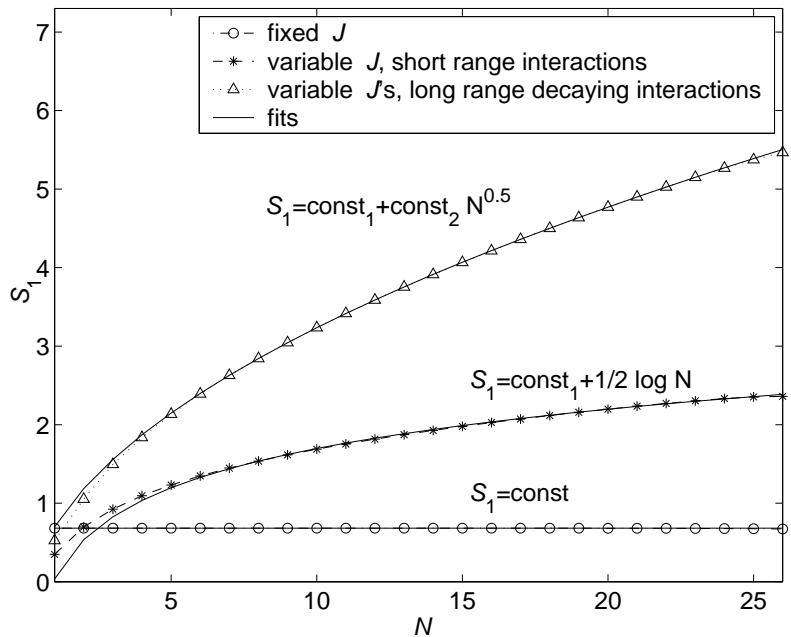}}
\caption{Subextensive part of the entropy as a function of the word
  length.}
\label{entr_subext}
\end{figure} 
As we see in Fig.~(\ref{entr_subext}), the three investigated chains
then exhibit {\em qualitatively} different features: for the first one,
$S_1$ is constant; for the second one, it is logarithmic; and, for the
third one, it clearly shows a power--law behavior. 

What is the significance of this observation? Of course, the
differences must be related to the ways we chose $J$'s for the
simulations. In the first case, $J$ is fixed, and there is not much
one can learn from observing the spin chain. For the second chain, $J$
changes, and the statistics of the spin--words is different in
different parts of the sequence. By looking at this statistics, one
can thus estimate coupling at the current position.  Finally, in the
third case there are many coupling constants that can be learned.  In
principle, as $N$ increases one becomes sensitive to correlations
caused by interactions over larger and larger distances, and, since
the variance of the couplings decays with the distance, interactions
of longer range do not interfere with learning short--scale
properties.  So, intuitively, the qualitatively different behavior of
$S_1(N)$ for the three plotted cases is due to a different character
of learning tasks involved in understanding the spin chains.  Much of
this Chapter can be seen as expanding on and quantifying this
intuitive observation.  \footnote{Note again that we are dealing here
  with subextensive properties of systems. These are the properties
  that are ignored in most problems in statistical mechanics.}


\section{Fundamentals}

The problem of prediction comes in various forms, as noted above.
Information theory allows us to treat the different notions of
prediction on the same footing.  The first step is to recognize that
all predictions are probabilistic---even if we can predict the
temperature at noon tomorrow, we should provide error bars or
confidence limits on our prediction. The next step is to remember
that, even before we look at the data, we know that certain futures
are more likely than others, and we can summarize this knowledge by a
prior probability distribution for the future.  Our observations on
the past lead us to a new, more tightly concentrated distribution, the
distribution of futures conditional on the past data. Different kinds
of predictions are different slices through or averages over this
conditional distribution, but information theory quantifies the
``concentration'' of the distribution without making any commitment as
to which averages will be most interesting.

Imagine that we observe a stream of data $x(t)$ over a time interval
$-T < t < 0$; let all of these past data be denoted by the shorthand
$x_{\rm past}$.  We are interested in saying something about the
future, so we want to know about the data $x(t)$ that will be observed
in the time interval $0 < t < T'$; let these future data be called
$x_{\rm future}$.  In the absence of any other knowledge, futures are
drawn from the probability distribution $P(x_{\rm future})$, while
observations of particular past data $x_{\rm past}$ tell us that
futures will be drawn from the conditional distribution $P(x_{\rm
  future} | x_{\rm past})$. The greater concentration of the
conditional distribution can be quantified by the fact that it has
smaller entropy than the prior distribution, and this reduction in
entropy is Shannon's definition of the information that the past
provides about the future.  We can write the average of this {\em
  predictive information} as
\begin{eqnarray}
{\cal I}_{\rm pred} (T,T') &=& 
{\Bigg\langle} \log_2 \left[ {{P(x_{\rm future}| x_{\rm past})} 
\over{P(x_{\rm future})}}\right]\Bigg\rangle
  \\ 
&=& -\langle\log_2 P(x_{\rm future})\rangle 
- \langle\log_2 P( x_{\rm past})\rangle
\nonumber\\
&&\,\,\,\,\,\,\,\,\,\, 
-\left[-\langle\log_2 P(x_{\rm future}, x_{\rm past})\rangle\right]\,,
\label{ents}
\end{eqnarray}
where $\langle \cdots \rangle$ denotes an average over the joint
distribution of the past and the future, $P(x_{\rm future} , x_{\rm
  past})$.

Each of the terms in Eq. (\ref{ents}) is an entropy. Since we are
interested in predictability or generalization, which are associated
with some features of the signal persisting forever, we may assume
stationarity or invariance under time translations. Then the entropy
of the past data depends only on the duration of our observations, so
we can write $ -\langle\log_2 P( x_{\rm past})\rangle = S(T) $, and by
the same argument $-\langle\log_2 P( x_{\rm future})\rangle = S(T')$.
Finally, the entropy of the past and the future taken together is the
entropy of observations on a window of duration $T+T'$, so that $
-\langle\log_2 P(x_{\rm future} , x_{\rm past})\rangle = S(T+T')$.
Putting these equations together, we obtain
\begin{equation}
{\cal I}_{\rm pred}(T,T') = S(T) +S(T') - S(T+T') . \label{IpredandST}
\end{equation}

In the same way that the entropy of a gas at fixed density is
proportional to the volume, the entropy of a time series
(asymptotically) is proportional to its duration, so that
$\lim_{T\rightarrow\infty} {{S(T)}/ T} = {\cal S}_0$; entropy is an
extensive quantity.  But from Eq. (\ref{IpredandST}) any extensive
component of the entropy cancels in the computation of the predictive
information: {\em predictability is a deviation from extensivity}.  If
we write $S(T) = {\cal S}_0 T +S_1(T)$, then Eq. (\ref{IpredandST})
tells us that the predictive information is related {\em only} to the
nonextensive term $S_1(T)$.

We know two general facts about the behavior of $S_1(T)$.  First, the
corrections to extensive behavior are positive, $S_1(T) \geq 0$.
Second, the statement that entropy is extensive is the statement that
the limit
\begin{eqnarray}
\lim_{T\rightarrow\infty} {{S(T)}\over T} = {\cal S}_0 
\end{eqnarray}
exists, and for this to be true we must also have
\begin{eqnarray}
\lim_{T\rightarrow\infty} {{S_1(T)}\over T} = 0.
\end{eqnarray}
Thus the nonextensive terms in the entropy must be {\em sub}extensive,
that is they must grow with $T$ less rapidly than a linear function.
Taken together, these facts guarantee that the predictive information
is positive and subextensive.  Further, if we let the future extend
forward for a very long time, $T' \rightarrow \infty$, then we can
measure the information that our sample provides about the entire
future,
\begin{equation}
I_{\rm pred} (T) = \lim_{T' \rightarrow \infty} {\cal I}_{\rm
pred}(T,T')
= S_1 (T).
\label{ipredlim}
\end{equation}

If we have been observing a time series for a (long) time $T$, then
the total amount of data we have taken in is measured by the entropy
$S(T)$, and at large $T$ this is given approximately by ${\cal S}_0
T$.  But the predictive information that we have gathered cannot grow
linearly with time, even if we are making predictions about a future
which stretches out to infinity. As a result, of the total information
we have taken in by observing $x_{\rm past}$, only a vanishing
fraction is of relevance to the prediction:
\begin{equation}
\lim_{T\rightarrow\infty} {{\rm Predictive\ Information} \over{\rm
Total\ Information}} = {{I_{\rm pred} (T)} \over {S(T)}}
\rightarrow 0. \label{chuck}
\end{equation}
In this precise sense, most of what we observe is irrelevant to the
problem of predicting the future. We can think of Eq.  (\ref{chuck})
as a law of diminishing returns: although we collect data in
proportion to our observation time $T$, a smaller and smaller fraction
of this information is useful in the problem of prediction.  Note that
these diminishing returns are not due to a limited lifetime, since we
calculate the predictive information assuming that we have a future
extending forward to infinity.

Now consider the case where time is measured in discrete steps, so
that we have seen $N$ time points $x_1, x_2 , \cdots , x_N$. How much
have we learned about the underlying pattern in these data?  The more
we know, the more effectively we can predict the next data point
$x_{N+1}$ and hence the fewer bits we will need to describe the
deviation of this data point from our prediction: our accumulated
knowledge about the time series is measured by the degree to which we
can compress the description of new observations.  On average, the
length of the code word required to describe the point $x_{N+1}$,
given that we have seen the previous $N$ points, is given by
\begin{equation}
\ell(N) = -\langle \log_2 P(x_{N+1} | x_1 , x_2 , \cdots , x_N )
\rangle \,\,{\rm bits},
\label{condentropy}
\end{equation}
where the expectation value is taken over the joint distribution of
all the $N+1$ points, $P(x_1, x_2, \cdots , x_N, x_{N+1})$.  It is
easy to see that
\begin{eqnarray}
\ell(N) = S(N+1) - S(N) \approx 
{{\partial S (N)}\over {\partial N}} .
\end{eqnarray}
As we observe for longer times, we learn more and this word length
decreases.  It is natural to define a learning curve that measures
this improvement.  Usually we define learning curves by measuring the
frequency or costs of errors; here the cost is that our encoding of
the point $x_{N+1}$ is longer than it could be if we had perfect
knowledge.  This ideal encoding has a length which we can find by
imagining that we observe the time series for an infinitely long time,
$\ell_{\rm ideal} = \lim_{N\rightarrow\infty} \ell(N)$, but this is
just another way of defining the extensive component of the entropy
${\cal S}_0$. Thus we can define a learning curve
\begin{eqnarray}
\Lambda (N) &\equiv& \ell(N) - \ell_{\rm ideal} \\ 
&=& S(N+1) - S(N) - {\cal S}_0 \nonumber\\ 
&=& S_1(N+1) - S_1(N) \nonumber\\ 
&\approx&
{{\partial S_1 (N)}\over {\partial N}} 
= {{\partial I_{\rm pred}(N)}\over{\partial N}},
\label{unilcurve}
\end{eqnarray}
and we see once again that the extensive component of the entropy
cancels.

It is well known that the problems of prediction and compression are
related, and what we have done here is to illustrate one aspect of
this connection.  Specifically, if we ask how much one segment of a
time series can tell us about the future, the answer is contained in
the subextensive behavior of the entropy.  If we ask how much we are
learning about the structure of the time series, then the natural and
universally defined learning curve is related again to the
subextensive entropy: the learning curve is the derivative of the
predictive information.

This universal learning curve is connected to the more conventional
learning curves in specific contexts.  As an example (cf. Section
\ref{testcase}), consider fitting a set of data points $\{ x_{\rm n} ,
y_{\rm n}\}$ with some class of functions $y = f(x; {\bgm \alpha}),$
where the $\bgm\alpha$ are unknown parameters that need to be learned;
we also allow for some Gaussian noise in our observation of the
$y_{\rm n}$.  Here the natural learning curve is the evolution of
$\chi^2$ for generalization as a function of the number of examples.
Within the approximations discussed below, it is straightforward to
show that as $N$ becomes large,
\begin{eqnarray}
\langle \chi^2(N) \rangle = {1\over {\sigma^2}} \langle
\left[ y - f(x;{\bgm \alpha})\right]^2\rangle \rightarrow 
2 \ln2 \,\Lambda (N) + 1,
\end{eqnarray}
where $\sigma^2$ is the variance of the noise. Thus a more
conventional measure of performance at learning a function is equal to
the universal learning curve defined purely by information theoretic
criteria. In other words, if a learning curve is measured in the right
units, then its integral represents the amount of the useful
information accumulated. Since one would expect any learning curve to
decrease to zero eventually, we again obtain the `law of diminishing
returns'.

Different quantities related to the subextensive entropy have been
discussed in several contexts. For example, the code length $\ell(N)$
has been defined as a learning curve in the specific case of neural
networks (Opper and Haussler 1995) and has been termed the
``thermodynamic dive'' (Crutchfield and Shalizi 1998) and ``$N^{\rm
  th}$ order block entropy'' (Grassberger 1986). Mutual information
between all of the past and all of the future (both semi--infinite) is
known also as the ``excess entropy,'' ``effective measure
complexity,'' ``stored information,'' and so on [see Shalizi and
Crutchfield (1999) and references therein, as well as the discussion
below]. If the data allow a description by a model with a finite
number of parameters, then mutual information between the data and the
parameters is of interest, and this is also the predictive information
about all of the future; some special cases of this problem have been
discussed by Opper and Haussler (1995) and by Herschkowitz and Nadal
(1999).  What is important is that {\em the predictive information or
  subextensive entropy is related to all these quantities}, and that
{\em it can be defined for any process without a reference to a class
  of models}.  It is this universality that we find appealing, and
this universality is strongest if we focus on the limit of long
observation times.  Qualitatively, in this regime
($T\rightarrow\infty$) we expect the predictive information to behave
in one of three different ways: it may either stay finite, or grow to
infinity together with $T$; in the latter case the rate of growth may
be slow (logarithmic) or fast (sublinear power).

The first possibility, $\lim_{T\rightarrow\infty} I_{\rm pred} (T) = $
constant, means that no matter how long we observe we gain only a
finite amount of information about the future. This situation
prevails, for example, when the dynamics are too regular: for a purely
periodic system, complete prediction is possible once we know the
phase, and if we sample the data at discrete times this is a finite
amount of information; longer period orbits intuitively are more
complex and also have larger $I_{\rm pred}$, but this doesn't change
the limiting behavior $\lim_{T\rightarrow\infty} I_{\rm pred} (T) =$
constant.

Alternatively, the predictive information can be small when the
dynamics are irregular but the best predictions are controlled only by
the immediate past, so that the correlation times of the observable
data are finite [see, for example, Crutchfield and Feldman (1997) and the fixed short--range interactions plot on Fig.~(\ref{entr_subext})].
Imagine, for example, that we observe $x(t)$ at a series of discrete
times $\{t_{\rm n}\}$, and that at each time point we find the value
$x_{\rm n}$. Then we can always write the joint distribution of the
$N$ data points as a product,
\begin{eqnarray}
P(x_1 , x_2 , \cdots , x_N ) &=& P(x_1 ) P(x_2 | x_1) 
P(x_3 | x_2 , x_1) \cdots . 
\end{eqnarray}
For Markov processes, what we observe at $t_{\rm n}$ depends only on
events at the previous time step $t_{\rm n-1}$, so that
\begin{eqnarray}
P(x_{\rm n} | \{x_{\rm 1\leq i \leq n-1}\}) &=& 
P(x_{\rm n} | x_{\rm n-1}) ,
\end{eqnarray}
and hence the predictive information reduces to
\begin{eqnarray}
I_{\rm pred} = \Bigg\langle \log_2 \left[ {{P(x_{\rm n}|x_{\rm n-1})}
\over{P(x_{\rm n})}} \right] \Bigg\rangle .
\end{eqnarray}
The maximum possible predictive information in this case is the
entropy of the distribution of states at one time step, which in turn
is bounded by the logarithm of the number of accessible states. To
approach this bound the system must maintain memory for a long time,
since the predictive information is reduced by the entropy of the
transition probabilities. Thus systems with more states and longer
memories have larger values of $I_{\rm pred}$.

More interesting are those cases in which $I_{\rm pred}(T)$ diverges
at large $T$. In physical systems we know that there are critical
points where correlation times become infinite, so that optimal
predictions will be influenced by events in the arbitrarily distant
past. Under these conditions the predictive information can grow
without bound as $T$ becomes large; for many systems the divergence is
logarithmic, $I_{\rm pred} (T\rightarrow\infty) \propto \ln T$, as for
the variable $J_{\rm ij}$, short range Ising model of
Figs.~(\ref{entr_lin}, \ref{entr_subext}). Long range correlation also
are important in a time series where we can learn some underlying
rules.  It will turn out that when the set of possible rules can be
described by a finite number of parameters, the predictive information
again diverges logarithmically, and the coefficient of this divergence
counts the number of parameters.  Finally, a faster growth is also
possible, so that $I_{\rm pred} (T\rightarrow\infty) \propto
T^\alpha$, as for the variable $J_{\rm ij}$ long range Ising model,
and we shall see that this behavior emerges from, for example,
nonparametric learning problems.


\section{Learning and predictability}

Learning is of interest precisely in those situations where
correlations or associations persist over long periods of time. In the
usual theoretical models, there is some rule underlying the observable
data, and this rule is valid forever; examples seen at one time inform
us about the rule, and this information can be used to make
predictions or generalizations. The predictive information quantifies
the average generalization power of examples, and we shall see that
there is a direct connection between the predictive information and
the complexity of the possible underlying rules.

\subsection{A test case}
\label{testcase}

Let us begin with a simple example already mentioned above.  We
observe two streams of data $x$ and $y$, or equivalently a stream of
pairs $(x_{\rm 1} , y_{\rm 1})$, $(x_{\rm 2} , y_{\rm 2})$, $\cdots$ ,
$(x_{\rm N} , y_{\rm N})$.  Assume that we know in advance that the
$x$'s are drawn independently and at random from some distribution
$P(x)$, while the $y$'s are noisy versions of some function acting on
$x$,
\begin{eqnarray}
y_{\rm n} = f(x_{\rm n} ; {\bgm\alpha} ) + \eta_{\rm n} ,
\end{eqnarray}
where $f(x; {\bgm\alpha})$ is a class of functions parameterized by
$\bgm\alpha$, and $\eta_{\rm n}$ is some noise which for simplicity we
will assume is Gaussian with some known standard deviation $\sigma$.
We can even start with a {\em very} simple case, where the function
class is just a linear combination of some basis functions, so that
\begin{eqnarray}
f(x; {\bgm\alpha}) = \sum_{\rm \mu =1}^K \alpha_\mu \phi_\mu (x) .
\end{eqnarray}
The usual problem is to estimate, from $N$ pairs $\{x_{\rm i} , y_{\rm
  i}\}$, the values of the parameters $\bgm\alpha$; in favorable cases
such as this we might even be able to find an effective regression
formula.  We are interested in evaluating the predictive information,
which means that we need to know the entropy $S(N)$.  We go through
the calculation in some detail because it provides a model for the
more general case.

To evaluate the entropy $S(N)$ we first construct the probability
distribution $P(x_{\rm 1}, y_{\rm 1}, x_{\rm 2}, y_{\rm 2}, \cdots ,
x_{\rm N} , y_{\rm N })$. The same set of rules apply to the whole
data stream, which here means that the same parameters $\bgm\alpha$
apply for all pairs $\{x_{\rm i}, y_{\rm i}\}$, but these parameters
are chosen at random from a distribution ${\cal P}({\bgm\alpha})$ at
the start of the stream. Thus we write
\begin{eqnarray}
&&P(x_{\rm 1}, y_{\rm 1}, x_{\rm 2} , y_{\rm 2}, \cdots , x_{\rm
N} , y_{\rm N }) 
\nonumber\\
&&\,\,\,\,\,= \int d^K \alpha \,
P(x_{\rm 1}, y_{\rm 1}, x_{\rm 2} , y_{\rm 2}, \cdots , x_{\rm N}
, y_{\rm N }  | {\bgm\alpha}) {\cal P}({\bgm\alpha}) \,,
\label{start} 
\end{eqnarray}
and now we need to construct the conditional distributions for fixed
$\bgm\alpha$. By hypothesis each $x$ is chosen independently, and once
we fix $\bgm\alpha$ each $y_{\rm i}$ is correlated only with the
corresponding $x_{\rm i}$, so that we have
\begin{eqnarray}
P(x_{\rm 1}, y_{\rm 1}, x_{\rm 2} , y_{\rm 2}, \cdots , x_{\rm N}
, y_{\rm N } | {\bgm\alpha}) = \prod_{\rm i =1}^{\rm N} \left[
P(x_{\rm i})\, P(y_{\rm i} | x_{\rm i} ; {\bgm\alpha})\right]\,.
\end{eqnarray}
Further, with the simple assumptions above about the class of
functions and Gaussian noise, the conditional distribution of $y_{\rm
  i}$ has the form
\begin{eqnarray}
P(y_{\rm i} | x_{\rm i} ; {\bgm\alpha}) = 
{1\over\sqrt{2 \pi\sigma^2}} 
\exp\left[ -{1\over{2\sigma^2}} \left( y_{\rm i} - \sum_{\mu =1}^K 
\alpha_\mu \phi_\mu(x_{\rm i})\right)^2 \right] \,.
\end{eqnarray}
Putting all these factors together,
\begin{eqnarray}
&&P(x_{\rm 1}, y_{\rm 1}, x_{\rm 2} , y_{\rm 2}, \cdots , x_{\rm
N} , y_{\rm N }) 
\nonumber\\ 
&&\,\,\,\,\,= \left[ \prod_{\rm i =1}^{\rm N} P(x_{\rm i}) \right] 
\left( {1\over \sqrt{2\pi\sigma^2}}\right)^{N} \int d^K \alpha \, 
{\cal P}({\bgm\alpha})
\exp\left[
-{1\over{2\sigma^2}} \sum_{\rm i=1}^{\rm N} y_{\rm i}^2\right] 
\nonumber\\
&&\,\,\,\,\,\,\,\,\,\,\times 
\exp\left[
-{N\over 2} \sum_{\mu , \nu =1}^K A_{\mu\nu} (\{x_{\rm i}\})
\alpha_\mu\alpha_\nu 
+ N \sum_{\mu=1}^K B_\mu(\{x_{\rm i} , y_{\rm i}\}) 
\alpha_\mu \right] ,
\end{eqnarray}
where
\begin{eqnarray}
A_{\mu\nu} (\{x_{\rm i}\}) &=& {1\over{\sigma^2 N}}
\sum_{\rm i=1}^{\rm N} \phi_\mu(x_{\rm i} ) \phi_\nu (x_{\rm i}) ,
\,\,\,\,{\rm and}
\\
B_\mu(\{x_{\rm i} , y_{\rm i}\}) &=& {1\over{\sigma^2 N}}\sum_{\rm
i=1}^{\rm N} y_{\rm i} \phi_\mu(x_{\rm i} ) .
\end{eqnarray}
Our placement of the factors of $N$ means that both $A_{\mu\nu}$ and
$B_\mu$ are of order unity as $N\rightarrow\infty$.  These quantities
are empirical averages over the samples $\{ x_{\rm i} , y_{\rm i}\}$,
and if the $\phi_\mu$ are well behaved we expect that these empirical
means converge to expectation values for most realizations of the
series $\{x_{\rm i}\}$:
\begin{eqnarray}
\lim_{N\rightarrow\infty} A_{\mu\nu} (\{x_{\rm i}\})
 &=&A_{\mu\nu}^\infty = {1\over{\sigma^2 }}\int dx P(x) \phi_\mu(x  )
\phi_\nu (x )\,,
\\
\lim_{N\rightarrow\infty} B_\mu(\{x_{\rm i} , y_{\rm i}\})
&=&B_\mu^\infty  = \sum_{\nu =1}^K A_{\mu\nu}^\infty
\bar{\alpha}_\nu\, ,
\end{eqnarray}
where $\bar{\bgm\alpha}$ are the parameters that actually gave rise to
the data stream $\{ x_{\rm i} , y_{\rm i}\}$.  In fact we can make the
same argument about the terms in $\sum y_{\rm i}^2$,
\begin{eqnarray}
\lim_{N\rightarrow\infty} \sum_{\rm i=1}^{\rm N} y_{\rm i}^2 = N
\sigma^2 \left[ \sum_{\mu,\nu =1}^K {\bar\alpha}_\mu
A_{\mu\nu}^\infty {\bar\alpha}_\nu + 1 \right].
\end{eqnarray}
Conditions for this convergence of empirical means to expectation
values are at the heart of learning theory.  Our approach here is
first to assume that this convergence works, then to examine the
consequences for the predictive information, and finally to address
the conditions for and implications of this convergence breaking down.

Putting the different factors together, we obtain
\begin{eqnarray}
&&P(x_{\rm 1}, y_{\rm 1}, x_{\rm 2} , y_{\rm 2}, \cdots , x_{\rm
N} , y_{\rm N }) 
\nonumber
\\ 
&&\,\,\,\,\,{\widetilde\rightarrow} \left[ \prod_{\rm i =1}^{\rm N} 
P(x_{\rm i}) \right] \left( {1\over \sqrt{2\pi\sigma^2}}\right)^{N} 
\int d^K \alpha {\cal P}({\bgm\alpha}) 
\exp\left[
-N E_N({\bgm\alpha} ; \{x_{\rm i} , y_{\rm i}\}) \right] ,
\label{integral2}
\nonumber\\
&&
\end{eqnarray}
where the effective ``energy'' per sample is given by
\begin{eqnarray}
E_N({\bgm\alpha} ; \{x_{\rm i} , y_{\rm i}\}) =
{1\over 2}  + 
{1\over 2} \sum_{\mu , \nu =1}^K (\alpha_\mu - {\bar\alpha}_\mu )
A_{\mu\nu}^\infty (\alpha_\nu - {\bar\alpha}_\nu ) .
\end{eqnarray}
Here we use the symbol ${\widetilde\to}$ to indicate that we not only
take the limit of large $N$, but also neglect the fluctuations. Note
that in this approximation the dependence on the sample points
themselves is hidden in the definition of $\bar{\bgm\alpha}$ as being
the parameters that generated the samples.

The integral that we need to do in Eq.~(\ref{integral2}) involves an
exponential with a large factor $N$ in the exponent; the free energy
$F_N$ is of order unity as $N\rightarrow\infty$.  This suggests that
we evaluate the integral by a saddle point or steepest descent
approximation [similar analyses were performed by Clarke and Barron
(1990), by MacKay (1992), and by Balasubramanian (1997)]:
\begin{eqnarray}
&&
\int d^K \alpha {\cal P}({\bgm\alpha})\exp
\left[-N E_N({\bgm\alpha} ; \{x_{\rm i}, y_{\rm i}\}) \right] 
\nonumber 
\approx 
{\cal P}({\bgm\alpha}_{\rm cl})
\\
&&\,\,\,\,
\times \exp \left[-NE_N({\bgm\alpha}_{\rm cl} ; 
\{x_{\rm i} , y_{\rm i}\})
-{K\over 2} \ln {N\over {2\pi}}  - 
{1\over2} \ln\det {\cal F}_N + \cdots \right],
\end{eqnarray}
where ${\bgm\alpha}_{\rm cl}$ is the ``classical'' value of
${\bgm\alpha}$ determined by the extremal conditions
\begin{eqnarray}
{{\partial E_N({\bgm\alpha} ; \{x_{\rm i} , y_{\rm i}\})}
\over{\partial\alpha_\mu}}\Bigg|_{{\bgm\alpha} ={\bgm\alpha}_{\rm
cl}} = 0 ,
\end{eqnarray}
the matrix ${\cal F}_N$ consists of the second derivatives of $E_N$,
\begin{eqnarray}
{\cal F}_N = 
{{\partial^2 E_N({\bgm\alpha} ; \{x_{\rm i} , y_{\rm i}\})}
\over{\partial\alpha_\mu\partial\alpha_\nu}}\Bigg|_{{\bgm\alpha}
={\bgm\alpha}_{\rm cl}} ,
\end{eqnarray}
and $\cdots$ denotes terms that vanish as $N\rightarrow\infty$.  If we
formulate the problem of estimating the parameters $\bgm\alpha$ from
the samples $\{x_{\rm i}, y_{\rm i}\}$, then as $N\rightarrow\infty$
the matrix $N{\cal F}_N$ is the Fisher information matrix (Cover and
Thomas 1991); the eigenvectors of this matrix give the principal axes
for the error ellipsoid in parameter space, and the (inverse)
eigenvalues give the variances of parameter estimates along each of
these directions. The classical ${\bgm\alpha}_{\rm cl}$ differs from
$\bar{\bgm\alpha}$ only in terms of order $1/N$; we neglect this
difference and further simplify the calculation of leading terms as
$N$ becomes large.  After a little more algebra, then, we find the
probability distribution we have been looking for:
\begin{eqnarray}
&&P(x_{\rm 1}, y_{\rm 1}, x_{\rm 2} , y_{\rm 2}, \cdots , x_{\rm N} , 
y_{\rm N }) 
\nonumber
\\ 
&&\,\,\,\,\,\widetilde\rightarrow 
\left[ \prod_{\rm i =1}^{\rm N} P(x_{\rm i}) \right] 
{1\over {Z_A}} {\cal P}(\bar{\bgm\alpha}) \exp \left[ -{N\over 2} 
\ln(2\pi {\rm e}\sigma^2) - {K\over 2} \ln N + \cdots \right],
\label{finalp}
\end{eqnarray}
where the normalization constant
\begin{eqnarray}
{Z_A} = \sqrt{(2\pi)^K\det A^\infty}.
\end{eqnarray}
Again we note that the sample points $\{ x_{\rm i}, y_{\rm i} \}$ are
hidden in the value of $\bar{\bgm\alpha}$ that gave rise to these
points.\footnote{We emphasize again that there are two approximations
  leading to Eq.~(\ref{finalp}).  First, we have replaced empirical
  means by expectation values, neglecting fluctuations associated with
  the particular set of sample points $\{ x_{\rm i} , y_{\rm i}\}$.
  Second, we have evaluated the average over parameters in a saddle
  point approximation.  At least under some condition, both of these
  approximations would become increasingly accurate as $N\rightarrow
  \infty$, so that this approach should yield the asymptotic behavior
  of the distribution and hence the subextensive entropy at large $N$.
  Although we give a more detailed analysis below, it is worth noting
  here how things can go wrong. The two approximations are
  independent, and we could imagine that fluctuations are important
  but saddle point integration still works, for example. Controlling
  the fluctuations turns out to be exactly the question of whether our
  finite parameterization captures the true dimensionality of the
  class of models, as discussed in the classic work of Vapnik,
  Chervonenkis, and others [see Vapnik (1998) for a review].  The
  saddle point approximation can break down because the saddle point
  becomes unstable or because multiple saddle points become important.
  It will turn out that instability is exponentially improbable as
  $N\rightarrow\infty$, while multiple saddle points are a real
  problem in certain classes of models, again when counting parameters
  doesn't really measure the complexity of the model class.}

To evaluate the entropy $S(N)$ we need to compute the expectation
value of the (negative) logarithm of the probability distribution in
Eq.~(\ref{finalp}); there are three terms.  One is constant, so
averaging is trivial.  The second term depends only on the $x_{\rm
  i}$, and because these are chosen independently from the
distribution $P(x)$ the average again is easy to evaluate.  The third
term involves $\bar{\bgm\alpha}$, and we need to average this over the
joint distribution $P(x_{\rm 1}, y_{\rm 1}, x_{\rm 2} , y_{\rm 2},
\cdots , x_{\rm N} , y_{\rm N })$. As above, we can evaluate this
average in steps: first we choose a value of the parameters
$\bar{\bgm\alpha}$, then we average over the samples given these
parameters, and finally we average over parameters.  But because
$\bar{\bgm\alpha}$ is defined as the parameters that generate the
samples, this stepwise procedure simplifies enormously. The end result
is that
\begin{equation}
S(N) = N\left[ S_x +{1\over 2} \log_2 (2\pi{\rm e}\sigma^2)\right]
+ {K\over 2} \log_2 N + S_{\bgms\alpha} + \langle \log_2 {Z_A}
\rangle_{\bgms \alpha} + \cdots ,
\label{Stoy}
\end{equation}
where $\langle\cdots\rangle_{\bgms \alpha}$ means averaging over
parameters, $S_x$ is the entropy of the distribution of $x$,
\begin{eqnarray}
S_x = -\int dx \,P(x) \log_2 P(x) ,
\end{eqnarray}
and similarly for the entropy of the distribution of parameters,
\begin{eqnarray}
S_{\bgms\alpha} = -\int d^K \alpha \, {\cal P}({\bgm\alpha}) \log_2
{\cal P}({\bgm\alpha}) .
\end{eqnarray}

The different terms in the entropy Eq.~(\ref{Stoy}) have a
straightforward interpretation. First we see that the extensive term
in the entropy,
\begin{eqnarray}
{\cal S}_0 = S_x +{1\over 2} \log_2 (2\pi{\rm e}\sigma^2) ,
\label{extens1}
\end{eqnarray}
reflects contributions from the random choice of $x$ and from the
Gaussian noise in $y$; these extensive terms are independent of the
variations in parameters $\bgm\alpha$, and these would be the only
terms if the parameters were not varying (that is, if there were
nothing to learn).  There also is a term which reflects the entropy of
variations in the parameters themselves, $S_{\bgms\alpha}$.  This
entropy is not invariant with respect to coordinate transformations in
the parameter space, but the term $\langle\log_2
Z_A\rangle_{\bgms\alpha}$ compensates for this noninvariance.
Finally, and most interestingly for our purposes, the subextensive
piece of the entropy is dominated by a logarithmic divergence,
\begin{equation}\label{S1example}
S_1(N) \rightarrow {K\over 2}\log_2 N \;\;\;{\rm (bits)}.
\end{equation}
The coefficient of this divergence counts the number of parameters
independent of the coordinate system that we choose in the parameter
space. Furthermore, this result does not depend on the set of basis
functions $\{\phi_{\mu}(x)\}$. This is a hint that the result in
Eq.~(\ref{S1example}) is more universal than our simple example.

\subsection{Learning a parameterized distribution}
\label{learn_distr_sec}

The problem discussed above is an example of supervised learning: we
are given examples of how the points $x_{\rm n}$ map into $y_{\rm n}$,
and from these examples we are to induce the association or functional
relation between $x$ and $y$.  An alternative view is that pair of
points $(x,y)$ should be viewed as a vector $\vec x$, and what we are
learning is the distribution of this vector.  The problem of learning
a distribution usually is called unsupervised learning, but in this
case supervised learning formally is a special case of unsupervised
learning; if we admit that all the functional relations or
associations that we are trying to learn have an element of noise or
stochasticity, then this connection between supervised and
unsupervised problems is quite general.

Suppose a series of random vector variables $\{{\vec x}_{\rm i}\}$ are
drawn independently from the same probability distribution $Q({\vec x}
| {\bgm\alpha})$, and this distribution depends on a (potentially
infinite dimensional) vector of parameters $\bgm{\alpha}$.  As above,
the parameters are unknown, and before the series starts they are
chosen randomly from a distribution ${\cal P} (\bgm{\alpha})$. With no
constraints on the densities ${\cal P} (\bgm{\alpha})$ or $Q({\vec x}
| {\bgm\alpha})$ it is impossible to derive any regression formulas
for parameter estimation, but one can still calculate the leading
terms in the entropy of the data series and thus the predictive
information. 

We begin with the definition of entropy
\begin{equation}
S(N)\equiv S[\left\{ \vec{x}_{\rm i}\right\}]=
-\int d{\vec x}_{\rm 1}\cdots d{\vec x}_{\rm N} \,
P({\vec x}_{\rm 1} , {\vec x}_{\rm 2} , \cdots , {\vec x}_{\rm N}) 
\,\log_2 P({\vec x}_{\rm 1} , {\vec x}_{\rm 2} , \cdots , {\vec x}_{\rm N}).
\end{equation}
By analogy with Eq.~(\ref{start}) we then write
\begin{eqnarray}
P({\vec x}_{\rm 1} , {\vec x}_{\rm 2} , \cdots , {\vec x}_{\rm N})
= \int d^K\alpha {\cal P} ({\bgm\alpha}) \prod_{\rm i=1}^{\rm N} 
Q({\vec x}_{\rm i} | {\bgm\alpha}) .
\end{eqnarray}
Next, combining the last two equations and rearranging the order of
integration, we can rewrite $S(N)$ as
\begin{eqnarray}
S(N) &=& 
-\int d^K\bar{\bgm\alpha} {\cal P} (\bar{\bgm\alpha}) \left\{
\int d{\vec x}_{\rm 1}\cdots d{\vec x}_{\rm N} \,
\prod_{\rm j=1}^{\rm N} Q({\vec x}_{\rm j} | \bar{\bgm\alpha})
\;\log_2 P(\{{\vec x}_{\rm i}\})
\right\}.
\label{Srewritten}
\end{eqnarray}

Eq.~(\ref{Srewritten}) allows an easy interpretation. There is the
`true' set of parameters $\bar{\bgm\alpha}$ that gave rise to the data
sequence ${\vec x}_{\rm 1}\cdots {\vec x}_{\rm N}$ with the
probability $\prod_{\rm j=1}^{\rm N} Q({\vec x}_{\rm j} |
\bar{\bgm\alpha})$. We need to average $\log_2 P({\vec x}_{\rm
  1}\cdots {\vec x}_{\rm N})$ first over all possible realizations of
the data keeping the true parameters fixed, and then over the
parameters $\bar{\bgm\alpha}$ themselves. With this interpretation in
mind, the joint probability density, the logarithm of which is being
averaged, can be rewritten in the following useful way:
\begin{eqnarray}
P({\vec x}_{\rm 1} , \cdots , {\vec x}_{\rm N})
&=& \prod_{\rm j=1}^{\rm N} Q({\vec x}_{\rm j} | \bar{\bgm\alpha})
\int d^K\alpha {\cal P} ({\bgm\alpha}) \prod_{\rm i=1}^{\rm N} \left[
{{Q({\vec x}_{\rm i} | {\bgm\alpha})}\over {Q({\vec x}_{\rm i} |
\bar{\bgm\alpha})}} \right] 
\nonumber
\\ 
&=& \prod_{\rm j=1}^{\rm N}
Q({\vec x}_{\rm j} | \bar{\bgm\alpha}) 
\int d^K\alpha {\cal P} ({\bgm\alpha})\exp\left[
-N{\cal E}_N({\bgm\alpha} ; \{{\vec x}_{\rm i}\} )\right],
\label{pfn}
\\
{\cal E}_N({\bgm\alpha} ; \{{\vec x}_{\rm i}\} ) 
&=&  - {1\over N} \sum_{\rm i=1}^{\rm N} \ln\left[
{{Q({\vec x}_{\rm i} | {\bgm\alpha})}\over {Q({\vec x}_{\rm i} |
\bar{\bgm\alpha})}} \right] .
\end{eqnarray}
Since, by our interpretation, $\bar{\bgm\alpha}$ are the true
parameters that gave rise to the particular data $\{{\vec x}_{\rm
  i}\}$, we may expect empirical means to converge to expectation
values, so that
\begin{equation}\label{fn}
{\cal E}_N({\bgm\alpha} ; \{{\vec x}_{\rm i}\} ) = 
 - \int d^Dx Q(x|\bar{\bgm\alpha})  \ln\left[
{{Q({\vec x} | {\bgm\alpha})}\over {Q({\vec x} |
\bar{\bgm\alpha})}} \right] 
-\psi ({\bgm\alpha},\bar{\bgm\alpha} ; \{ x_{\rm i} \} ),
\end{equation}
where $\psi \rightarrow 0$ as $N\rightarrow \infty$; here we neglect
$\psi$, and return to this term below.

The first term on the right hand side of Eq.~(\ref{fn}) is the
Kullback--Leibler divergence, $D_{\rm KL}
({\bgm{\bar\alpha}}||{\bgm\alpha})$, between the true distribution
characterized by parameters $\bar{\bgm\alpha}$ and the possible
distribution characterized by $\bgm\alpha$.  Thus at large $N$ we have
\begin{eqnarray}
P({\vec x}_{\rm 1} , {\vec x}_{\rm 2} , \cdots , {\vec x}_{\rm N})
\widetilde\rightarrow \prod_{\rm j=1}^{\rm N} Q({\vec x}_{\rm j} |
\bar{\bgm\alpha}) \int d^K\alpha {\cal P} ({\bgm\alpha})
\exp\left[-ND_{\rm KL} (\bar{\bgm\alpha}||{\bgm\alpha}) \right] ,
\end{eqnarray}
where again the notation $\widetilde\rightarrow$ reminds us that we
are not only taking the limit of large $N$ but also making another
approximation in neglecting fluctuations.  By the same arguments as
above we can proceed (formally) to compute the entropy of this
distribution, and we find
\begin{eqnarray}
S(N) &\approx& {\cal S}_0 \cdot N + S^{({\rm a})}_1(N),
\\ 
{\cal S}_0 &=& \int d^K\alpha {\cal P}({\bgm\alpha}) \left[
-\int d^Dx Q({\vec x}|{\bgm\alpha}) \log_2 Q({\vec
x}|{\bgm\alpha}) \right], \,\,\,\,{\rm and}
\label{S0}
\\ 
S^{({\rm a})}_1(N)
&=& -\int d^K {\bar\alpha} {\cal P}(\bar{\bgm\alpha}) \log_2 \left[ \int
d^K\alpha P({\bgm\alpha}) {\rm e}^{-ND_{\rm KL}
(\bar{\bgms\alpha}||{\bgms\alpha})} \right] . 
\label{S1annealed}
\end{eqnarray}
Here $S^{({\rm a})}_1$ is an approximation to $S_1$ that neglects
fluctuations $\psi$. This is the same as the annealed approximation in
the statistical mechanics of disordered systems, as has been used
widely in the study of supervised learning problems (Seung et al.\ 
1992). Thus we can identify the data sequence ${\vec x}_{\rm
  1}\cdots{\vec x}_{\rm N}$ with the disorder, ${\cal
  E}_N({\bgm\alpha} ; \{{\vec x}_{\rm i}\} )$ with the energy of the
quenched system, and $D_{\rm KL} (\bar{\bgm\alpha}||{\bgm\alpha})$
with its annealed analogue.

The extensive term ${\cal S}_0$, Eq.~(\ref{S0}), is the average
entropy of a distribution in our family of possible distributions,
generalizing the result of Eq. (\ref{extens1}). The subextensive terms
in the entropy are controlled by the $N$ dependence of the partition
function
\begin{equation}\label{Z}
Z ({\bar{\bgm\alpha}} ;N) = \int d^K\alpha {\cal P} ({\bgm\alpha})
\exp\left[ -ND_{\rm KL} ( \bar{\bgm\alpha} || {\bgm\alpha}
)\right] ,
\end{equation}
and $S_1(N) = -\langle\log_2 Z ({\bar{\bgm\alpha}}
;N)\rangle_{\bar{\bgms\alpha}}$ is analogous to the free energy. Since
what is important in this integral is the Kullback--Leibler (KL)
divergence between different distributions, it is natural to ask about
the density of models that are KL divergence $D$ away from the target
$\bar{\bgm\alpha}$,
\begin{eqnarray}
\rho (D;{\bar{\bgm\alpha}}) = \int d^K\alpha  {\cal P} ({\bgm\alpha})
\delta [D - D_{\rm KL} ( \bar{\bgm\alpha} || {\bgm\alpha} )];
\end{eqnarray}
note that this density could be very different for different targets.
The density of divergences is normalized because the original
distribution over parameter space, $P({\bgm\alpha})$, is normalized,
\begin{eqnarray}
\int dD \rho (D;{\bar{\bgm\alpha}}) = \int d^K\alpha
{\cal P} ({\bgm\alpha}) = 1.
\label{normalize}
\end{eqnarray}
Finally, the partition function takes the simple form
\begin{equation}\label{ZD}
Z ({\bar{\bgm\alpha}} ;N) = \int dD \rho (D;{\bar{\bgm\alpha}})
\exp[-ND] .
\end{equation}

We recall that in statistical mechanics the partition function is
given by
\begin{eqnarray}
Z(\beta ) = \int dE \rho(E) \exp[-\beta E],
\end{eqnarray}
where $\rho(E)$ is the density of states that have energy $E$, and
$\beta$ is the inverse temperature.  Thus the subextensive entropy in
our learning problem is analogous to a system in which energy
corresponds to the Kullback--Leibler divergence relative to the target
model, and temperature is inverse to the number of examples.  As we
increase the length $N$ of the time series we have observed, we
``cool'' the system and hence probe models which approach the target;
the dynamics of this approach is determined by the density of low
energy states, that is the behavior of $\rho (D;{\bar{\bgm\alpha}})$
as $D\rightarrow 0$.

The structure of the partition function is determined by a competition
between the (Boltzmann) exponential term, which favors models with
small $D$, and the density term, which favors values of $D$ that can
be achieved by the largest possible number of models. Because there
(typically) are many parameters, there are very few models with $D
\rightarrow 0$.  This picture of competition between the Boltzmann
factor and a density of states has been emphasized in previous work on
supervised learning (Haussler et al. 1996).

The behavior of the density of states, $\rho (D;{\bar{\bgm\alpha}})$,
at small $D$ is related to the more intuitive notion of
dimensionality. In a parameterized family of distributions, the
Kullback--Leibler divergence between two distributions with nearby
parameters is approximately a quadratic form,
\begin{eqnarray}
D_{\rm KL} ( \bar{\bgm\alpha} || {\bgm\alpha} ) \approx {1\over 2}
\sum_{\mu\nu} (\bar{\alpha}_\mu  -\alpha_\mu) {\cal F}_{\mu\nu}
(\bar{\alpha}_\nu  -\alpha_\nu) + \cdots ,
\end{eqnarray}
where ${\cal F}$ is the Fisher information matrix.  Intuitively, if we
have a reasonable parameterization of the distributions, then similar
distributions will be nearby in parameter space, and more importantly
points that are far apart in parameter space will never correspond to
similar distributions; Clarke and Barron (1990) refer to this
condition as the parameterization forming a ``sound'' family of
distributions.  If this condition is obeyed, then we can approximate
the low $D$ limit of the density $\rho (D;{\bar{\bgm\alpha}})$:
\begin{eqnarray}
\rho (D;{\bar{\bgm\alpha}}) &=& \int d^K\alpha {\cal P} ({\bgm\alpha})
\delta [D - D_{\rm KL} ( \bar{\bgm\alpha} || {\bgm\alpha} )]
\nonumber
\\
&\approx& \int d^K\alpha {\cal P} ({\bgm\alpha}) \delta \left[D - {1\over
2} \sum_{\mu\nu} (\bar{\alpha}_\mu  -\alpha_\mu) {\cal F}_{\mu\nu}
(\bar{\alpha}_\nu  -\alpha_\nu)\right]
\nonumber
\\ &=& \int d^K\alpha
{\cal P} (\bar{\bgm\alpha} + {\cal U}\cdot{\bgm\xi}) \delta\left[ D -
{1\over 2} \sum_\mu \Lambda_\mu\xi_\mu^2\right] ,
\label{rhoxi}
\end{eqnarray}
where $\cal U$ is a matrix that diagonalizes $\cal F$,
\begin{eqnarray}
({\cal U}^T \cdot {\cal F} \cdot {\cal U})_{\mu\nu} = \Lambda_\mu
\delta_{\mu\nu} .
\end{eqnarray}
The delta function restricts the components of $\bgm\xi$ in
Eq.~(\ref{rhoxi}) to be of order $\sqrt{D}$ or less, and so if
$P({\bgm\alpha})$ is smooth we can make a perturbation expansion.
After some algebra the leading term becomes
\begin{eqnarray}
\rho (D\rightarrow 0;{\bar{\bgm\alpha}}) \approx
{\cal P} ({\bar{\bgm\alpha}}) \frac{2 \pi^{K/2}}{\Gamma(K/2)} 
\left(\det \cal F\right)^{-1/2} D^{(K-2)/2} . 
\label{rho-finiteKclass}
\end{eqnarray}
Here, as before, $K$ is the dimensionality of the parameter vector.
Computing the partition function from Eq. (\ref{ZD}), we find
\begin{equation}
Z ({\bar{\bgm\alpha}} ;N\rightarrow\infty) \approx
f({\bar{\bgm\alpha}}) \cdot {{\Gamma(K/2)}\over{N^{K/2}}} ,
\label{Zfinite}
\end{equation}
where $f({\bar{\bgm\alpha}})$ is some function of the target parameter
values. Finally, this allows us to evaluate the subextensive entropy,
from Eqs. (\ref{S1annealed}, \ref{Z}):
\begin{eqnarray}
S^{({\rm a})}_1(N) &=& - \int d^K{\bar\alpha} {\cal P}
(\bar{\bgm\alpha}) \log_2
Z(\bar{\bgm\alpha} ; N )
\label{s1thruZ}
\\
&\rightarrow& {K\over2}
\log_2 N + \cdots\;\;\; {\rm (bits)},
\label{s1distr}
\end{eqnarray}
where $\cdots$ are finite as $N \rightarrow \infty$. Thus, general
$K$--parameter model classes have the same subextensive entropy as for
the simplest example considered in the previous section.  To the
leading order, this result is independent even of the prior
distribution ${\cal P} ({\bgm\alpha})$ on the parameter space, so that
the predictive information seems to count the number of parameters
under some very general conditions [cf.\ Fig.~(\ref{entr_subext}) for
a numerical example of the logarithmic behavior].

Although Eq.~(\ref{s1distr}) is true under a wide range of conditions,
this cannot be the whole story.  Much of modern learning theory is
concerned with the fact that counting parameters is not quite enough
to characterize the complexity of a model class; the naive dimension
of the parameter space $K$ should be viewed in conjunction with the
Vapnik--Chervonenkis (VC) dimension $d_{\rm VC}$ (also known as the
pseudodimension) and the phase space dimension $d$. The phase space
dimension is defined in the usual way through the scaling of volumes
in the model space (see, for example, Opper 1994). On the other hand,
$d_{\rm VC}$ measures not volumes, but capacity of the model class,
and its definition is a bit trickier: for a set of binary (indicator)
functions $F(\vec{x},{\bgm\alpha})$, VC dimension is defined as the
maximal number of vectors $\vec{x}_1 \cdots \vec{x}_{d_{\rm VC}}$ that
can be classified into two different classes in all $2^{d_{\rm VC}}$
possible ways using this set of functions. Similarly, for real--valued
functions $F(\vec{x},{\bgm\alpha})$ one can first define a complete
set of indicators using step functions, $\theta\left[
  F(\vec{x},{\bgm\alpha})-\beta\right]$, and then the VC dimension of
this set is the VC dimension of the real--valued functions (Vapnik
1998). Separation of a vector in all possible ways is called
shattering, and hence another name for the VC dimension---the
shattering dimension.

Both $d$ and $d_{\rm VC}$ can differ from the number of parameters in
several ways. One possibility is that $d_{\rm VC}$ is infinite when
the number of parameters is finite, a problem discussed below.
Another possibility is that the determinant of ${\cal F}$ is zero, and
hence $d_{\rm VC}$ and $d$ are both smaller than the number of
parameters because we have adopted a redundant description.  It is
possible that this sort of degeneracy occurs over a finite fraction
but not all of the parameter space, and this is one way to generate an
effective fractional dimensionality.  One can imagine multifractal
models such that the effective dimensionality varies continuously over
the parameter space, but it is not obvious where this would be
relevant. Finally, models with $d<d_{\rm VC}<\infty$ are also possible
[see, for example, Opper (1994)], and this list probably is not
exhaustive.

The calculation above, Eq.~(\ref{rho-finiteKclass}), lets us actually
{\em define} the phase space dimension through the exponent in the
small $D_{\rm KL}$ behavior of the model density,
\begin{eqnarray}
\rho(D\rightarrow 0; \bgm{\bar\alpha}) \propto D^{(d-2)/2} ,
\end{eqnarray}
and then $d$ appears in place of $K$ as the coefficient of the log
divergence in $S_1(N)$ (Clarke and Barron 1990, Opper 1994). However,
this simple conclusion can fail in two ways. First, it can happen that
a macroscopic weight gets accumulated at some nonzero value of $D_{\rm
  KL}$, so that the small $D_{\rm KL}$ behavior is irrelevant for the
large $N$ asymptotics.  Second, the fluctuations neglected here may be
uncontrollably large, so that the asymptotics are never reached. Since
controllability of fluctuations is a function of $d_{\rm VC}$ (see
Vapnik 1998 and later in this paper), we may summarize this in the
following way. Provided that the small $D_{\rm KL}$ behavior of the
density function is the relevant one, the coefficient of the
logarithmic divergence of $I_{\rm pred}$ measures the phase space or
the scaling dimension $d$ and nothing else. This asymptote is valid,
however, only for $N \gg d_{\rm VC}$. It is still an open question
whether the two pathologies that can violate this asymptotic behavior
are related.

\subsection{Learning a parameterized process}
\label{learn_proc_sec}

Consider a process where samples are not independent, and our task is
to learn their joint distribution $Q( {\vec x}_1 , \cdots, {\vec
  x}_{\rm N}| {\bgm\alpha})$. Again, ${\bgm\alpha}$ is an unknown
parameter vector which is chosen randomly at the beginning of the
series. If ${\bgm\alpha}$ is a $K$ dimensional vector, then one still
tries to learn just $K$ numbers and there are still $N$ examples, even
if there are correlations. Therefore, although such problems are much
more general than those considered above, it is reasonable to expect
that the predictive information is still measured by $(K/2)\log_2 N$
provided that some conditions are met.

One might suppose that conditions for simple results on the predictive
information are very strong, for example that the distribution $Q$ is
a finite order Markov model.  In fact all we really need are the
following two conditions:
\begin{eqnarray}
S\left[\{ {\vec x}_{\rm i}\}|{\bgm\alpha}\right] &\equiv& - \int
d^N{\vec x}\, Q(\{{\vec x}_{\rm i}\} | {\bgm\alpha})\, \log_2
Q(\{{\vec x}_{\rm i}\}| {\bgm\alpha})\nonumber\\
 &\to& N 
{\cal S}_0 + {\cal S}^*_0; \;\;\;\;\;\;\; {\cal S}^*_0 = O(1)\;
\label{S0markov},
\\ 
D_{\rm KL} \left[ Q(\{{\vec x}_{\rm i}\} |
\bar{\bgm\alpha})||Q(\{{\vec x}_{\rm i}\} | {\bgm\alpha})\right]
&\to & N {\cal D}_{\rm KL}\left( \bar{\bgm\alpha}||{\bgm\alpha}\right) +
o(N)\;.
\label{Dmarkov}
\end{eqnarray}
Here the quantities ${\cal S}_0$, ${\cal S}^*_0$, and ${\cal D}_{\rm
  KL}\left( \bar{\bgm\alpha}||{\bgm\alpha}\right)$ are defined by
taking limits $N\to\infty$ in both equations.  The first of the
constraints limits deviations from extensivity to be of order unity,
so that if ${\bgm\alpha}$ is known there are no long range
correlations in the data---all of the long range predictability is
associated with learning the parameters.\footnote{Suppose that we
  observe a Gaussian stochastic process and we try to learn the power
  spectrum.  If the class of possible spectra includes ratios of
  polynomials in the frequency (rational spectra) then this condition
  is met.  On the other hand, if the class of possible spectra
  includes $1/f$ noise, then the condition may not be met.  For more
  on long range correlations, see below.}  The second constraint, Eq.
(\ref{Dmarkov}), is a less restrictive one, and it ensures that the
``energy'' of our statistical system is an extensive quantity.

With these conditions it is straightforward to show that the results
of the previous subsection carry over virtually unchanged. With the
same cautious statements about fluctuations and the distinction
between $K$, $d$, and $d_{\rm VC}$, one arrives at the result:
\begin{eqnarray}
S(N) &=& {\cal S}_0 \cdot N + S^{({\rm a})}_1(N)\;,
\\
S^{({\rm a})}_1(N)&=& \frac{K}{2} \log_2 N + \cdots \;\;\;
({\rm bits}) ,
\label{s1mark}
\end{eqnarray}
where $\cdots$ stands for terms of order one. Note again that for the
results Eq.~(\ref{s1mark}) to be valid, the process considered is not
required to be a finite order Markov process.  Memory of all previous
outcomes may be kept, provided that the accumulated memory does not
contribute a divergent term to the subextensive entropy.

It is interesting to ask what happens if the condition in Eq.
(\ref{S0markov}) is violated, so that there are long range
correlations even in the conditional distribution $Q( {\vec x}_1 ,
\cdots, {\vec x}_{\rm N}| {\bgm\alpha})$.  Suppose, for example, that
${\cal S}^*_0=(K^*/2) \log_2 N$. Then the subextensive entropy becomes
\begin{equation}
S^{({\rm a})}_1(N)= \frac{K+K^*}{2} \log_2 N + \cdots \;\;\;
({\rm bits}) .
\label{s1markcorr}
\end{equation}
We see the that the subextensive entropy makes no distinction between
pre\-dic\-ta\-bil\-ity that comes from unknown parameters and
predictability that comes from intrinsic correlations in the data; in
this sense, two models with the same $K+K^*$ are equivalent.  This,
actually, {\em must} be so.  As an example, consider a chain of Ising
spins with long range interactions in one dimension.  This system can
order (magnetize) and exhibit long range correlations, and so the
predictive information will diverge at the transition to ordering.  In
one view, there is no global parameter analogous to $\bgm \alpha$,
just the long range interactions. On the other hand, there are regimes
in which we can approximate the effect of these interactions by saying
that all the spins experience a mean field which is constant across
the whole length of the system, and then formally we can think of the
predictive information as being carried by the mean field itself.  In
fact there are situations in which this is not just an approximation,
but an exact statement.  Thus we can trade a description in terms of
long range interactions ($K^* \neq 0$, but $K=0$) for one in which
there are unknown parameters describing the system but given these
parameters there are no long range correlations ($K\neq 0, \, K^*
=0$).  The two descriptions are equivalent, and this is captured by
the subextensive entropy.\footnote{There are a number of interesting
  questions about how the coefficients in the diverging predictive
  information relate to the usual critical exponents, and we hope to
  return to this problem in a later paper.}


\subsection{Taming the fluctuations: finite $d_{\rm VC}$ case}

The preceding calculations of the subextensive entropy $S_{\rm 1}$ are
worthless unless we prove that the fluctuations $\psi$ are
controllable. In this subsection we are going to discuss when and if
this, indeed, happens. We limit the discussion to the analysis of
fluctuations in the case of finding a probability density (Section
\ref{learn_distr_sec}); the case of learning a process (Section
\ref{learn_proc_sec}) is very similar.

Clarke and Barron (1990) solved essentially the same problem. They did
not make a separation into the annealed and the fluctuation term, and
the quantity they were interested in was a bit different from ours,
but, interpreting loosely, they proved that, modulo some reasonable
technical assumptions on differentiability of functions in question,
the fluctuation term always approaches zero. However, they did not
investigate the speed of this approach, and we believe that, by doing
so, they missed some important qualitative distinctions between
different problems that can arise due to a difference between $d$ and
$d_{\rm VC}$. In order to illuminate these distinctions, we here go
through the trouble of analyzing fluctuations all over again.

Returning to Eqs.~(\ref{pfn}, \ref{fn}) and the definition of entropy,
we can write the entropy $S(N)$ exactly as
\begin{eqnarray}
S(N)&=& - \int d^K\bar{\alpha} {\cal P} (\bar{\bgm\alpha})\int \prod_{j=1}^N 
\left[ d {\vec x}_{\rm j}\,
Q({\vec x}_{\rm j} | \bar{\bgm\alpha})\right]\nonumber\\
&&\times \log_2 \left[ \prod_{i=1}^N Q({\vec x}_{\rm i}|\bar{\bgm
\alpha}) 
\int  d^K\alpha
{\cal P} ({\bgm\alpha})\; {\rm e}^{-ND_{\rm KL}
(\bar{\bgms\alpha}||{\bgms\alpha}) + N \psi({\bgms
\alpha},\bar{\bgms\alpha};\{{\vec x}_{\rm i}\})}\right] .
\end{eqnarray}
This expression can be decomposed into the terms identified above,
plus a new contribution to the subextensive entropy that comes from
the fluctuations alone, $S_1^{\rm (f)}(N)$:
\begin{eqnarray}
S(N) &=& {\cal S}_0 \cdot N + S_1^{\rm (a)}(N) + S_1^{\rm (f)}(N),
\\
S^{({\rm f})}_1 &=&  - \int d^K {\bar\alpha}
{\cal P} (\bar{\bgm\alpha})
\,\prod_{j=1}^N \left[ 
d{\vec x}_{\rm j} \,Q({\vec x}_{\rm j}|\bar{\bgm \alpha})\right]
\nonumber
\\
&&\times\log_2 \left[ \int \frac{d^K\alpha
{\cal P} ({\bgm\alpha})}{Z(\bar{\bgm\alpha};N)} {\rm e}^{-ND_{\rm KL}
(\bar{\bgms\alpha}||{\bgms\alpha}) + N \psi({\bgms
\alpha},\bar{\bgms\alpha};\{{\vec x}_{\rm i}\})}\right] ,
\label{Sf}
\end{eqnarray}
where $\psi$ is defined as in Eq.~(\ref{fn}), and $Z$ as in
Eq.~(\ref{Z}).

Some loose but useful bounds can be established.  First, the
predictive information is a positive (semidefinite) quantity, and so
the fluctuation term may not be smaller than the value of $- S_{\rm
  1}^{\rm (a)}$ as calculated in Eqs. (\ref{s1distr}, \ref{s1mark}).
Second, since fluctuations make it more difficult to generalize from
samples, the predictive information should {\em always} be reduced by
fluctuations, so that $S^{({\rm f})}$ is negative.  This last
statement corresponds to the fact that for the statistical mechanics
of disordered systems, the annealed free energy always is less than
the average quenched free energy, and may be proven rigorously by
applying Jensen's inequality to the (concave) logarithm function in
Eq.~(\ref{Sf}); essentially the same argument was given by Opper and
Haussler (1995).  A related Jensen's inequality argument allows us to
show that the total $S_1(N)$ is bounded,
\begin{eqnarray}
S_1(N) &\leq& N \int d^K \alpha \int d^K \bar\alpha  
{\cal P} ({\bgm \alpha}) {\cal P} ({\bgm {\bar\alpha}}) 
D_{\rm KL} ( {\bgm{\bar\alpha}}||{\bgm\alpha}) 
\nonumber
\\
&\equiv& 
\langle D_{\rm KL} ( {\bgm{\bar\alpha}}||{\bgm\alpha})  
\rangle_{{\bgms{\bar\alpha}},{\bgms\alpha}},
\label{boundbymeanKL}
\end{eqnarray}
so that if we have a class of models (and a prior ${\cal P} ({\bgm
  \alpha})$) such that the average Kullback--Leibler divergence among
pairs of models is finite, then the subextensive entropy is
necessarily properly defined.  Note that $\langle D_{\rm KL}
({\bgm{\bar\alpha}}||{\bgm\alpha})
\rangle_{{\bgms{\bar\alpha}},{\bgms\alpha}}$ includes ${\cal S}_0$ as
one of its terms, so that usually ${\cal S}_0$ and $S_1$ are well-- or
ill--defined together.

Tighter bounds require nontrivial assumptions about the classes of
distributions considered. The fluctuation term would be zero if $\psi$
were zero, and $\psi$ is the difference between an expectation value
(KL divergence) and the corresponding empirical mean. There is a broad
literature that deals with this type of difference (see, for example,
Vapnik 1998).

We start with the case when the pseudo-dimension ($d_{\rm VC}$) of the
set of probability densities $\{Q({\vec x}|{\bgm\alpha})\}$ is finite.
Then for any reasonable function $F({\vec x}; \beta)$, deviations of
the empirical mean from the expectation value can be bounded by
probabilistic bounds of the form
\begin{eqnarray}
&&P\left\{ \sup_{\beta} \left| \frac{\frac{1}{N} \sum_{\rm j}
F({\vec x}_{\rm j}; \beta) - \int d {\vec x}\, Q({\vec
x}|\bar{\bgm \alpha}) \,F({\vec x}; \beta)}{L[F]} \right| >
\epsilon \right\}
\nonumber
\\
&&\,\,\,\,\,\,\;\;\;\;\;\,\,\,\,\,\,
< M(\epsilon, N,d_{\rm VC}) {\rm e}^{-c
N\epsilon^2}\, ,
\label{GC}
\end{eqnarray}
where $c$ and $L[F]$ depend on the details of the particular bound
used.  Typically, $c$ is a constant of order one, and $L[F]$ is either
some moment of $F$ or the range of its variation. In our case, $F$ is
the log--ratio of two densities, so that $L[F]$ may be assumed bounded
for almost all $\beta$ without loss of generality in view of
Eq.~(\ref{boundbymeanKL}).  In addition, $M(\epsilon, N, d_{\rm VC})$
is finite at zero, grows at most subexponentially in its first two
arguments, and depends exponentially on $d_{\rm VC}$.  Bounds of this
form may have different names in different contexts:
Glivenko--Cantelli, Vapnik--Chervonenkis, Hoeffding, Chernoff, ...;
for review see Vapnik (1998) and the references therein.

To start the proof of finiteness of $S^{({\rm f})}_1$ in this case, we
first show that only the region ${\bgm\alpha}\approx\bar{\bgm\alpha}$
is important when calculating the inner integral in Eq.~(\ref{Sf}).
This statement is equivalent to saying that at large values of
${\bgm\alpha} - \bar{\bgm\alpha}$ the KL divergence almost always
dominates the fluctuation term, that is, the contribution of sequences
of $\{{\vec x}_{\rm i}\}$ with atypically large fluctuations is
negligible (atypicality is defined as $\psi\ge \delta$, where $\delta$
is some small constant independent of $N$). Since the fluctuations
decrease as $1/\sqrt{N}$ [see Eq. (\ref{GC})], and $D_{\rm KL}$ is of
order one, this is plausible. To show this, we bound the logarithm in
Eq.~(\ref{Sf}) by $N$ times the supremum value of $\psi$.  Then we
realize that the averaging over $\bar{\bgm\alpha}$ and $\{{\vec
  x}_{\rm i}\}$ is equivalent to integration over all possible values
of the fluctuations. The worst case density of the fluctuations may be
estimated by differentiating Eq.~(\ref{GC}) with respect to $\epsilon$
(this brings down an extra factor of $N\epsilon$). Thus the worst case
contribution of these atypical sequences is
\begin{eqnarray}
S^{({\rm f}),{\rm atypical}}_1 \sim \int_{\delta}^{\infty}
d\epsilon\, N^2 \epsilon^2 M(\epsilon) {\rm e}^{-c N \epsilon^2}
\sim {\rm e}^{-cN\delta^2} \ll 1\; {\rm for \;large\; } N .
\end{eqnarray}

This bound lets us focus our attention on the region
${\bgm\alpha}\approx\bar{\bgm\alpha}$. We expand the exponent of the
integrand of Eq. (\ref{Sf}) around this point and perform a simple
Gaussian integration. In principle, large fluctuations might lead to
an instability (positive or zero curvature) at the saddle point, but
this is atypical and therefore is accounted for already. Curvatures at
the saddle points of both numerator and denominator are of the same
order, and throwing away unimportant additive and multiplicative
constants of order unity, we obtain the following result for the
contribution of typical sequences:
\begin{eqnarray}
S^{({\rm f}),{\rm typical}}_1 &\sim& \int d^K {\bar\alpha}
{\cal P} (\bar{\bgm\alpha}) \,d^N{\vec x} \prod_{\rm j} Q({\vec x}_{\rm
j}|\bar{\bgm \alpha})\; N\; ({\bf B} \,{\cal A}^{-1}\, {\bf
B})\;;\label{chi2}
\\
B_{\mu}&=&\frac{1}{N}\sum_{\rm i} \frac{\partial \log Q({\vec x}_{\rm
i}|\bar{\bgm\alpha})}{\partial
\bar{\alpha}_{\mu}}\;,\;\;\; \langle{\bf B}\rangle_{\vec x}=0\;;
\nonumber
\\
({\cal A})_{\mu \nu}&=&\frac{1}{N}\sum_{\rm i}
\frac{\partial^2 \log Q({\vec x}_{\rm
i}|\bar{\bgm\alpha})}{\partial \bar{\alpha}_{\mu} \partial
\bar{\alpha}_{\nu}} \;,\;\;\; \langle{\cal A}\rangle_{\vec x}={\cal
F}\;.
\nonumber
\end{eqnarray}
Here $\langle \cdots \rangle_{\vec x}$ means an averaging with respect
to all ${\vec x}_{\rm i}$'s keeping $\bar{\bgm\alpha}$ constant. One
immediately recognizes that ${\bf B}$ and ${\cal A}$ are,
respectively, first and second derivatives of the empirical KL
divergence that was in the exponent of the inner integral in Eq.
(\ref{Sf}).

We are dealing now with typical cases. Therefore, large deviations of
${\cal A}$ from ${\cal F}$ are not allowed, and we may bound Eq.
(\ref{chi2}) by replacing ${\cal A}^{-1}$ with ${\cal F}^{-1}
(1+\delta)$, where $\delta$ again is independent of $N$.  Now we have
to average a bunch of products like
\begin{eqnarray}
\frac{\partial \log Q({\vec x}_{\rm i}|\bar{\bgms \alpha})}
{\partial \bar{\alpha}_{\mu}} ({\cal F}^{-1})_{\mu\nu} \frac{\partial
  \log Q({\vec x}_{\rm j}|\bar{\bgms\alpha})} {\partial
  \bar{\alpha}_{\nu}}
\end{eqnarray}
over all ${\vec x}_{\rm i}$'s. Only the terms with ${\rm i}={\rm j}$
survive the averaging. There are $K^2N$ such terms, each contributing
of order $N^{-1}$. This means that the total contribution of the
typical fluctuations is bounded by a number of order one and does not
grow with $N$. This concludes the proof of controllability of
fluctuations for $d_{\rm VC}<\infty$.

\subsection{Taming the fluctuations: the role of the prior}
\label{fluct_prior_sec}

One may notice that we never used the specific form of
$M(\epsilon,N,d_{\rm VC})$, which is the only thing dependent on the
precise value of the dimension.  Actually, a more thorough look at the
proof shows that we do not even need the strict uniform convergence
enforced by the Glivenko--Cantelli bound. With some modifications the
proof should still hold if there exist some {\em a priori} improbable
values of ${\bgm\alpha}$ and $\bar{\bgm\alpha}$ that lead to violation
of the bound. That is, if the prior ${\cal P} ({\bgm\alpha})$ has
sufficiently narrow support, then we may still expect fluctuations to
be unimportant even for VC--infinite problems.

To see this, consider two examples.  A variable $x$ is distributed
according to the following probability density functions:
\begin{eqnarray}
Q(x|\alpha) &=& \frac{1}{\sqrt{2\pi}} \exp \left[ -\frac{1}{2}
\left(x-\alpha\right)^2\right]\;, \;\; x\in (-\infty;+\infty)\;;
\\
Q(x|\alpha) &=& \frac{\exp \left( - \sin \alpha x
  \right)}{\int_0^{2\pi} dx\,\exp \left( - \sin \alpha x
  \right)}\;,\;\;
x\in [0; 2\pi)\;.
\label{expsin}
\end{eqnarray}
Learning the parameter in the first case is a $d_{\rm VC}=1$ problem,
while in the second case $d_{\rm VC}=\infty$.  In the first example,
as we have shown above, one may construct a uniform bound on
fluctuations irrespective of the prior ${\cal P} ({\bgm\alpha})$. The
second one does not allow this. Indeed, suppose that the prior is
uniform in a box $0<\alpha < \alpha_{\rm max}$, and zero elsewhere,
with $\alpha_{\rm max}$ rather large.  Then for not too many sample
points $N$, the data would be better fitted not by some value in the
vicinity of the actual parameter, but by some much larger value, for
which almost all data points are at the crests of $-\sin \alpha x$.
Adding a new data point would not help, until that best, but wrong,
parameter estimate is less than $\alpha_{\rm
  max}$.\footnote{Interestingly, since for the model
  Eq.~(\ref{expsin}) KL divergence is bounded from below and {\em
    above}, for $\alpha_{\rm max} \to \infty$ the weight in
  $\rho(D;{\bar{\bgm\alpha}})$ at small $D_{\rm KL}$ vanishes, and a
  finite weight accumulates at some nonzero value of $D$. Thus, even
  putting the fluctuations aside, the asymptotic behavior based on the
  phase space dimension is invalidated, as mentioned above.} So the
fluctuations are large, and the predictive information is small in
this case.  Eventually, however, data points would overwhelm the box
size, and the best estimate of $\alpha$ would swiftly approach the
actual value. At this point the argument of Clarke and Barron (1990)
would become applicable, and the leading behavior of the subextensive
entropy would converge to its asymptotic value of $(1/2) \log N$. On
the other hand, there is no uniform bound on the value of $N$ for
which this convergence will occur---it is guaranteed only for $N\gg
d_{\rm VC}$, which is never true if $d_{\rm VC}=\infty$. For some
sufficiently wide priors this asymptotically correct behavior would be
never reached in practice.  Further, if we imagine a thermodynamic
limit where the box size and the number of samples both become large,
then by analogy with problems in supervised learning (Seung et al.
1992, Haussler et al. 1996) we expect that there can be sudden changes
in performance as a function of the number of examples.  The arguments
of Clarke and Barron cannot encompass these phase transitions or
``aha!'' phenomena.

Following the intuition inferred from this example, we can now proceed
with a more formal analysis. As the above argument about the smallness
of fluctuations in the finite $d_{\rm VC}$ case paralleled the
discussion of the Empirical Risk Minimization (ERM) approach (Vapnik
1998), this present argument closely resembles some statements of the
Structural Risk Minimization (SRM) theory (Vapnik 1998), which deals
with the case of $d_{\rm VC}=\infty$ or, equivalently, $N/d_{\rm VC} <
1$. While ERM solves the problem of uniform non--Bayesian learning,
there seems to be a general agreement that SRM theory is a solution to
the problem of learning with a prior. However, to our knowledge, no
explicit identification of why this is so has been done, so we try
to do it here.

Suppose that, as in the above example, admissible solutions of a
learning problem belong to some subset $C_1$ of the whole
$K$--dimensional parameter space $C$.  Suppose also that for any finite
$C_1$ the VC dimension of the corresponding learning problem, $d_{\rm
  VC} (C_1)$, is finite, but $d_{\rm VC} (C)=\infty$.  In SRM theory a
nested set of such subspaces $C_1 \subset C_2 \subset C_3 \subset
\cdots$ is called a {\em structure} ${\cal C}$ if $C= \bigcup C_{\rm
  n}$. Each $C_n$ is known as a {\em structure element}. Since the
subsets are nested, $d_{\rm VC}(C_1) \le d_{\rm VC}(C_2) \le d_{\rm
  VC}(C_3) \le \cdots$.  We know that these are the large VC
dimensions and, therefore, parameters that belong to the large
structure elements $C_n,\,n\to\infty$, that are responsible for large
fluctuations. But in view of Eq.~(\ref{normalize}), for any properly
defined prior ${\cal P} ({\bgm\alpha})$, very large values of
${\bgm\alpha}$ are a priori improbable. Thus the fight between the
prior and the data may result in an effective cutoff $n^*$, so that
all $C_n, n>n^*,$ contribute little to $S_1^{(f)}$, and the
fluctuations are controlled.

Indeed, let's form a structure by assigning all ${\bgm\alpha}$'s for
which $-\log{\cal P} ({\bgm\alpha}) + \max\log{\cal P} \le n$ to the
element $C_n$ ($n$ is not necessarily integer). This imposes an a
priori probability $\nu(n)$ on the elements themselves. Now we can
bound the internal integral in Eq.~(\ref{Sf}) by replacing
$\psi({\bgm\alpha},\bar{\bgm\alpha},\{\vec{x}_{\rm i}\})$ with
$\psi_n(\bar{\bgm\alpha},\{\vec{x}_{\rm i}\})$---its maximal value on
the smallest element $C_n$ that includes ${\bgm\alpha}$. If the
logarithm of the a priori probability $\nu(n)$ falls off faster than
$N\psi_n(\bar{\bgm\alpha},\{\vec{x}_{\rm i}\})$ increases as $n$
grows, then one can select a particular $n^*$, for which the integral
over all $C_n,\, n>n^*,$ is smaller than any predefined $\delta$.
Effectively $n^*$ then serves as a cutoff. Note that, since
fluctuations enter multiplied by $N$, $n^* (N)$ is a nondecreasing
function. If it grows in a way such that $d_{\rm VC} (C_{n^*})$ is
sublinear in $N$ ($\sim N/ \log N$ suffices), then
$M(\epsilon,N,d_{\rm VC})$ is still subexponential, and we can use the
proofs of the preceding section to show that the fluctuations are
controllable.  The only difference that occurs is that the
contribution of typical fluctuations is dominated by a saddle point
near ${\bgm\alpha}_{\rm cl}$, which solves the equation
\begin{equation}
\left.\frac{\partial}{\partial \alpha_{\mu}}
\right|_{{\bgm\alpha}_{\rm cl}}
\left[
 -\log {\cal P} ({\bgm\alpha})  + 
N D(\bar{\bgm\alpha}||{\bgm\alpha}) 
\right]=0.
\end{equation}
If $\bar{\bgm\alpha}$ is only in very large structure elements that
contribute little to the internal integral of Eq.~(\ref{Sf}), then
${\bgm\alpha}_{\rm cl}$ may be quite far from $\bar{\bgm\alpha}$. That
is, the best estimate of $\bar{\bgm\alpha}$ may be imprecise at any
finite $N$. This is particularly important in the case of
nonparametric learning (see Sections~\ref{nonparam_ex_sec},
\ref{sm_scale_sec}).

In finite dimensional cases similar to the above example, every $C_n,
\, n<\infty,$ has finite VC dimension $d_{\rm VC}$, and this dimension
is bounded from above by the phase space dimension $d$.  The magnitude
of fluctuations depends mostly on $d_{\rm VC}$. Therefore, beyond some
$n^*(N)$ for which $d_{\rm VC}(C_{n^*})=d$, the fluctuations will
practically stop growing. This means that any proper prior ${\cal P}$,
however slowly decreasing at infinities, is enough to impose a finite
cutoff and render fluctuations finite. This is in complete agreement
with Clark and Barron---but prior-dependent.

We want to emphasize again that, in general, fluctuations are
controlled only if two related, but not equivalent, assumptions are
true. First, for any finite $N$ there has to be a finite cutoff
$n^*(N)$. This means that ${\cal P} ({\bgm\alpha})$ is narrow enough
to define a valid structure.  Second, for the fluctuations within
$C_{n^*}$ to be small, $d_{\rm VC}(C_{n^*(N)})$ must grow sublinearly
in $N$.~\footnote{Actually, the $n^*$-dependence of the factors similar
  to $L[F]$, defined above, may require a different, yet slower,
  growth [see Vapnik (1998) for details]. But this is outside the
  scope of this discussion.}  In this case the number of samples
eventually outgrows the current VC dimension by an arbitrarily large
factor, and determination of parameters is possible to any precision.
Both of these conditions are well known in SRM theory (Vapnik 1998).

In the classical SRM theory, only selection of the law $n^*=n^*(N)$ is
a part of the problem, and the structure is usually assumed to be
given.  Ideally, this law is selected by minimizing the expected error
of learning, which consists of uncertainties due to the limited set of
allowed solutions ($n^* <\infty$) and due to the fluctuations within
this set. These uncertainties behave oppositely as $n^*$ increases. If
calculating the expected error is difficult, people may be content
with even preselecting the law $n^*=n^*(N)$, and then every law for
which the VC dimension grows sublinearly does the job---better or
worse---just as we have shown above.  In our current treatment the
structure {\em and} the law of the VC dimension growth are both a
result of the prior. If the prior is appropriate, then so are the
structure and the law. If not, then learning with this prior is
impossible. On general grounds, we know that when the prior correctly
embodies the a priori knowledge, it results in the fastest average
learning possible.  Therefore we are {\em guaranteed} that, on
average, the law $n^*=n^*(N)$ is optimal if this law is imposed by the
prior (see Sections~\ref{wrong_pr_sec}, \ref{sm_scale_sec} for more on
this).

Summarizing, we note that while much of learning theory has properly
focused on problems with finite VC dimension, it might be that the
conventional scenario in which the number of examples eventually
overwhelms the number of parameters or dimensions is too weak to deal
with many real world problems.  Certainly in the present context there
is not only a quantitative, but also a qualitative difference between
reaching the asymptotic regime in just a few measurements, or in many
millions of them. Finitely parameterizable models with finite or
infinite $d_{\rm VC}$ fall in {\em essentially different universality
  classes} with respect to the predictive information.


\subsection{Beyond finite parameterization: general considerations}
\label{nonparam_gen_sec}

The previous sections have considered learning from time series where
the underlying class of possible models is described with a finite
number of parameters.  If the number of parameters is not finite then
in principle it is impossible to learn anything unless there is some
appropriate regularization of the problem.  If we let the number of
parameters stay finite but become large, then there is {\em more} to
be learned and correspondingly the predictive information grows in
proportion to this number, as in Eq.~(\ref{s1distr}). On the other
hand, if the number of parameters becomes infinite without
regularization, then the predictive information should go to zero
since nothing can be learned.  We should be able to see this happen in
a regularized problem as the regularization weakens: eventually the
regularization would be insufficient and the predictive information
would vanish.  The only way this can happen is if the subextensive
term in the entropy grows more and more rapidly with $N$ as we weaken
the regularization, until finally it becomes extensive at the point
where learning becomes impossible. More precisely, if this scenario
for the breakdown of learning is to work, there must be situations in
which the predictive information grows with $N$ more rapidly than the
logarithmic behavior found in the case of finite parameterization.

Subextensive terms in the entropy are controlled by the density of
models as function of their Kullback--Leibler divergence to the target
model.  If the models have finite VC and phase space dimensions then
this density vanishes for small divergences as $\rho \sim
D^{(d-2)/2}$.  Phenomenologically, if we let the number of parameters
increase, the density vanishes more and more rapidly. We can imagine
that beyond the class of finitely parameterizable problems there is a
class of regularized infinite dimensional problems in which the
density $\rho (D \rightarrow 0)$ vanishes more rapidly than any power
of $D$.  As an example, we could have
\begin{eqnarray}
\rho (D \rightarrow 0) \approx A \exp\left[
-{B\over{D^\mu}}\right] ,\,\,\,\,\,\,\, \mu>0 ;
\end{eqnarray}
that is, an essential singularity at $D=0$. For simplicity we assume
that the constants $A$ and $B$ can depend on the target model, but
that the nature of the essential singularity ($\mu$) is the same
everywhere.  Before providing an explicit example, let us explore the
consequences of this behavior.

From Eq.~(\ref{ZD}) above, we can write the partition function as
\begin{eqnarray}
Z (\bar{\bgm\alpha} ; N ) &=& \int dD \rho (D;{\bar{\bgm\alpha}})
\exp[-ND]
\nonumber
\\
&\approx& A (\bar{\bgm\alpha}) \int dD\exp \left[
-{{B(\bar{\bgm\alpha}) }\over{D^\mu}} - ND \right]
\nonumber
\\
&\approx&
{\tilde A}(\bar{\bgm\alpha}) \exp\left[ -{1\over 2}
{{\mu+2}\over{\mu+1}}\ln N - C(\bar{\bgm\alpha}) N^{\mu/(\mu+1)}
\right] ,
\label{Zinfgeneral}
\end{eqnarray}
where in the last step we use a saddle point or steepest descent
approximation which is accurate at large $N$, and the coefficients are
\begin{eqnarray}
{\tilde A}(\bar{\bgm\alpha}) &=& A(\bar{\bgm\alpha}) \left({{2\pi
\mu^{1/(\mu+1)}} \over{\mu+1}} \right)^{1/2}\cdot
[B(\bar{\bgm\alpha})]^{1/(2\mu+2)}
\\
C(\bar{\bgm\alpha}) &=& [B(\bar{\bgm\alpha})]^{1/(\mu+1)} \left(
{1\over{\mu^{\mu/(\mu+1)}}} + \mu^{1/(\mu+1)} \right).
\end{eqnarray}
Finally we can use Eqs.~(\ref{s1thruZ}, \ref{Zinfgeneral}) to compute
the subextensive term in the entropy, keeping only the dominant term
at large $N$,
\begin{equation}
S_1^{({\rm a})} (N) \rightarrow {1\over{\ln 2}} 
\langle C (\bar{\bgm\alpha})
\rangle_{\bar{\bgms\alpha}} N^{\mu/(\mu+1)} \;\;\; ({\rm bits}),
\label{s1power}
\end{equation}
where $\langle \cdots \rangle_{\bar{\bgms\alpha}}$ denotes an average
over all the target models.

This behavior of the first subextensive term is qualitatively
different from everything we have observed so far. A power law
divergence is much stronger than a logarithmic one. Therefore, a lot
more predictive information is accumulated in an ``infinite
parameter'' (or nonparametric) system; the system is much richer and
more complex, both intuitively and quantitatively.

Subextensive entropy also grows as a power law in a finitely
parameterizable system with a growing number of parameters [compare to the spin chain with decaying interactions on Fig.~(\ref{entr_subext})].  For
example, suppose that we approximate the distribution of a random
variable by a histogram with $K$ bins, and we let $K$ grow with the
quantity of available samples as $K \sim N^\nu$.  Equation
(\ref{s1distr}) suggests that in a $K$--parameter system, the $N^{\rm
  th}$ sample point contributes $\sim K/2N$ bits to the subextensive
entropy. If $K$ changes as mentioned, the $N^{\rm th}$ example then
carries $\sim N^{\nu-1}$ bits. Summing this up over all samples, we
find $S^{({\rm a})}_1 \sim N^{\nu}$, and if we let $\nu = \mu/(\mu
+1)$ we obtain Eq.~(\ref{s1power}).  Note that the growth of the
number of parameters is slower than $N$ ($\nu = \mu/(\mu +1) < 1$),
which makes sense both intuitively and within the framework of the
above SRM fluctuation analysis.  Indeed, this growing number of
parameters is nothing but expanding structure elements, and $d_{\rm
  VC}$ increasing with them, $d_{\rm VC}(C_{n^*(N)}) \equiv d_{\rm
  VC}(N)$.  Therefore, sublinear growth is needed for the fluctuation
control.

Power law growth of the predictive information illustrates the point
made earlier about the transition from learning more to finally
learning nothing as the class of investigated models becomes more
complex. As $\mu$ increases, the problem becomes richer and more
complex, and this is expressed in the stronger divergence of the first
subextensive term of the entropy; for fixed large $N$, the predictive
information increases with $\mu$.  However, if $\mu\rightarrow\infty$
the problem is too complex for learning---in our model example the
number of bins grows in proportion to the number of samples, which
means that we are trying to find too much detail in the underlying
distribution.  As a result, the subextensive term becomes extensive
and stops contributing to predictive information. Thus, at least to
the leading order, predictability is lost, as promised.

\subsection{Beyond finite parameterization: example}
\label{nonparam_ex_sec}

The discussion in the previous section suggests that we should look
for power--law behavior of the predictive information in learning
problems where rather than learning ever more precise values for a
fixed set of parameters, we learn a progressively more detailed
description---effectively increasing the number of parameters---as we
collect more data. One example of such a problem is learning the
distribution $Q(x)$ for a continuous variable $x$, but rather than
writing a parametric form of $Q(x)$ we assume only that this function
itself is chosen from some distribution that enforces a degree of
smoothness. There are some natural connections of this problem to the
methods of quantum field theory (Bialek, Callan, and Strong 1996)
which we can exploit to give a complete calculation of the predictive
information, at least for a class of smoothness constraints.

We write $Q(x)= (1/l_0) \exp [-\phi(x)]$ so that positivity of the
distribution is automatic, and then smoothness may be expressed by
saying that the `energy' (or action) associated with a function
$\phi(x)$ is related to an integral over its derivatives, like the
strain energy in a stretched string.  The simplest possibility
following this line of ideas is that the distribution of functions is
given by
\begin{equation}
\label{prior_nonparam}
{\cal P}[\phi(x)]= \frac{1}{{\cal Z}} \exp \left[-\frac{l}{2}\int dx
\left(\frac{\partial \phi}{\partial x}\right)^2\right] \delta
\left[\frac{1}{l_0}\int dx\, {\rm e}^{- \phi(x)} -1 \right]\; ,
\end{equation}
where ${\cal Z}$ is the normalization constant for ${\cal P}[\phi]$,
the delta function insures that each distribution $Q(x)$ is
normalized, and $l$ sets a scale for smoothness. If distributions are
chosen from this distribution, then the optimal Bayesian estimate of
$Q(x)$ from a set of samples $x_1 , x_2, \cdots , x_N$ converges to
the correct answer, and the distribution at finite $N$ is nonsingular,
so that the regularization provided by this prior is strong enough to
prevent the development of singular peaks at the location of observed
data points (Bialek, Callan, and Strong 1996) \footnote{We caution the
  reader that our discussion in this section is less self--contained
  than in other sections.  Since the crucial steps exactly parallel
  those in the earlier work, here we just give references. To
  compensate for this, we compiled a summary of the original results
  by Bialek et al.~in the Appendix~\ref{bcs_app}.}.  Further
developments of the theory, including alternative choices of ${\cal
  P}[\phi(x)]$, have been given by Periwal (1997, 1998), Holy (1997),
and Aida (1998). We chose the original formulation for our analysis
because our goal here is to be illustrative rather than exhaustive.

From the discussion above we know that the predictive information is
related to the density of Kullback--Leibler divergences, and that the
power--law behavior we are looking for comes from an essential
singularity in this density function.  To illustrate this point, we
calculate the predictive information using the density, even though an
easier direct way exists.

With $Q(x)= (1/l_0) \exp [-\phi(x)]$, we can write the KL divergence
as
\begin{equation}
D_{\rm KL} [\bar\phi (x) ||\phi (x) ]
= {1\over{l_0}}\int dx \exp[-\bar\phi(x)] [\phi(x) - \bar\phi(x)]\, .
\end{equation}
We want to compute the density,
\begin{eqnarray}
\rho (D; \bar\phi) &=& \int [d\phi(x)] {\cal P}[\phi(x)]
\delta\left( D - D_{\rm KL} [\bar\phi (x) ||\phi (x) ]\right)\\
&=&
M \int [d\phi(x)] {\cal P} [\phi(x)]
\delta\left( MD - MD_{\rm KL} [\bar\phi (x) ||\phi (x) ]\right),
\end{eqnarray}
where we introduce a factor $M$ which we will allow to become large so
that we can focus our attention on the interesting limit $D\rightarrow
0$.  To compute this integral over all functions $\phi(x)$, we
introduce a Fourier representation for the delta function, and then
rearrange the terms:
\begin{eqnarray}
\rho (D; \bar\phi) &=&
M \int {{dz}\over{2\pi}} \exp(izMD) \int [d\phi(x)] {\cal P} [\phi(x)]
\exp(-izMD_{\rm KL})\\
&=&
M \int {{dz}\over{2\pi}} \exp\left(izMD
+ {{izM}\over{l_0}} \int dx \bar\phi(x)\exp[-\bar\phi(x)]\right)
\nonumber\\
&&\times \int [d\phi(x)] {\cal P} [\phi(x)]\exp\left(
-{{izM}\over{l_0}}\int dx \phi(x)\exp[-\bar\phi(x)]
\right) .
\end{eqnarray}
The inner integral over the functions $\phi(x)$ is exactly the
integral which was evaluated in the original discussion of this
problem (Bialek, Callan and Strong 1996); in the limit that $zM$ is
large we can use a saddle point approximation, and standard field
theoretic methods allow us to compute the fluctuations around the
saddle point.  The result is that
[cf.~Eqs.~(\ref{dist_bcs})--(\ref{stationary_bcs})]
$$
\int [d\phi(x)] {\cal P} [\phi(x)]\exp\left( -{{izM}\over{l_0}}\int dx
  \phi(x)\exp[-\bar\phi (x)] \right)
$$
\begin{eqnarray}
&=&  
\exp\left(
-{{izM}\over{l_0}}\int dx \phi_{\rm cl}(x)\exp[-\bar\phi(x)]
- S_{\rm eff}[\phi_{\rm cl}(x);zM]\right),
\\
S_{\rm eff}[\phi_{\rm cl};zM]&=&
\frac{l}{2}\int dx
\left(\frac{\partial \phi_{\rm cl}}{\partial x}\right)^2
+\frac{1}{2} \left({{izM}\over{l l_0}}\right)^{1/2} 
\int dx \exp[-\phi_{\rm cl} (x)/2] ,
\label{action_nonparam}
\end{eqnarray} 
\begin{equation}
l \partial^2_x \phi_{\rm cl}(x) + \frac{izM}{l_0}
\exp [-\phi_{\rm cl}(x)]
= \frac{izM}{l_0}\exp [ -\bar\phi(x)]\,.
\label{stationary_nonparametric}
\end{equation}
Now we can do the integral over $z$, again by a saddle point method.
The two saddle point approximations are both valid in the limit that
$D\rightarrow 0$ and $MD^{3/2} \rightarrow\infty$; we are interested
precisely in the first limit, and we are free to set $M$ as we wish,
so this gives us a good approximation for $\rho(D\rightarrow 0 ;
\bar\phi)$.  Also, since $M$ is arbitrarily large, $\phi_{\rm
  cl}(x)=\bar\phi(x)$. This results in 
\begin{eqnarray}
\rho(D\to 0 ; \bar\phi) &=& A[\bar\phi(x)] D^{-3/2}
\exp\left( - {{B [\bar\phi(x)]}\over D}\right) ,\\
A[\bar\phi(x)]
&=&
{1\over{\sqrt{16\pi l l_0}}}
\exp\left[-\frac{l}{2}\int dx
\left(\partial_x \bar\phi \right)^2\right]
\int {{dx}}
\exp[-\bar\phi(x)/2 ]\\
B [\bar\phi(x)]
&=&
{1\over{16 l l_0}}
\left(\int {{dx}}
\exp[-\bar\phi(x)/2]\right)^2 \, .
\end{eqnarray}
Except for the factor of $D^{-3/2}$, this is exactly the sort of
essential singularity that we considered in the previous section, with
$\mu =1$.  The $D^{-3/2}$ prefactor does not change the leading large
$N$ behavior of the predictive information, and we find that
\begin{equation}
S^{({\rm a})}_{\rm 1}(N) \sim
{1\over{2\ln 2\sqrt{l l_0}}}
\Bigg\langle
\int dx \exp[-\bar\phi(x)/2]\Bigg\rangle_{\bar\phi}
N^{1/2} ,
\label{almostdone}
\end{equation}
where $\langle \cdots \rangle_{\bar\phi}$ denotes an average over the
target distributions $\bar\phi(x)$ weighted once again by ${\cal
  P}[\bar\phi(x)]$.  Notice that if $x$ is unbounded then the average
in Eq.~(\ref{almostdone}) is infrared divergent; if instead we let the
variable $x$ range from $0$ to $L$ then this average should be
dominated by the uniform distribution.  Replacing the average by its
value at this point, we obtain the approximate result
\begin{equation}\label{S1infbox}
S^{({\rm a})}_{\rm 1}(N)\sim \frac{1}{2\ln 2} \,\sqrt{N}\,
\left(\frac{L}{l}\right)^{1/2} \,{\rm bits.}
\end{equation}

To understand the result in Eq.~(\ref{S1infbox}), we recall that this
field theoretic approach is more or less equivalent to an adaptive
binning procedure in which we divide the range of $x$ into bins of
local size $\sqrt{l/NQ(x)}$ (Bialek, Callan, and Strong 1996, see also
Appendix~\ref{bcs_app}).  From this point of view,
Eq.~(\ref{S1infbox}) makes perfect sense: the predictive information
is directly proportional to the number of bins that can be put in the
range of $x$. This also is in direct accord with a comment from the
previous subsection that power law behavior of predictive information
arises from the number of parameters in the problem depending on the
number of samples. More importantly, since learning a distribution
consisting of $\sim\sqrt{NL/l}$ bins is, certainly, a $d_{\rm VC}
\sim\sqrt{NL/l}$ problem, we can refer back to our discussion of
fluctuations in prior controlled learning scenarios
(Section~\ref{fluct_prior_sec}) to infer that fluctuations pose no
threat to this nonparametric learning setup.

One thing which remains troubling is that the predictive information
depends on the arbitrary parameter $l$ describing the scale of
smoothness in the distribution.  In the original work it was proposed
that one should integrate over possible values of $l$ (Bialek, Callan
and Strong 1996).  Numerical simulations demonstrate that this
parameter can be learned from the data itself (see
Chapter~\ref{numeric}), but perhaps even more interesting is a
formulation of the problem by Periwal (1997, 1998) which recovers
complete coordinate invariance by effectively allowing $l$ to vary
with $x$.  In this case the whole theory has no length scale, and
there also is no need to confine the variable $x$ to a box (here of
size $L$).  We expect that this coordinate invariant approach will
lead to a universal coefficient multiplying $\sqrt{N}$ in the analog
of Eq.~(\ref{S1infbox}), but this remains to be shown.

In summary, the field theoretic approach to learning a smooth
distribution in one dimension provides us with a concrete, calculable
example of a learning problem with power--law growth of the predictive
information.  The scenario is exactly as suggested in the previous
section, where the density of KL divergences develops an essential
singularity.  Heuristic considerations (Bialek, Callan, and Strong
1996; Aida 1999) suggest that different smoothness penalties [for
example, replacing the kinetic term in the prior,
Eq.~(\ref{prior_nonparam}), by $(\partial_x^{\eta} \phi)^2$] and
generalizations to higher dimensional problems ($ \dim {\vec x} =
\zeta$) will have sensible effects on the predictive information
\begin{equation}
S_1(N) \sim N^{\zeta/2\eta}.
\label{S1general}
\end{equation}
This shows a power--law growth. Smoother functions have smaller powers
(less to learn) and higher dimensional problems have larger powers
(more to learn)---but real calculations for these cases remain
challenging.


\section{$I_{\rm pred}$ as a measure of complexity}

The problem of quantifying complexity is very old.  Solomonoff (1964),
Kolmogorov (1965), and Chaitin (1975) investigated a mathematically
rigorous notion of complexity that measures (roughly) the minimum
length of a computer program that simulates the observed time series
[see also Li and Vit{\'a}nyi (1993)].  Unfortunately there is no
algorithm that can calculate the Kolmogorov complexity of any data
set. In addition, algorithmic or Kolmogorov complexity is closely
related to the Shannon entropy, which means that it measures something
closer to our intuitive concept of randomness than to the intuitive
concept of complexity [as discussed, for example, by Bennett (1990)].
These problems have fueled continued research along two different
paths, representing two major motivations for defining complexity.
First, one would like to make precise an impression that some
systems---such as life on earth or a turbulent fluid flow---evolve
toward a state of higher complexity, and one would like to be able to
classify those states. Second, in choosing among different models that
describe an experiment, one wants to quantify a preference for simpler
explanations or, equivalently, provide a penalty for complex models
that can be weighed against the more conventional goodness of fit
criteria. We bring our readers up to date with some developments in
both of these directions, and then discuss the role of predictive
information as a measure of complexity. This also gives us an
opportunity to discuss more carefully the relation of our results to
previous work.

\subsection{Complexity of statistical models}

The construction of complexity penalties for model selection is a
statistics problem. As far as we know, however, the first discussions
of complexity in this context belong to philosophical literature. Even
leaving aside the early contributions of William of Occam on the need
for simplicity, Hume on the problem of induction, and Popper on
falsifiability, Kemeney (1953) suggested explicitly that it would be
possible to create a model selection criterion that balances goodness
of fit versus complexity. Wallace and Burton (1968) hinted that this
balance may result in the model with ``the briefest recording of all
attribute information.'' Even though he probably had a somewhat
different motivation, Akaike (1974a, 1974b) made the first
quantitative step along these lines. His {\em ad hoc} complexity term
was independent of the number of data points and was proportional to
the number of free independent parameters in the model.

These ideas were rediscovered and developed systematically by Rissanen
in a series of papers starting from 1978. He has emphasized strongly
(Rissanen 1984, 1986, 1987) that fitting a model to data represents an
encoding of those data, or predicting future data, and that in
searching for an efficient code we need to measure not only the number
of bits required to describe the deviations of the data from the
model's predictions (goodness of fit), but also the number of bits
required to specify the parameters of the model (complexity). This
specification has to be done to a precision supported by the
data.\footnote{Within this framework Akaike's suggestion can be seen
  as coding the model to (suboptimal) fixed precision.}  Rissanen
(1984) and Clarke and Barron (1990) in full generality were able to
prove that the optimal encoding of a model requires a code with length
asymptotically proportional to the number of independent parameters
and logarithmically dependent on the number of data points we have
observed.  The minimal amount of space one needs to encode a data
string (minimum description length or MDL) within a certain assumed
model class was termed by Rissanen {\em stochastic complexity,} and in
recent work he refers to the piece of the stochastic complexity
required for coding the parameters as the {\em model complexity}
(Rissanen 1996). This approach was further strengthened by a recent
result (Vit{\'a}nyi and Li 2000) that an estimation of parameters
using the MDL principle is equivalent to Bayesian parameter
estimations with a ``universal'' prior (Li and Vit{\'a}nyi 1993).

There should be a close connection between Rissanen's ideas of
encoding the data stream and the subextensive entropy.  We are
accustomed to the idea that the average length of a code word for
symbols drawn from a distribution $P$ is given by the entropy of that
distribution; thus it is tempting to say that an encoding of a stream
$x_1 , x_2, \cdots , x_N$ will require an amount of space equal to the
entropy of the joint distribution $P(x_1 , x_2, \cdots , x_N )$.  The
situation here is a bit more subtle, because the usual proofs of
equivalence between code length and entropy rely on notions of
typicality and asymptotics as we try to encode sequences of many
symbols; here we already have $N$ symbols and so it doesn't really
make sense to talk about a stream of streams. One can argue, however,
that atypical sequences are not truly random within a considered
distribution since their coding by the methods optimized for the
distribution is not optimal. So atypical sequences are better
considered as typical ones coming from a different distribution [a
point also made by Grassberger (1986)]. This allows us to identify
properties of an observed (long) string with the properties of the
distribution it comes from, as was done by Vit{\'a}nyi and Li (2000).
If we accept this identification of entropy with code length, then
Rissanen's stochastic complexity should be the entropy of the
distribution $P(x_1 , x_2, \cdots , x_N)$.

As emphasized by Balasubramanian (1996), the entropy of the joint
distribution of $N$ points can be decomposed into pieces that
represent noise or errors in the model's local predictions---an
extensive entropy---and the space required to encode the model itself,
which must be the subextensive entropy.  Since in the usual
formulation all long--term predictions are associated with the
continued validity of the model parameters, the dominant component of
the subextensive entropy must be this parameter coding, or model
complexity in Rissanen's terminology.  Thus the subextensive entropy
should be the model complexity, and in simple cases where we can
describe the data by a $K$--parameter model both quantities are equal
to $(K/2)\log_2 N$ bits to the leading order.

The fact that the subextensive entropy or predictive information
agrees with Rissanen's model complexity suggests that $I_{\rm pred}$
provides a reasonable measure of complexity in learning problems.  On
the other hand, this agreement might lead the reader to wonder if all
we have done is to rewrite the results of Rissanen et al.~in a
different notation. To calm these fears we recall again that our
approach distinguishes infinite VC problems from finite ones and
treats nonparametric cases as well. Indeed, the predictive information
is defined without reference to the idea that we are learning a model,
and thus we can make a link to physical aspects of the problem.

\subsection{Complexity of dynamical systems}

There is a strong prejudice that the complexity of physical systems
should be measured by quantities that are at least related to more
conventional thermodynamic quantities (temperature, entropy, $\dots$),
since this is the only way one will be able to calculate complexity
within the framework of statistical mechanics. Most proposals define
complexity as an entropy--like quantity, but an entropy of some
unusual ensemble.  For example, Lloyd and Pagels (1988) identified
complexity as {\em thermodynamic depth}, the entropy of the state
sequences that lead to the current state. The idea is clearly in the
same spirit as the measurement of predictive information, but this
depth measure does not completely discard the extensive component of
the entropy (Crutchfield and Shalizi 1999) and thus fails to resolve
the essential difficulty in constructing complexity measures for
physical systems: distinguishing genuine complexity from randomness
(entropy), the complexity should be zero both for purely regular and
for purely random systems.

New definitions of complexity that try to satisfy these criteria
(Lopez--Ruiz et al.~1995, Gell--Mann and Lloyd 1996, Shiner et
al.~1999, Sole and Luque 1999, Adami and Cerf 2000) and criticisms of
these proposals (Crutchfield et al.~1999, Feldman and Crutchfield
1998, Sole and Luque 1999) continue to emerge even now. Aside from the
obvious problems of not actually eliminating the extensive component
for all or a part of the parameter space or not expressing complexity
as an average over a physical ensemble, the critiques often are based
on a clever argument first mentioned explicitly by Feldman and
Crutchfield (1998). In an attempt to create a universal measure, the
constructions can be made {\em over--universal}: many proposed
complexity measures depend only on the entropy density ${\cal S}_0$
and thus are functions only of disorder---not a desired feature. In
addition, many of these and other definitions are flawed because they
fail to distinguish among the richness of classes beyond some very
simple ones.

In a series of papers, Crutchfield and coworkers identified {\em
  statistical complexity} with the entropy of {\em causal states,}
which in turn are defined as all those microstates (or histories) that
have the same conditional distribution of futures (Crutchfield and
Young 1989, Shalizi and Crutchfield 1999).  The causal states provide
an optimal description of a system's dynamics in the sense that these
states make as good a prediction as the histories themselves.
Statistical complexity is very similar to predictive information, but
Shalizi and Crutchfield (1999) define a quantity which is even closer
to the spirit of our discussion: their {\em excess entropy} is exactly
the mutual information between the semi--infinite past and future.
Unfortunately, by focusing on cases in which the past and future are
infinite but the excess entropy is finite, their discussion is limited
to systems for which (in our language) $I_{\rm
  pred}(T\rightarrow\infty) = {\rm constant}$.

In our view, Grassberger (1986) has made the clearest and the most
appealing definitions. He emphasized that the slow approach of the
entropy to its extensive limit is a sign of complexity, and has
proposed several functions to analyze this slow approach. His {\em
  effective measure complexity} is the subextensive entropy term of an
infinite data sample.  Unlike Crutchfield et al., he allows this
measure to grow to infinity. As an example, for low dimensional
dynamical systems, the effective measure complexity is finite whether
the system exhibits periodic or chaotic behavior, but at the
bifurcation point that marks the onset of chaos, it diverges
logarithmically. More interestingly, Grassberger also notes that
simulations of specific cellular automaton models that are capable of
universal computation indicate that these systems can exhibit an even
stronger, power--law, divergence.

Grassberger (1986) also introduces the {\em true measure complexity},
which is the minimal information one needs to extract from the past in
order to provide optimal prediction. This quantity is exactly the
statistical complexity of Crutchfield et al., and the two approaches
are actually much closer than they seem.  The relation between the
true and the effective measure complexities, or between the
statistical complexity and the excess entropy, closely parallels the
idea of extracting or compressing relevant information (Tishby et
al.~1999, Bialek and Tishby, in preparation), as discussed below.

\subsection{A unique measure of complexity?}

We recall that entropy provides a measure of information that is
unique in satisfying certain plausible constraints (Shannon 1948). It
would be attractive if we could prove a similar uniqueness theorem for
the predictive information, or any part of it, as a measure of the
complexity or richness of a time dependent signal $x(0 < t <T)$ drawn
from a distribution $P[x(t)]$.  Before proceeding with such an
argument we have to ask, however, whether we want to attach measures
of complexity to a particular signal $x(t)$, or whether we are
interested in measures (like the entropy itself) which constitute an
average over the ensemble $P[x(t)]$.
 
In most cases, including the learning problems discussed above, it is
clear that we want to measure complexity of the dynamics underlying
the signal, or equivalently the complexity of a model that might be
used to describe the signal.  This is very different from trying to
define the complexity of a single realization, because there can be
atypical data streams.  Either we must treat atypicality explicitly,
arguing that atypical data streams from one source should be viewed as
typical streams from another source, as discussed by Vit{\'a}nyi and
Li (2000), or we have to look at average quantities.  Grassberger
(1986) in particular has argued that our visual intuition about the
complexity of spatial patterns is an ensemble concept, even if the
ensemble is only implicit [see also Tong in the discussion session of
Rissanen (1987)].  So we shall search for measures of complexity that
are averages over the distribution $P[x(t)]$.

Once we focus on average quantities, we can start by adopting
Shannon's postulates as constraints on a measure of complexity: if
there are $N$ equally likely signals, then the measure should be
monotonic in $N$; if the signal is decomposable into statistically
independent parts then the measure should be additive with respect to
this decomposition; and if the signal can be described as a leaf on a
tree of statistically independent decisions then the measure should be
a weighted sum of the measures at each branching point. We believe
that these constraints are as plausible for complexity measures as for
information measures, and it is well known from Shannon's original
work that this set of constraints leaves the entropy as the only
possibility.  Since we are discussing a time dependent signal, this
entropy depends on the duration of our sample, $S(T)$.  We know of
course that this cannot be the end of the discussion, because we need
to distinguish between randomness (entropy) and complexity.  The path
to this distinction is to introduce other constraints on our measure.

First we notice that if the signal $x$ is continuous, then the entropy
is not invariant under transformations of $x$.  It seems reasonable to
ask that complexity be a function of the process we are observing and
not of the coordinate system in which we choose to record our
observations. The examples above show us, however, that it is not the
whole function $S(T)$ which depends on the coordinate system for
$x$;\footnote{Here we consider instantaneous transformations of $x$,
  not filtering or other transformations that mix points at different
  times.} it is only the extensive component of the entropy that has
this noninvariance.  This can be seen more generally by noting that
subextensive terms in the entropy contribute to the mutual information
among different segments of the data stream (including the predictive
information defined here), while the extensive entropy cannot; mutual
information is coordinate invariant, so all of the noninvariance must
reside in the extensive term.  Thus, any measure complexity that is
coordinate invariant must discard the extensive component of the
entropy.

The fact that extensive entropy cannot contribute to complexity is
discussed widely in the physics literature (Bennett 1990), as our
short review above shows. To statisticians and computer scientists,
who are used to Kolmogorov's ideas, this is less obvious.  However,
Rissanen (1986, 1987) also talks about ``noise'' and ``useful
information'' in a data sequence, which is similar to splitting
entropy into its extensive and the subextensive parts. His ``model
complexity,'' aside from not being an average as required above, is
essentially equal to the subextensive entropy. Similarly, Whittle [in
the discussion of Rissanen (1987)] talks about separating the
predictive part of the data from the rest.

If we continue along these lines, we can think about the asymptotic
expansion of the entropy at large $T$.  The extensive term is the
first term in this series, and we have seen that it must be discarded.
What about the other terms?  In the context of learning a
parameterized model, most of the terms in this series depend in detail
on our prior distribution in parameter space, which might seem odd for
a measure of complexity.  More generally, if we consider
transformations of the data stream $x(t)$ that mix points within a
temporal window of size $\tau$, then for $T >> \tau$ the entropy
$S(T)$ may have subextensive terms which are constant, and these are
not invariant under this class of transformations.  On the other hand,
if there are divergent subextensive terms, these {\em are} invariant
under such temporally local transformations.\footnote{Throughout this
  discussion we assume that the signal $x$ at one point in time is
  finite dimensional.  There are subtleties if we allow $x$ to
  represent the configuration of a spatially infinite system.}  So if
we insist that measures of complexity be invariant not only under
instantaneous coordinate transformations, but also under temporally
local transformations, then we can discard both the extensive and the
finite subextensive terms in the entropy, leaving only the divergent
subextensive terms as a possible measure of complexity.

An interesting example of these arguments is provided by the
statistical mechanics of polymers.  It is conventional to make models
of polymers as random walks on a lattice, with various interactions or
self avoidance constraints among different elements of the polymer
chain.  If we count the number $\cal N$ of walks with $N$ steps, we
find that ${\cal N}(N) \sim A N^\gamma z^N$ (de Gennes 1979). Now the
entropy is the logarithm of the number of states, and so there is an
extensive entropy ${\cal S}_0 = \log_2 z$, a constant subextensive
entropy $\log_2 A$, and a divergent subextensive term $S_1(N)
\rightarrow \gamma \log_2 N$.  Of these three terms, only the
divergent subextensive term (related to the critical exponent
$\gamma$) is universal, that is independent of the detailed structure
of the lattice.  Thus, as in our general argument, it is only the
divergent subextensive terms in the entropy that are invariant to
changes in our description of the local, small scale dynamics.

We can recast the invariance arguments in a slightly different form
using the relative entropy.  We recall that entropy is defined cleanly
only for discrete processes, and that in the continuum there are
ambiguities.  We would like to write the continuum generalization of
the entropy of a process $x(t)$ distributed according to $P[x(t)]$ as
\begin{eqnarray}
S_{\rm cont} = -\int Dx(t) \,P[x(t)]\log_2 P[x(t)] ,
\end{eqnarray}
but this is not well defined because we are taking the logarithm of a
dimensionful quantity.  Shannon gave the solution to this problem: we
use as a measure of information the relative entropy or KL divergence
between the distribution $P[x(t)]$ and some reference distribution
$Q[x(t)]$,
\begin{eqnarray}
S_{\rm rel} = -\int Dx(t) \,P[x(t)]\log_2
\left({{P[x(t)]}\over{Q[x(t)]}} \right) ,
\end{eqnarray}
which is invariant under changes of our coordinate system on the space
of signals.  The cost of this invariance is that we have introduced an
arbitrary distribution $Q[x(t)]$, and so really we have a family of
measures.  We can find a unique complexity measure within this family
by imposing invariance principles as above, but in this language we
must make our measure invariant to different choices of the reference
distribution $Q[x(t)]$.

The reference distribution $Q[x(t)]$ embodies our expectations for the
signal $x(t)$; in particular, $S_{\rm rel}$ measures the extra space
needed to encode signals drawn from the distribution $P[x(t)]$ if we
use coding strategies that are optimized for $Q[x(t)]$. If $x(t)$ is a
written text, two readers who expect different numbers of spelling
errors will have different $Q$s, but to the extent that spelling
errors can be corrected by reference to the immediate neighboring
letters we insist that any measure of complexity be invariant to these
differences in $Q$. On the other hand, readers who differ in their
expectations about the global subject of the text may well disagree
about the richness of a newspaper article. This suggests that
complexity is a component of the relative entropy that is invariant
under some class of local translations and misspellings.

Suppose that we leave aside global expectations, and construct our
reference distribution $Q[x(t)]$ by allowing only for short ranged
interactions---certain letters tend to follow one another, letters
form words, and so on, but we bound the range over which these rules
are applied. Models of this class cannot embody the full structure of
most interesting time series (including language), but in the present
context we are not asking for this.  On the contrary, we are looking
for a measure that is invariant to differences in this short ranged
structure.  In the terminology of field theory or statistical
mechanics, we are constructing our reference distribution $Q[x(t)]$
from local operators.  Because we are considering a one dimensional
signal (the one dimension being time), distributions constructed from
local operators cannot have any phase transitions as a function of
parameters; again it is important that the signal $x$ at one point in
time is finite dimensional.  The absence of critical points means that
the entropy of these distributions (or their contribution to the
relative entropy) consists of an extensive term (proportional to the
time window $T$) plus a constant subextensive term, plus terms that
vanish as $T$ becomes large.  Thus, if we choose different reference
distributions within the class constructible from local operators, we
can change the extensive component of the relative entropy, and we can
change constant subextensive terms, but the divergent subextensive
terms are invariant.

To summarize, the usual constraints on information measures in the
continuum produce a family of allowable complexity measures, the
relative entropy to an arbitrary reference distribution. If we insist
that all observers who choose reference distributions constructed from
local operators arrive at the same measure of complexity, or if we
follow the first line of arguments presented above, then this measure
must be the divergent subextensive component of the entropy or,
equivalently, the predictive information. We have seen that this
component is connected to learning, quantifying the amount that can be
learned about dynamics that generate the signal, and to measures of
complexity that have arisen in statistics and in dynamical systems
theory.


\section{Discussion}
\label{pred_discus_sec}

We have presented predictive information as a {\em characterization}
of data streams.  In the context of learning, predictive information
is related directly to generalization.  More generally, the structure
or order in a time series or a sequence is related almost by
definition to the fact that there is predictability along the
sequence.  The predictive information measures the amount of such
structure, but doesn't exhibit the structure in a concrete form.
Having collected a data stream of duration $T$, what are the features
of these data that carry the predictive information $I_{\rm pred}(T)$?
From Eq.~(\ref{chuck}) we know that most of what we have seen over the
time $T$ must be irrelevant to the problem of prediction, so that the
predictive information is a small fraction of the total information;
can we separate these predictive bits from the vast amount of
nonpredictive data?

The problem of separating predictive from nonpredictive information is
a special case of the problem discussed recently (Tishby et al.~1999,
Bialek and Tishby, in preparation): given some data $x$, how do we
compress our description of $x$ while preserving as much information
as possible about some other variable $y$?  Here we identify $x =
x_{\rm past}$ as the past data and $y= x_{\rm future}$ as the future.
When we compress $x_{\rm past}$ into some reduced description $\hat
x_{\rm past}$ we keep a certain amount of information about the past,
$I(\hat x_{\rm past} ; x_{\rm past})$, and we also preserve a certain
amount of information about the future, $I(\hat x_{\rm past} ; x_{\rm
  future})$.  There is no single correct compression $x_{\rm past}
\rightarrow \hat x_{\rm past}$; instead there is a one parameter
family of strategies which trace out an optimal curve in the plane
defined by these two mutual informations, $I(\hat x_{\rm past} ;
x_{\rm future})$ vs.  $I(\hat x_{\rm past} ; x_{\rm past})$.

The predictive information preserved by compression must be less than
the total, so that $I(\hat x_{\rm past} ; x_{\rm future}) \leq I_{\rm
  pred}(T)$.  Generically no compression can preserve all of the
predictive information so that the inequality will be strict, but
there are interesting special cases where equality can be achieved. If
prediction proceeds by learning a model with a finite number of
parameters, we might have a regression formula that specifies the best
estimate of the parameters given the past data; using the regression
formula compresses the data but preserves all of the predictive power.
In cases like this (more generally, if there exist sufficient
statistics for the prediction problem) we can ask for the minimal set
of $\hat x_{\rm past}$ such that $I(\hat x_{\rm past} ; x_{\rm
  future}) = I_{\rm pred}(T)$.  The entropy of this minimal $\hat
x_{\rm past}$ is the true measure complexity defined by Grassberger
(1986) or the statistical complexity defined by Crutchfield and Young
(1989) [in the framework of the causal states theory a very similar
comment was made recently by Shalizi and Crutchfield (2000)].

In the context of statistical mechanics, long range correlations are
characterized by computing the correlation functions of order
parameters, which are coarse--grained functions of the system's
microscopic variables.  When we know something about the nature of the
order parameter (e.~g., whether it is a vector or a scalar), then
general principles allow a fairly complete classification and
description of long range ordering and the nature of the critical
points at which this order can appear or change.  On the other hand,
defining the order parameter itself remains something of an art.  For
a ferromagnet, the order parameter is obtained by local averaging of
the microscopic spins, while for an antiferromagnet one must average
the staggered magnetization to capture the fact that the ordering
involves an alternation from site to site, and so on.  Since the order
parameter carries all the information that contributes to long range
correlations in space and time, it might be possible to define order
parameters more generally as those variables that provide the most
efficient compression of the predictive information, and this should
be especially interesting for complex or disordered systems where the
nature of the order is not obvious intuitively; a first try in this
direction was made by Bruder (1998). At critical points the predictive
information will diverge with the size of the system, and the
coefficients of these divergences should be related to the standard
scaling dimensions of the order parameters, but the details of this
connection need to be worked out.

If we compress or extract the predictive information from a time
series we are in effect discovering ``features'' that capture the
nature of the ordering in time.  Learning itself can be seen as an
example of this, where we discover the parameters of an underlying
model by trying to compress the information that one sample of $N$
points provides about the next, and in this way we address directly
the problem of generalization (Bialek and Tishby, in preparation).
The fact that (as mentioned above) nonpredictive information is
useless to the organism suggests that one crucial goal of neural
information processing is to separate predictive information from the
background.  Perhaps rather than providing an efficient representation
of the current state of the world---as suggested by Attneave (1954),
Barlow (1959, 1961), and others (Atick 1992)---the nervous system
provides an efficient representation of the predictive
information.\footnote{If, as seems likely, the stream of data reaching
  our senses has diverging predictive information then the space
  required to write down our description grows and grows as we observe
  the world for longer periods of time.  In particular, if we can
  observe for a very long time then the amount that we know about the
  future will exceed, by an arbitrarily large factor, the amount that
  we know about the present.  Thus representing the predictive
  information may require many more neurons than would be required to
  represent the current data.  If we imagine that the goal of primary
  sensory cortex is to represent the current state of the sensory
  world, then it is difficult to understand why these cortices have so
  many more neurons than they have sensory inputs.  In the extreme
  case, the region of primary visual cortex devoted to inputs from the
  fovea has nearly 30,000 neurons for each photoreceptor cell in the
  retina (Hawken and Parker 1991); although much remains to be learned
  about these cells, it is difficult to imagine that the activity of
  so many neurons constitutes an efficient representation of the
  current sensory inputs.  But if we live in a world where the
  predictive information in the movies reaching our retina diverges,
  it is perfectly possible that an efficient representation of the
  predictive information available to us at one instant requires
  thousands of times more space than an efficient representation of
  the image currently falling on our retina.}  It should be possible
to test this directly by studying the encoding of reasonably natural
signals and asking if the information which neural responses provide
about the future of the input is close to the limit set by the
statistics of the input itself, given that the neuron only captures a
certain number of bits about the past.  Thus we might ask if, under
natural stimulus conditions, a motion sensitive visual neuron captures
features of the motion trajectory that allow for optimal prediction or
extrapolation of that trajectory; by using information theoretic
measures we both test the ``efficient representation'' hypothesis
directly and avoid arbitrary assumptions about the metric for errors
in prediction.  For more complex signals such as communication sounds,
even identifying the features that capture the predictive information
is an interesting problem.

It is natural to ask if these ideas about predictive information could
be used to analyze experiments on learning in animals or humans.  We
have emphasized the problem of learning probability distributions or
probabilistic models rather than learning deterministic functions,
associations or rules.  It is known that the nervous system adapts to
the statistics of its inputs, and this adaptation is evident in the
responses of single neurons (Smirnakis et al.~1996, Brenner et
al.~2000); these experiments provide a simple example of the system
learning a parameterized distribution.  When making saccadic eye
movements, human subjects alter their distribution of reaction times
in relation to the relative probabilities of different targets, as if
they had learned an estimate of the relevant likelihood ratios
(Carpenter and Williams 1995).  Humans also can learn to discriminate
almost optimally between random sequences (fair coin tosses) and
sequences that are correlated or anticorrelated according to a Markov
process; this learning can be accomplished from examples alone, with
no other feedback (Lopes and Oden 1987).  Acquisition of language may
require learning the joint distribution of successive phonemes,
syllables, or words, and there is direct evidence for learning of
conditional probabilities from artificial sound sequences, both by
infants and by adults (Saffran et al.~1996; 1999).  These examples,
which are not exhaustive, indicate that the nervous system can learn
an appropriate probabilistic model,\footnote{As emphasized above, many
  other learning problems, including learning a function from noisy
  examples, can be seen as the learning of a probabilistic model.
  Thus we expect that this description applies to a much wider range
  of biological learning tasks.} and this offers the opportunity to
analyze the dynamics of this learning using information theoretic
methods: What is the entropy of $N$ successive reaction times
following a switch to a new set of relative probabilities in the
saccade experiment?  How much information does a single reaction time
provide about the relevant probabilities?  Following the arguments
above, such analysis could lead to a measurement of the universal
learning curve $\Lambda (N)$.

The learning curve $\Lambda (N)$ exhibited by a human observer is
limited by the predictive information in the time series of stimulus
trials itself.  Comparing $\Lambda (N)$ to this limit defines an
efficiency of learning in the spirit of the discussion by Barlow
(1983); while it is known that the nervous system can make efficient
use of available information in signal processing tasks [cf. Chapter 4
of Rieke et al.~(1997)], it is not known whether the brain is an
efficient learning machine in the analogous sense.  Given our
classification of learning tasks by their complexity, it would be
natural to ask if the efficiency of learning were a critical function
of task complexity: perhaps we can even identify a limit beyond which
efficient learning fails, indicating a limit to the complexity of the
internal model used by the brain during a class of learning tasks.  We
believe that our theoretical discussion here at least frames a clear
question about the complexity of internal models, even if for the
present we can only speculate about the outcome of such experiments.

An important result of our analysis is the characterization of time
series or learning problems beyond the class of finitely
parameterizable models, that is the class with power--law divergent
predictive information.  Qualitatively this class is more complex than
{\em any} parametric model, no matter how many parameters there may
be, because of the more rapid asymptotic growth of $I_{\rm pred}(N)$.
On the other hand, with a finite number of observations $N$, the
actual amount of predictive information in such a nonparametric
problem may be {\em smaller} than in a model with a large but finite
number of parameters.  Specifically, if we have two models, one with
$I_{\rm pred}(N)\sim AN^\nu$ and one with $K$ parameters so that
$I_{\rm pred}(N)\sim (K/2)\log_2 N$, the infinite parameter model has
less predictive information for all $N$ smaller than some critical
value
\begin{equation}
N_c \sim \left[ {K\over{2A\nu}}\log_2\left(
{K\over{2A}}\right)\right]^{1/\nu} .
\label{Nc}
\end{equation}
In the regime $N \ll N_c$, it is possible to achieve more efficient
prediction by trying to learn the (asymptotically) more complex model,
as we illustrate concretely in numerical simulations of the density
estimation problem, Section~\ref{num_help_sec}.  Even if there are a
finite number of parameters---such as the finite number of synapses in
a small volume of the brain---this number may be so large that we
always have $N \ll N_c$, so that it may be more effective to think of
the many parameter model as approximating a continuous or
nonparametric one.

It is tempting to suggest that the regime $N << N_c$ is the relevant
one for much of biology.  If we consider, for example, 10 mm$^2$ of
inferotemporal cortex devoted to object recognition (Logothetis and
Sheinberg 1996), the number of synapses is $K\sim 5\times 10^9$.  On
the other hand, object recognition depends on foveation, and we move
our eyes roughly three times per second throughout perhaps 15 years of
waking life during which we master the art of object recognition.
This limits us to at most $N\sim 10^9$ examples.  Remembering that we
must have $\nu < 1$, even with large values of $A$ Eq.~(\ref{Nc})
suggests that we operate with $N < N_c$.  One can make similar
arguments about very different brains, such as the mushroom bodies in
insects (Capaldi, Robinson and Fahrbach 1999). If this identification
of biological learning with the regime $N << N_c$ is correct, then the
success of learning in animals must depend on strategies that
implement sensible priors over the space of possible models.

There is one clear empirical hint that humans can make effective use
of models that are beyond finite parameterization (in the sense that
predictive information diverges as a power--law), and this comes from
studies of language.  Long ago, Shannon (1951) used the knowledge of
native speakers to place bounds on the entropy of written English, and
his strategy made explicit use of predictability.  Shannon showed
$N$--letter sequences to native speakers (readers), asked them to
guess the next letter, and recorded how many guesses were required
before they got the right answer.  Thus each letter in the text is
turned into a number, and the entropy of the distribution of these
numbers is an upper bound on the conditional entropy $\ell (N)$ [cf.
Eq.~(\ref{condentropy})].  Shannon himself thought that the
convergence as $N$ becomes large was rather quick, and quoted an
estimate of the extensive entropy per letter ${\cal S}_0$.  Many years
later, Hilberg (1990) reanalyzed Shannon's data and found that the
approach to extensivity in fact was very slow: certainly there is
evidence for a large component $S_1(N) \propto N^{1/2}$, and this may
even dominate the extensive component for accessible $N$.  Ebeling and
P{\"o}schel (1994; see also P\"oschel, Ebeling, and Ros\'e 1995)
studied the statistics of letter sequences in long texts (like {\em
  Moby Dick}) and found the same strong subextensive component.  It
would be attractive to repeat Shannon's experiments with a slightly
different design that emphasizes the detection of subextensive terms
at large $N$.\footnote{Associated with the slow approach to
  extensivity is a large mutual information between words or
  characters separated by long distances, and several groups have
  found that this mutual information declines as a power law. Cover
  and King (1978) criticize such observations by noting that such
  behavior is impossible in Markov chains of arbitrary order. While it
  is possible that existing mutual information data have not reached
  asymptotia, the criticism of Cover and King misses the possibility
  that language is {\em not} a Markov process.  Of course it cannot be
  Markovian if it has a power--law divergence in the predictive
  information.}

In summary, we believe that our analysis of predictive information
solves the problem of measuring the complexity of time series.  This
analysis unifies ideas from learning theory, coding theory, dynamical
systems, and statistical mechanics.  In particular we have focused
attention on a class of processes that are qualitatively more complex
than those treated in conventional learning theory, and there are
several reasons to think that this class includes many examples of
relevance to biology and cognition.



\chapter{Learning continuous distributions: \\
Simulations with field theoretic priors}
\label{numeric}

\section{Occam factors in statistics}

As we have discussed extensively in the preceding Chapter, one of the
central problems in learning is to balance `goodness of fit' criteria
against the complexity of models.  An important development in the
Bayesian approach thus was the realization that there does not need to
be any extra penalty for model complexity: if we compute the total
probability that data are generated by a model, there is a factor from
the volume in parameter space---the ``Occam factor''---that
discriminates against more complex models (MacKay 1992,
Balasubramanian 1997).  This works remarkably well for systems with a
finite number of parameters and creates a complexity ``razor'' (named
after ``Occam's razor'') that is equivalent to the {\em model
  complexity} of the celebrated Minimal Description Length (MDL)
principle (Rissanen 1989, 1996). It is not clear, however, what
happens if we leave this finite dimensional setting and consider
nonparametric problems such as the estimation of a smooth probability
density.

As we have emphasized, the behavior of the predictive information,
Eq.~(\ref{ipredlim}), is controlled by the density of models, and
therefore the predictive information is closely related to the Occam
factor. Since the density and consequently $I_{\rm pred}$ are well
defined for the finite parameter as well as for the nonparametric
cases (cf.~Sections~\ref{learn_distr_sec} and \ref{nonparam_gen_sec})
one can hope that a nonparametric analogue of the Occam factor exists
and can do its job of punishing complexity. However, since in these
two cases the densities of models are very different, the Occam factor
details certainly must be different too.

The 1996 formulation of nonparametric learning by Bialek, Callan, and
Strong, which we have summarized in Appendix~\ref{bcs_app} and
investigated further in Section~\ref{nonparam_ex_sec}, may serve as a
good example in which to study infinite dimensional Occam factors. In
this Bayesian quantum field theory formulation, standard field theory
methods may be used not only to find a nowhere singular estimate of a
continuous density, but also to compute the infinite dimensional
analog of the Occam factor, at least asymptotically for large numbers
of samples. This factor, which we also call the fluctuation
determinant, is the second term of the effective Hamiltonian
Eqs.~(\ref{action_bcs},~\ref{action_nonparam})
\begin{equation}
R=\frac{1}{2}\left(\frac{N}{l l_0}\right)^{1/2} 
\int dx \exp [-\phi_{\rm cl} (x)/2].
\label{occam_num}
\end{equation}

Intuitively, smaller values of $l$ allow more rapidly varying and thus
more complex [as measured by the predictive information,
Eqs.~(\ref{almostdone},~\ref{S1infbox})] estimates of the density.
Correspondingly, the infinite dimensional Occam factor is bigger and
thus exponentially punishes more complex models. As Bialek et al.~have
speculated, $l$, the only free parameter of their theory, can be
determined by a fight between the log--likelihood goodness of fit and
the Occam factor to provide for the shortest total description
(highest probability) of the data, much like in the finite parameter
MDL theory. However, their proposed scaling for $l^*$ (the best value
of the parameter) as a function of $N$, $l^* \sim N^{1/3}$ seems to be
over--universal and requires further analysis.

There are more questions not clearly answered either by the original
work, or its further developments (Periwal 1997, 1998, Holy 1997, Aida
1999). Can this method be implemented in practice? Can we really use
the infinite dimensional Occam factor to balance against the goodness
of fit? How does the algorithm's performance compare to the absolute
bounds set by the predictive information?  What happens if the
learning problem is strongly atypical of the prior distribution? And
what is the role of the Occam factor in this case?

To answer all of these questions we turn to numerical simulations.

\section{The algorithm}

To simplify the algorithm, maximize the speed of simulations, and
shorten our presentation, we do the numerical analysis only in the
framework of the original paper (Bialek, Callan, and Strong 1996, see
also Appendix~\ref{bcs_app} and Section~\ref{nonparam_ex_sec}).  This
may seem too specific, but we believe that our results are very
general and will hold for the alternative formulations of Periwal
(1997, 1998) and Holy (1997) since the mechanisms of regularization
and complexity control are everywhere the same.

Due to our most recent developments
(cf.~Section~\ref{nonparam_ex_sec}) and the specific questions we ask,
we need to modify the original setup slightly before proceeding.
First of all, we will investigate the performance of the method in
many different learning problems, some of them not characteristic of
the prior Eq.~(\ref{prior_bcs}). For these purposes we will take
densities at random from an `actual' a priori distribution that
minimally generalizes Eq.~(\ref{prior_bcs}),
\begin{equation}
{\cal P}[\phi(x)]= \frac{1}{\cal Z}
\exp \left[-\frac{l_a^{2\eta_a -1}}{2}
\int dx \left(\frac{\partial^{\eta_a} \phi}{\partial x^{\eta_a}}
\right)^2 \right]  \delta \left[\frac{1}{l_0} 
\int dx\, {\rm e}^{- \phi(x)} -1 \right].
\label{prior_numeric}
\end{equation}
Here $\eta_a>1/2$ to ensure UV convergence, ${\cal Z}$ is the
normalization constant, and the $\delta$--function enforces
normalization of $Q$. We refer to $l_a$ and $\eta_a$ as the {\em
  smoothness scale} and the {\em exponent}, respectively, and the
subscript {\em a} stands for `actual'. We will use non--subscripted
$\eta$ and $l$ to indicate the parameters the algorithm uses, that is,
the learning machine's own a priori expectations, and then
$\eta_a=\eta\equiv 1$ and $l_a=l$ reduces to the original formulation
of Bialek et al.

The other modification we make relates to the problem of the infrared
divergence of the predictive information, Eq.~(\ref{almostdone}), or,
equivalently, to the nonuniform convergence of the estimate $Q_{\rm
  est} (x)$ to the target $P(x)$, Eq.~(\ref{psi2_app}). To cure this
we can put the system in the box of size $L$, just like we did in
Section~\ref{nonparam_ex_sec}. Also, we realize that the variance of
fluctuations between the target and the estimate (Bialek et al.~1996)
is just an {\em ad hoc} measure of performance of the learning
machine. The {\em universal learning curve} $\Lambda(N)$,
Eq.~(\ref{unilcurve}), is a much better choice. For a proper Bayesian
learning with the prior Eq.~(\ref{prior_bcs}), using
Eqs.~(\ref{Z},~\ref{S1infbox}), we write
\begin{equation}
\Lambda (N) \equiv \left<  \left< D_{\rm
      KL}[P(x)||Q_{\rm est}(x)]\right>_{\{x_{\rm i}\}} 
\right>^{(0)} \sim  \frac{1}{4} \sqrt{\frac{L}{lN}},
\label{lcurve_num}
\end{equation}
where $\left< \cdots \right>^{(0)}$ means an average over the prior,
and the $\log 2$ factor is omitted because we choose to measure
entropies in nats (that is, use natural logarithms) in this Chapter.
Note that the coefficient in front of the square root is probably
meaningless since it is calculated here only to the zeroth order (see
Section~\ref{nonparam_ex_sec}).

After these modifications, the algorithm to implement the theory is
rather simple.  We need to solve the second order differential
equation [cf.~Eq.~(\ref{stationary_bcs})]
\begin{equation}
l \partial^2_x \phi_{\rm cl}(x) + \frac{N}{l_0} \,
\exp \left[ -\phi_{\rm cl}(x) \right]
= \sum_{j=1}^N \delta(x-x_j)\,.
\label{stationary_num}
\end{equation}
Normalization of $Q_{\rm cl}$ fixes one integration constant, and the
other is set by a periodicity constraint for $\phi_{\rm cl}$,
\begin{equation}
\phi(x)=\phi(x+L),
\end{equation}  
which is due to $x$ being in a box.  The resulting boundary value
problem is solved by a standard `relaxation' (or Newton) method of
iterative improvements to a guessed solution (see, for example, Press
et al.~1988). The target precision is always $10^{-5}$, which is
smaller than the smallest $D_{\rm KL}\sim 10^{-4}$ we intend to
investigate. It turns out that the method converges regardless of the
initial guess for all $l$ up to $\sim 5$.  However, convergence is not
uniform in $l$ and, as $l \to 0$, the number of iterations required to
reach the same precision grows almost quadratically in $1/l$. The
independent variable $x \in [0,L]$ is discretized in equal steps to
ensure stability of the method.  We expect the estimate distribution
to vary over a local length scale [Bialek et al.~1996,
cf.~Eq.~(\ref{lsize_bcs})]
\begin{equation}
\xi(x) \sim \left[l/ N Q_{\rm est}(x)\right]^{1/2} 
\approx \left[l/ N P(x)\right]^{1/2}\,.
\label{varscale_num}
\end{equation}
Empirically we see that, for small $l$, the maximal value of the
target distribution $P(x)$ grows approximately as $\sim l^{-1/2}$.
This means that for Figs.~(\ref{correct_ex}--\ref{incorrect_diff}),
where $N \le 10^5$ and $l \ge 0.05$, we are safe with $10^4$ grid
points. Similarly, for Figs.~(\ref{scale_select},~\ref{adaptive_D}),
we need $10^5$ discretization steps because $N=10^6$ and $l=0.01$
are present there.

Since the prior we use, Eq.~(\ref{prior_numeric}), is UV convergent,
we can generate random probability densities from it by replacing
$\phi$ with its Fourier series and truncating the latter at some
sufficiently high wavenumber $k_c$,
\begin{equation}
\phi(x)= \sum_{k=0}^{k_c} \left[ 
A_k \cos \frac{2 \pi k x }{L}
+B_k \sin \frac{2 \pi k x}{L}
\right].
\end{equation}
Then Eq.~(\ref{prior_numeric})
enforces the amplitudes of the $k$'th mode to be distributed normally
around zero with the standard deviation
\begin{equation}
\langle A_k^2 \rangle^{1/2} 
=\langle B_k^2 \rangle^{1/2}
=\frac{2^{1/2}}{l^{\eta-1/2}} 
\left(\frac{L}{2\pi k}\right)^{\eta}, \,\,\,\,\, k=1,2,\cdots .
\label{sigmak}
\end{equation}
In addition, the amplitude of the zeroth mode, $A_0$, is always set
by the normalization constraint.  For the same sets of figures, it is
enough to have $k_c=1000$ and $5000$ respectively, and then we should
see very little effects associated with the finiteness of $k_c$.

Coded in such a way, the simulations are extremely computationally
intensive. Therefore, the Monte Carlo averages given here are only
over 500 runs, and fluctuation determinants are calculated according
to Eq.~(\ref{occam_num}), but not using numerical path integration.

\section{Learning with the correct prior}

\begin{figure}[t]
  \centerline{\epsfxsize=0.7\hsize\epsffile{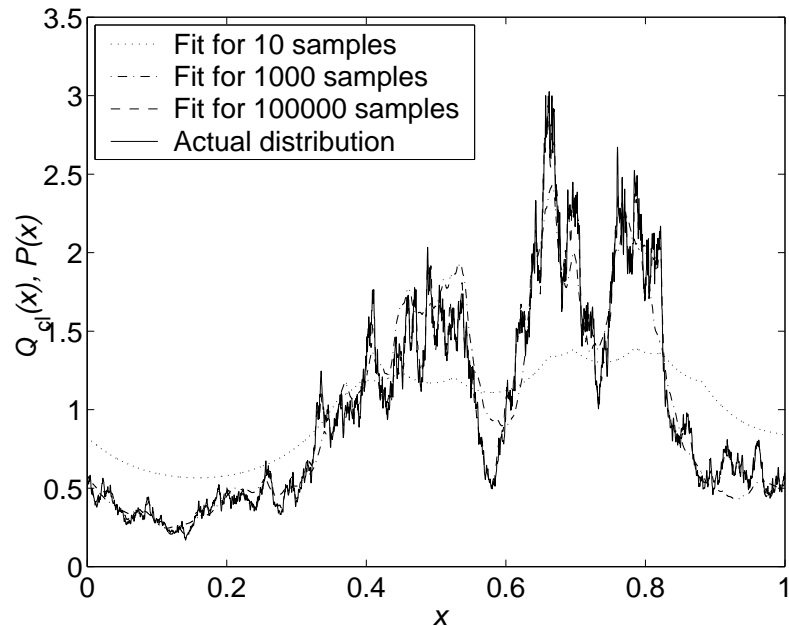}}
\caption{$Q_{\rm cl}$ found for different
  $N$ at $l=0.2$.}
\label{correct_ex}
\end{figure}

As an example of the algorithm's performance, Fig.~(\ref{correct_ex})
shows one particular learning run for $\eta = \eta_a = 1$ and
$l=l_a=0.2$. We see that singularities and overfitting are absent even
for $N$ as low as $10$. Moreover, the approach of $Q_{\rm cl}(x)$ to
the actual distribution $P(x)$ is remarkably fast: for $N=10$, they
are similar; for $N=1000$, very close; for $N=100000$, one needs to
look carefully to see the difference between the two.

\begin{figure}[ht]
  \centerline{\epsfxsize=0.7\hsize\epsffile{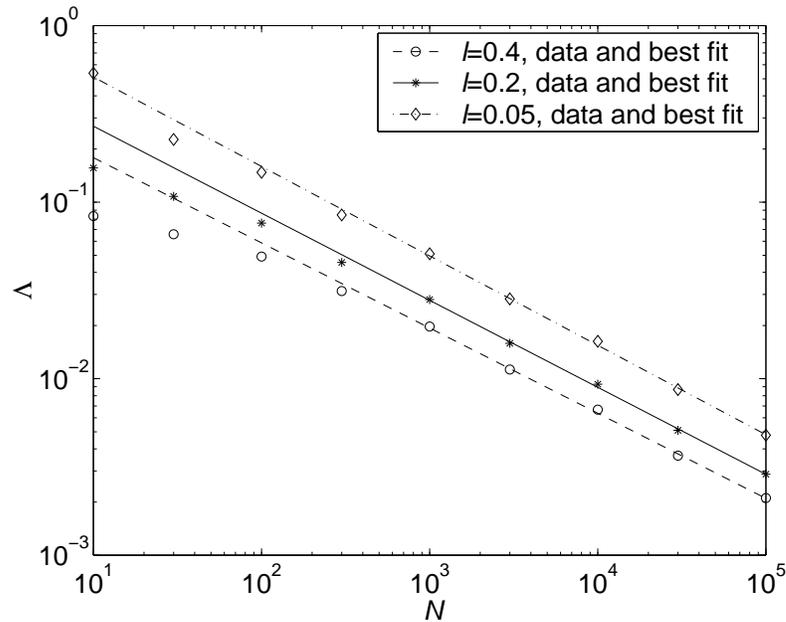}}
\caption[$\Lambda$ as a function of $N$ and $l$]
{$\Lambda$ as a function of $N$ and $l$.  The best fits are: for
  $l=0.4$, $\Lambda = (0.54 \pm 0.07) N^{-0.483 \pm 0.014}$; for
  $l=0.2$, $\Lambda = (0.83 \pm 0.08) N^{-0.493 \pm 0.09}$; for
  $l=0.05$, $\Lambda = (1.64 \pm 0.16) N^{-0.507 \pm 0.09}$.  }
\label{correct}
\end{figure}

The next question on our list is the behavior of the learning curves.
For the same $\eta$ and $l=0.4,\,0.2,\,0.05$, these are shown on
Fig.~(\ref{correct}).  One sees that the exponents are extremely close
to the expected $1/2$, and the ratios of the prefactors are within the
errors from the predicted scaling $\sim 1/\sqrt{l}$.  All of this
means that the proposed algorithm for finding densities not only works
as predicted, but is, at most, a constant factor away from being
optimal in using the predictive information of the sample set.
 
Note also that the data points approach their asymptotic regimes very
differently for different values of $l$: the bigger $l$ is, the lower
the data start compared to their respective fits. This is explainable
in view of the fact that smoother distributions usually vary over
smaller ranges. For example, for $l=0.4$ the target distribution
$P(x)$ usually takes values from $\sim 0.5$ to $\sim 2$. On the other
hand, for not too large $N$ the estimates are also smooth and close to
being uniform.  Therefore, the KL divergence usually comes out small
in this case. Thus the $l=0.4$ data starts so low not because we
manage to learn the distribution extremely well for $N=10$, but
because almost any guess is as good as any other at this level of
detail.

\section{Learning with `wrong' priors}
\label{wrong_pr_sec}

We stress first that there is no such thing as a {\em wrong prior}. If
one admits the possibility of a prior being wrong, then that prior
does not encode all of our a priori knowledge!  It does make sense,
however, to ask what happens if the distribution we are trying to
learn is an extreme outlier in the prior ${\cal P}[\phi(x)]$.  One way
to generate such an example is to choose a typical function from a
different prior ${\cal P}'[\phi(x) ]$, and this is what we mean by
`learning with an incorrect prior.' To study this we learn using
$\eta=1$ and $l$, but we choose our target distributions from the
prior Eq.~(\ref{prior_numeric}) with different values of $\eta_a$ and
$l_a$.

\begin{figure}[t]
  \centerline{
    \epsfxsize=0.7\hsize\epsffile{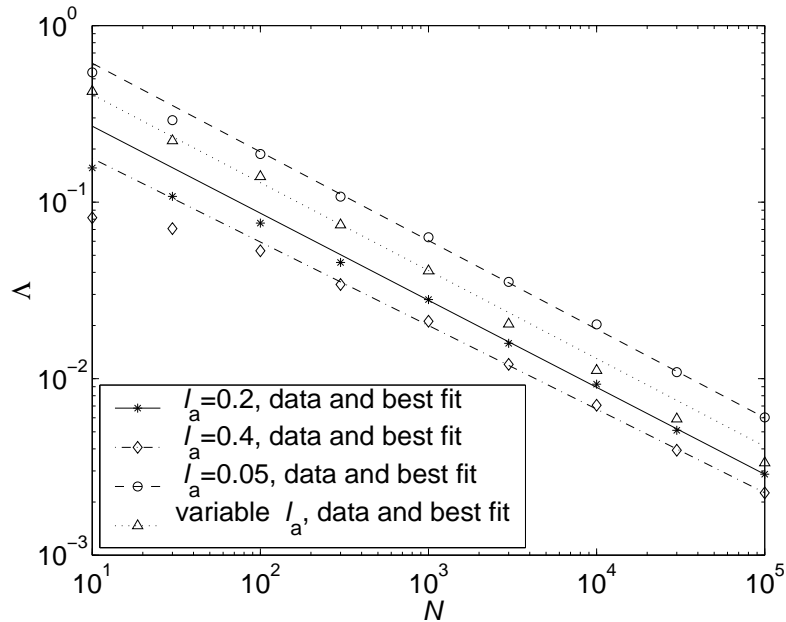}}
\caption[$\Lambda$ as a function of $N$ and $l_a$]
{$\Lambda$ as a function of $N$ and $l_a$. Best fits are: for
  $l_a=0.4$, $\Lambda = (0.56 \pm 0.08) N^{-0.477 \pm 0.015}$; for
  $l_a=0.05$, $\Lambda = (1.90 \pm 0.16) N^{-0.502 \pm 0.008}$; for
  variable $l_a$, $\Lambda = (1.28 \pm 0.13) N^{-0.498 \pm 0.016}$. In
  all cases we learn with $l=0.2$.  }
\label{incorrect_same}
\end{figure}

\begin{figure}[t]
  \centerline{
    \epsfxsize=0.7\hsize\epsffile{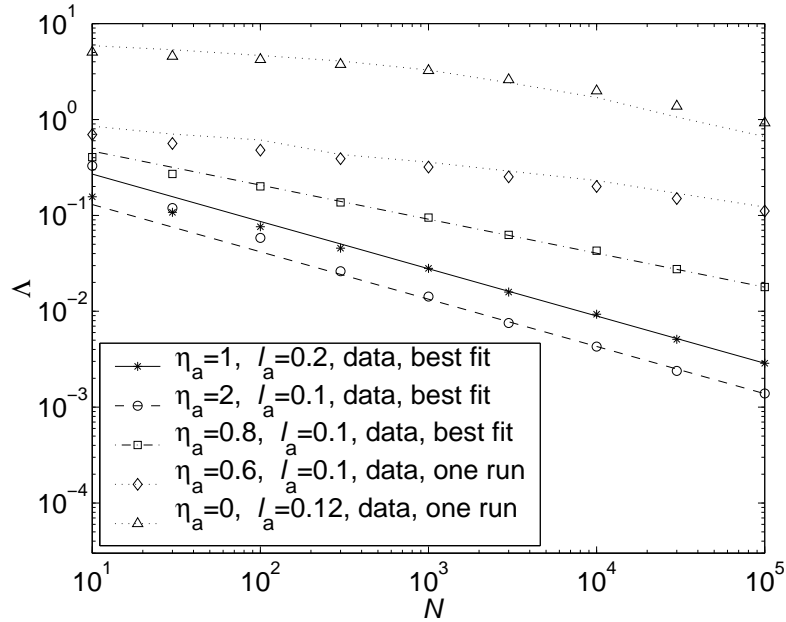}}
\caption[$\Lambda$ as a function of $N$, $\eta_a$ and $l_a$]
{$\Lambda$ as a function of $N$, $\eta_a$ and $l_a$.  Best fits: for
  $\eta_a=2$, $l_a=0.1$, $\Lambda = (0.40 \pm 0.05) N^{-0.493 \pm
    0.013}$, for $\eta_a=0.8$, $l_a=0.1$, $\Lambda = (1.06 \pm 0.08)
  N^{-0.355\pm 0.008}$.  $l=0.2$ for all graphs, but the one with
  $\eta_a=0$, for which $l=0.1$.}
\label{incorrect_diff}
\end{figure}

If the prior is wrong in the above sense, and the learning process is
as usual,
Eqs.~(\ref{solution_bcs},~\ref{dist_bcs}--\ref{stationary_bcs}), then
we still expect the asymptotic behavior, Eq.~(\ref{lcurve_num}), to
hold.  Indeed, once $\phi_{\rm cl}$ becomes close to $-\log P$, it
takes the same time to discover that the distribution's features at
the current relevant scale $\xi (N)$ are as expected, too big, or
almost absent.  Thus only the prefactors of $\Lambda$ should change,
and those must increase because there is an obvious advantage in
having the right prior. We illustrate these points in
Figs.~(\ref{incorrect_same}, \ref{incorrect_diff}).

Figure~(\ref{incorrect_same}) shows the learning curve for
distributions generated with the `actual' smoothness scale $l_a=
0.4,\, 0.05$ and studied using the `learning' smoothness scale $l=0.2$
(we show the case $l_a=l=0.2$ again as a reference). The $\Lambda \sim
1/\sqrt{N}$ behavior is seen unmistakably. The prefactors are a bit
larger (unfortunately, insignificantly) than the corresponding ones
from Fig.~(\ref{correct}), so we may expect that the `right' $l$,
indeed, provides better learning (see Section~\ref{sm_scale_sec} for a
detailed discussion).  Finally, the approach to the asymptotes again
is different for the different examples considered, but it is still
explainable by the argument we used for Fig.~(\ref{correct}).

To generate outliers that are even more uncommon than the ones just
discussed one may want to distort the $x$ axis (use different
parameterization), and this results in a variable smoothness scale
$l_a(x)$.  As an example, Fig.~(\ref{incorrect_same}) shows the
learning curve for $l_a=0.2$ distributions that have been rescaled
according to $x \to x-0.9 \sin(2\pi x/L)$. For the rescaled variable,
$l_a(x)$ varies from $0.02$ to $0.38$. Two separate straight line
fits---through the first five (shown) and the last four points---are
possible for this data.  Each of the fits separately agrees with the
prediction, but their prefactors are different. This is probably just
a numerical artifact because $1000$ Fourier modes used here feel like
much less in some places due to the rescaling, and this shows up at
large enough $N$.  Alternatively, this may be an indication that a
detailed analysis of the reparameterization invariant formulation
(Periwal 1997, 1998) is needed.

Finally, Fig.~(\ref{incorrect_diff}) illustrates learning when not
only $l$, but also $\eta$ is `wrong' in the sense defined above. We
illustrate this for $\eta_a=2,\,0.8,\,0.6,\,0$ (remember that only
$\eta_a>0.5$ removes UV divergences). Again, the inverse square root
decay of $\Lambda$ should be observed, and this is evident for
$\eta_a=2$. The $\eta_a=0.8,\,0.6,\, 0$ cases are different: even for
$N$ as high as $10^5$ the estimate of the distribution is far from the
target, thus the asymptotic regime is not reached. This is a crucial
observation for our subsequent analysis of the smoothness scale
determination from the data (Section~\ref{sm_scale_sec}). Remarkably,
$\Lambda$ (both averaged and in the single runs shown) is monotonic,
so even in the cases of {\it qualitatively} less smooth distributions
{\it there is still no overfitting}. On the other hand, $\Lambda$ is
well above the asymptote for $\eta=2$ and small $N$, which means that
initially too many details are expected and wrongfully introduced into
the estimate, but then they are almost immediately ($N \sim 300$)
eliminated by the data.

\section{Selecting the smoothness scale}
\label{sm_scale_sec}

Results presented in the last Figures already suggest that Occam
factors should work in this infinite dimensional case, and that it indeed
is possible to see this in numerical simulations. The competition
between the data and the Occam factor is equivalent to minimizing the
expression [cf.~Eq.~(\ref{dist_bcs},~\ref{action_bcs})]
\begin{equation}
H^*[\phi_{\rm cl}; \{x_{\rm i}\};l]  =
\int dx \frac{l}{2}(\partial_x \phi_{\rm cl})^2
+ \sum_{j=1}^N \phi_{\rm cl}(x_j)
+ \frac{1}{2}\sqrt{\frac{N}{l l_0}} 
\int dx 
\exp \left[-\phi_{\rm cl}(x)/2\right],
\label{tominimize}
\end{equation}
which makes the total probability of the data maximal, and thus the
length nee\-ded to code it minimal. How does the smoothness scale
$l^*$ that minimizes $H^*$ behave?  The data term [second in
Eq.~(\ref{tominimize})] on average is equal to $N D_{\rm KL} (P ||
Q_{\rm cl})$, and it can be small compared to the other terms for very
regular $P(x)$.  Then only the kinetic and the fluctuation terms
matter, and $l^* \sim N^{1/3}$, as was obtained by Bialek, Callan, and
Strong (1996).  For less regular distributions $P(x)$ [cf.~graphs for
$\eta_a=0,0.6,0.8$ on Fig.~(\ref{incorrect_diff})], this is not true.
Indeed, for $\eta=1$, $Q_{\rm cl}(x)$ approximates large scale
features of $P(x)$ very well, but details at scales smaller than
$\sim\sqrt{l/NL}$ are not present in it.  If $P(x)$ is taken from the
prior, Eq.~(\ref{prior_numeric}), characterized by some $\eta_a$ and
$l_a$, then according to Eq.~(\ref{sigmak}) the contribution of these
details falls off with the wave number $k$ as $\sim
(L/l_a)^{\eta_a-1/2} k^{-\eta_a}$.  Thus the expected data term is
\begin{equation}
ND_{\rm KL}(P||Q_{\rm cl}) \sim N 
\int_{\sqrt{NL/l}}^{\infty}\;\left({L \over {l_a}}\right)^{2\eta_a-1} 
k^{-2\eta_a}
= N \left({L \over {l_a}}\right)^{2\eta_a-1} 
\left({NL \over l}\right)^{-\eta_a+1/2},
\end{equation}
and this is not necessarily small. For $\eta_a <1.5$ it actually
dominates the kinetic term and competes with the Occam factor to
establish the relevant smoothness scale. Summarizing,
\begin{eqnarray}
l^* &\sim& N^{1/3}, \,\,\,\,\, \eta_a \ge 1.5
\label{l_dominant_bcs}
\\
l^* &\sim& N^{(\eta_a -1)/\eta_a},\,\,\,\,\, 0.5 <\eta_a < 1.5\,.
\label{l_dominant_my}
\end{eqnarray}
There are two noteworthy things about Eq.~(\ref{l_dominant_my}).
First, for $\eta_a=\eta=1$, $l^*$ stabilizes at some constant value,
which we expect to be equal to $l_a$. Second, even if $\eta_a \neq
\eta$, but $\eta_a<1.5$, then Eqs.~(\ref{lcurve_num},
\ref{l_dominant_my}) ensure that $\Lambda\sim N^{1/2 \eta_a -1}$, and
this asymptotic behavior will be reached almost immediately, unlike in
the $\eta_a=0,0.6$ examples from Fig.~(\ref{incorrect_diff}). This
performance is, at most, a constant factor away from the limits set by
heuristic calculations of predictive information,
Eq.~(\ref{S1general}), with the `right' priors $\eta_a=\eta \neq
1$---a remarkable result!

\begin{figure}[t]
  \centerline{\epsfxsize=.7\hsize\epsffile{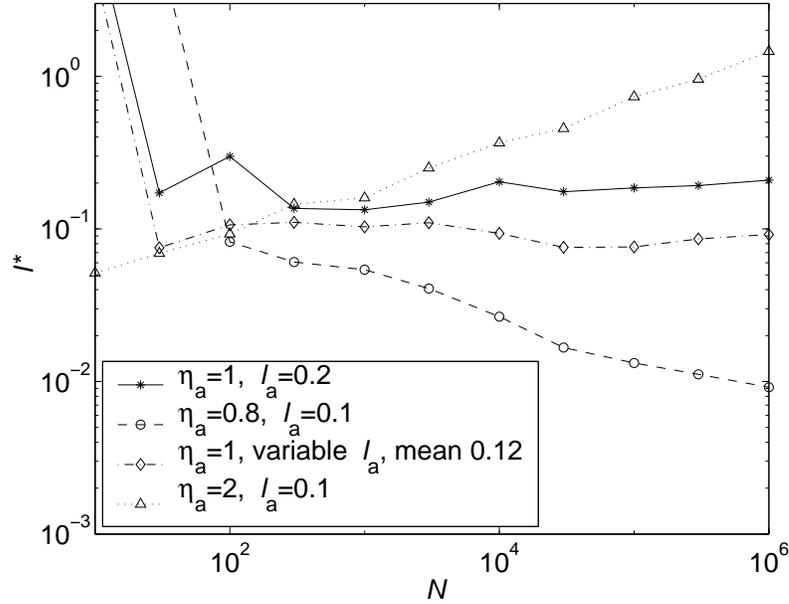}}
\caption[Smoothness scale selection by the data]
{Smoothness scale selection by the data. The lines that go off the
  axis for small $N$ symbolize that $H^*$ monotonically
  decreases as $l \to \infty$.}
\label{scale_select}
\end{figure}

\begin{figure}[t]
  \centerline{\epsfxsize=.7\hsize\epsffile{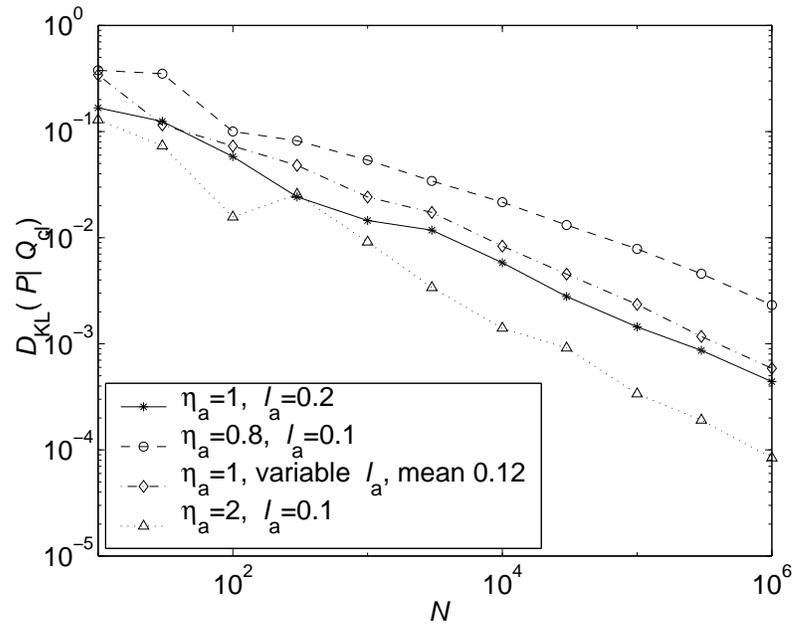}}
\caption{Learning with the data induced smoothness scale.} 
\label{adaptive_D}
\end{figure}

We present simulations relevant to these predictions in
Figs.~(\ref{scale_select},~\ref{adaptive_D}). Unlike in the previous
Figures, these results are not averaged due to extreme computational
costs, so all our further claims, which are inherently statistical,
have to be taken cautiously. On the other hand, selecting $l^*$ and
observing the effects associated with it in single runs has some
practical advantages since then we are able to ensure the best
possible learning for any given realization of the data, not just on
average.

Figure~(\ref{scale_select}) shows single learning runs for various
$\eta_a$ and $l_a$. In addition, to keep the Figure readable, we do
not show runs with $\eta_a=0.6,\,0.7,\,1.2,\,1.5,\,3$, and $\eta_a \to
\infty$, which is a finitely parameterizable distribution.  All of
these display a good agreement with the predicted scalings,
Eq.~(\ref{l_dominant_bcs},~\ref{l_dominant_my}).

\begin{figure}[t]
  \centerline{\epsfxsize=.7\hsize\epsffile{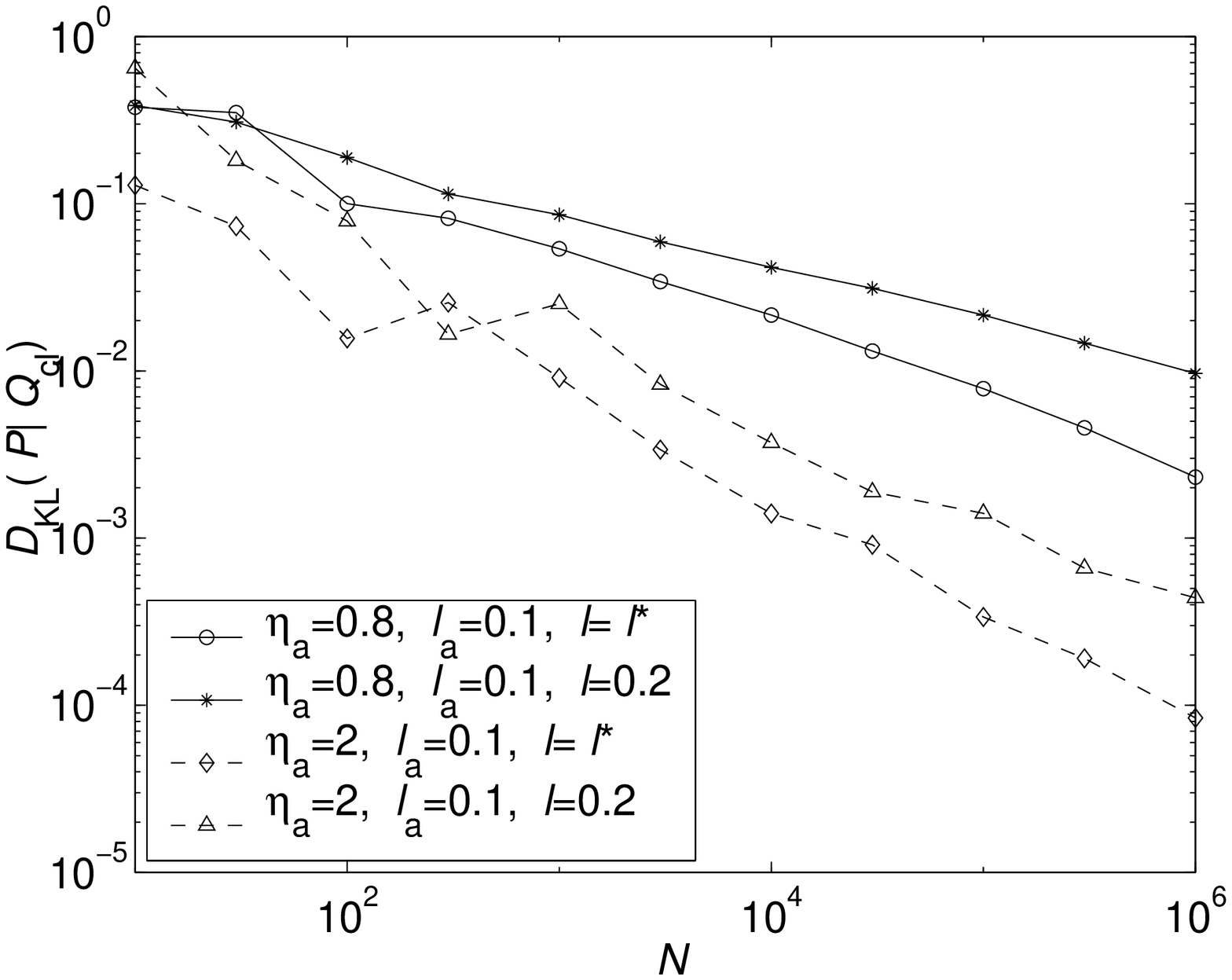}}
\caption[Learning speed comparison]
{Comparison of learning speeds for the same data sets with different a
  priori assumptions. In all runs we learn using the model with
  $\eta=1$, and $l$ is either predefined, or set by the Occam factor.}
\label{compare}
\end{figure}

Figure~(\ref{adaptive_D}) shows the KL divergence between the target
distribution and its classical estimate calculated at $l^*$; the
average of this divergence over the samples and the prior is the
learning curve. Again, the predictions are clearly fulfilled.  Note
that for all $\eta_a$ with exception of $\eta_a=\eta=1$ there is
indeed a {\em qualitative} advantage in using the data induced
smoothness scale. To illustrate this more clearly and ease the
comparison we replotted some of the curves with adaptive $l$ side by
side with their fixed $l$ analogues on Fig.~(\ref{compare}).

\section{Can the `wrong' prior help?}
\label{num_help_sec}

The last four Figures have illustrated some aspects of learning with
`wrong' priors. However, more importantly, all of our results may be
considered as belonging to the `wrong prior' class. Indeed, the actual
probability distributions we used were not nonparametric continuous
functions with smoothness constraints, but were composed of $k_c$
Fourier modes and thus had exactly $2k_c$ parameters. Usually it would
take well over $2k_c$ samples to even start to close down on the
actual value of the parameters, and many more to get accurate results.
However, using the wrong continuous parameterization [$\phi(x)$] we
were able to obtain good fits for as low as $1000$ samples
[cf.~Figs.~(\ref{correct_ex})] with the help of the prior
Eq.~(\ref{prior_bcs}). Moreover, learning happened continuously and
monotonically without huge chaotic jumps of overfitting that
necessarily accompany any brute force parameter estimation method at
low $N$. So, for some cases, a {\em seemingly more complex model} is
actually {\em easier} to learn!

We can summarize this by stating that, when data are scarce and the
parameters are abundant, one gains even by using the regularizing
powers of wrong priors. The priors select some large scale features
that are the most important to learn first and fill in the details as
more data become available. If the global features are dominant
(arguably, this is generic), one actually wins in the learning speed
[cf.~Figs.~(\ref{correct}, \ref{incorrect_same}, \ref{adaptive_D})].
If, however, small scale details are as important, then one is at
least guaranteed to avoid overfitting
[cf.~Fig.~(\ref{incorrect_diff})]. 

Thus we can argue that our numerical experiments support the
Occam--like claim we made in Section~\ref{pred_discus_sec}: if two
models provide equally good fits to data, {\em the simpler one should
  always be used}.  In particular, using Eq.~(\ref{Nc}) we see that
for our simulations, the nonparametric QFT model is simpler (as
characterized by the predictive information) than a finite dimensional
one for $N<N_c \sim (k_c \log k_c)^2$. We operate in this regime in
all of our simulations, and so we must learn faster and with less
overfitting if we use the wrong parameterization. 
Note, that these results are very much in the spirit of our whole
program: not only is the value of $l^*$ selected that simplifies the
description of the data, but the continuous parameterization itself
serves the same purpose. This is an unexpectedly neat generalization
of the MDL (Rissanen 1989) principle to nonparametric cases.

\section{Discussion}

The field theoretic approach to density estimation in principle not
only regularizes the learning process but also allows the
self--consistent selection of smoothness criteria through an infinite
dimensional version of the Occam factors. We have shown numerically
that this works, even more clearly than was conjectured. For
$\eta_a<1.5$, $Q_{\rm est}$ and the learning curve $\Lambda$ truly
become properties of the data and not of the Bayesian prior used for
learning: one can set a learning machine to work at $\eta=1$ and
be sure that this does not bias the estimates in any excessive
way. If we can extend these results to include $\eta_a > 1.5$ and
combine this work with the reparameterization invariant formulation of
Periwal (1997, 1998), this should give a complete theory of Bayesian
learning for one dimensional distributions, and this theory has no
arbitrary parameters. In addition, if this theory properly treats the
limit $\eta_a \to \infty$, we should be able to see how the
well--studied finite dimensional Occam razors and the MDL principle
arise from a more general nonparametric formulation.

These results also have some biological implications. First of all, it
may be that this smoothness scale adaptation mechanism is partly
responsible for a commonly known effect: children learn faster than
adults. More seriously, and more closely connected to the models
discussed here is the learning and development of smooth `maps' in the
nervous system (see, for example, Knudsen et al.~1987). These maps
become much less susceptible to the sensory inputs as time passes, and
this may be interpreted in terms of stiffening of the smoothness
constraint.  Indeed, starting from scratch, it is very easy to drift
the smoothness scale to such large values that susceptibility of the
learning machine (in other words, an animal) to the new data will be
extremely small.

Second, if our conclusions are correct, then a learning machine that
is programmed to solve problems at $\eta=1$ can {\em easily solve more
  complex problems} with any $\eta_a$, $1.5>\eta_a >0.5$, by
performing a simple averaging over the smoothness scales. At worst,
this procedure may lead to a constant multiplicative drop in
performance.  By analogy, we may expect that, once an animal is able
to treat a problem that falls in any power--law class with respect to
the predictive information, then it is able to treat any problem that
provides more predictive information with only a multiplicative
overhead.  Since, as we have already discussed (Hilberg 1990, Ebeling
and P{\"o}schel 1994, P\"oschel, Ebeling, and Ros\'e 1995), humans can
solve power--law problems, it is encouraging to know that there is no
learning task in this world that is, in principle, too difficult for
us (our lifetime is the only limiting factor). More seriously, if it
is, indeed, possible to construct a complete theory of one (and,
possibly, higher) dimensional learning, where both the smoothness
scale and the exponent can be self--consistently determined, then the
questions we asked in Section~\ref{pred_discus_sec} (like ``what
models do humans use for learning?'') may lose their meaning---any
model is almost as good as any other, and it is very difficult to look
for possible multiplicative differences between them. Surprisingly,
these questions are meaningful if such a theory cannot be constructed.
In this case, as we saw in
Eqs.~(\ref{l_dominant_bcs},~\ref{l_dominant_my}) and on
Figs.~(\ref{scale_select},~\ref{adaptive_D}), a complex learning
machine cannot effectively adapt to simple tasks. This again accords
with our subjective experience that it is sometime very hard to find a
simple solution when expecting a complex one. It should be possible to
devise an experiment that would quantify this failure to solve simple
problems in complex contexts.


\chapter{Learning discrete variables:\\
Information--theoretic regularization}
\label{it_reg}

\section{The general paradigm}

In Chapters \ref{predictive} and \ref{numeric} of this work we
discussed what we consider to be some of the most interesting problems
in modern statistics and learning theory. We defined predictive
information and complexity, studied the learning of nonparametric
distributions, and showed an example of how Occam factors generalized
to infinite dimensional problems. There is one tantalizing question
that followed us through this whole discussion: many problems require
a prior to regularize learning, but then how can one make sure that
the estimates and the conclusions are inferred from the data, rather
then being imposed by some a priori knowledge that the data do not
support? The results in the infinite dimensional generalization of the
Occam factors seemed especially interesting in this
respect---estimates can become almost insensitive even to the
qualitative choice of prior, at least in a broad range. This
conclusion, together with Periwal's (1997, 1998) work, are the only
results known to us that aim towards constructing a reparameterization
and prior invariant (or, better yet, ignorant) theory of nonparametric
Bayesian inference.

Even though we have concentrated on nonparametric problems, similar
difficulties also exist in seemingly simpler, parametric cases. The
choice of the prior for finite parameter learning scenarios is still a
hot topic, and various alternatives are proposed that make some
universal theoretical sense within the framework of information
theory, MDL theory, etc.  Examples include universal priors (Rissanen
1983, Lee and Vit\'anyi 1993) or Jeffreys' priors (Clarke and Barron
1994, Balasubramanian 1997).

We think, however, that the prior really should embody a priori
knowledge, and thus we cannot agree with the use of universal,
globally definable choices. On the other hand, there is an obvious
appeal for approaches based on more general theoretical principles.
For example, the problems of nonuniform convergence of the estimate to
the target, Eq.~(\ref{psi2_app}), and of the infrared divergence of
predictive information, Eq.~(\ref{almostdone}), are easily resolved if
one turns to reparameterization invariant priors (Periwal 1997, 1998).
These have a clear theoretical edge over non--invariant ones since
they estimate a true density, that is, a function that transforms like
a density under reparameterizations of the independent variable.

As we tried to emphasize in this work, learning is information
accrual, and, therefore, it is a part of Shannon's information theory.
Thus a general theoretical principle can be to construct priors solely
from information--theoretic quantities like entropy (or
self--information), Kullback--Leibler divergence (or relative
information), various mutual informations, etc., and with no other
constraints.  In addition, since information--related quantities are
formed from log--probabilities it makes sense to include them
exponentially into the priors, which are, after all, also
probabilities. In Sections~\ref{it_toytheory_sec} and
\ref{it_toynum_sec} we will present a simple example that illustrates
the use of this general principle---regularization with
information--theoretic quantities---and show that it is possible and
advantageous to learn with such priors.

\section{Discrete variables: a need for special attention}

When learning probability densities of continuous variables, except
for the very simple cases $N>>d_{\rm VC}$, priors are used to balance
the quality of fits to the data against the complexity of solutions
(cf.~Section~\ref{fluct_prior_sec} and Chapter~\ref{numeric}); this
smoothing of data prevents overfitting. It is easy to construct
smoothing priors for continuous variables: continuity implies a
metric, so locality is defined, and then one just needs to punish
distributions that exhibit large variations over small distances.
Such a `smoothness' prior will work as a regularizer.

The case when a variable is discrete, and the (discrete) metric is
impossible to define, presents a problem. Usually this case is not
considered interesting, because the law of large numbers guarantees
that, at large $N$, the frequencies of events estimate probabilities
well. However, if the number of examples $N$ is smaller or comparable
to the number of possible outcomes $K$, then statistical fluctuations
are large. This situation is not as uncommon as one might hope. For
example, it is possible that syntactic structures in a language can be
inferred from statistical arguments alone (see, for example, Pereira et
al.~1993, Manning and Sch\"{u}tze 1999).  Estimating probabilities of
occurrence of a few thousand common words is rather easy. It is even
feasible to construct conditional distributions of nouns given verbs
(Pereira et al.~1993).  However, it is totally unrealistic to expect
to build an accurate probability distribution of whole sentences from
the raw data without some a priori knowledge.

Similar problems appear in molecular biology. For example, gene
expression is governed by promoter regions in DNA molecules.  These
regions are usually thought to be constructed from two five base pair
long blocks (there are $4^5$ possible different realizations of these)
that appear anywhere inside `junk' genetic material, which is about a
hundred bases in length.  If one tries to find the meaningful $5+5$
structures by statistical methods (see, for example, van Helden et
al.\ 1998), then a probability distribution over $4^5*4^5\sim 10^6$
possibilities must be constructed. Many experiments are done on yeast,
and their genome is only a few millions of base pairs long. So getting
a full statistics is, at best, problematic.  Even worse, if one tries
to look for a possibility of promoters of a different length, then the
problem becomes totally hopeless.

To solve these and similar problems, one needs smoothing priors that
regularize fluctuations. However, now there is no notion of locality
to create them. It is not at all evident how to impose smoothness
conditions, or whether these conditions will speed up learning in any
way. It is clear though that any smoothing must be global---we cannot
talk about local changes, but global variability is well defined.
Recall that variability can be measured non--metrically by entropies,
and these have a very special meaning in information theory---they are
the unique measure of information (Shannon 1948).  Therefore,
learning a discrete variable may turn out to be an excellent example
of the general paradigm introduced above.

In the next Sections we present a toy model for learning a discrete
variable and show that it is possible to regularize and speed up such
learning with the help of information--theoretic priors. In general,
discrete calculus is much less developed than its continuous analogue
(precisely for the reason that the notion of locality is absent), so
analytical results can usually be achieved only for very simple
problems, and our example is like that. Nonetheless, this case is of
interest as it solves some real world problems and, more importantly,
because it develops techniques that later will be used in more
complicated tasks; the work on those is already in progress.

\section{Toy model: theory}
\label{it_toytheory_sec}

Consider the following `real world' problem. A US Census Bureau needs
a preliminary report on statistics of people in Trenton, NJ based on
the Census--2000 data.  Unfortunately, only a few thousand households
have filled in their Census cards, and this is clearly not enough to
sample adequately many classes $x,\, x=1\cdots K$ (we also call them
{\em possible outcomes} or {\em bins}) into which the people are
classified (ethnicity, marital status, educational level, size of the
household, etc.) Suppose now that Newark, NJ, perceived to have a
similar population statistics, has been quick to organize
door--to--door counting of people, so that a good sampling of Newark's
population, $Q^*(x)$, is available. Can the Census Bureau
statisticians use these data to answer their questions about the
Trenton people? An obvious solution would be to take a weighted
average of the (undersampled) Trenton counts and the (well sampled)
Newark probabilities. But how should the weights be set in order to
ensure that the Trenton estimate is just smoothed and not unfairly
biased by the Newark data?

We can offer a solution to the problem by choosing an a prior
probability density for $Q(x)$, the Trenton distribution, that is
biased towards the reference Newark distribution $Q^*(x)$. This may be
done in the following form
\begin{equation}
{\cal P} [Q(x), \lambda] = \frac{1}{{\cal Z}_Q(\lambda)}
\exp \left[-\lambda D (Q^*||Q) \right]
\, \delta \left( \sum_{x=1}^K Q (x) - 1 \right) \, {\cal P}(\lambda),
\label{prior_d}
\end{equation}
where ${\cal P} (\lambda)$ is some a priori normalized density for the
inverse temperature--like parameter $\lambda$, ${\cal
  Z}_Q(\lambda)\equiv\int [dQ(x)] {\cal P} [Q(x)||\lambda]$ is the
normalization of the a priori distribution of $Q(x)$ conditional on
$\lambda$, and $D$ is some measure of distance between the two
distributions, $Q^*$ and $Q$.  If we want to stick to our paradigm of
using information--theoretic quantities only, then we do not have much
freedom in selecting $D$ since the natural distance between any two
distributions in information theory is the familiar Kullback-Leibler
divergence,
\begin{equation}
D(Q^*||Q)\equiv D_{\rm KL}(Q^*||Q)=\sum_{x=1}^K Q^*(x) 
\log \frac{Q^*(x)}{Q(x)}.
\label{doneway}
\end{equation}
In the language of coding theory, this choice of $D$ means that we
optimize our coding strategies for $Q$, but we want $Q$'s such that
the data coming from the reference distribution can be coded compactly
also. We could have chosen to measure distances in the opposite
direction and switch arguments in the KL divergence.  Then the best
coding for $Q^*$ is fixed, and we look for $Q$ that is still coded
well.  We chose Eq.~(\ref{doneway}) over the other choice because this
allows an exact solution.

Now, similarly to Bialek et al.~(1996, see also
Appendix~\ref{bcs_app}), we apply the Bayes formula to get the
probability density for $Q(x)$ and $\lambda$ given the data $\{x_{\rm
  i} \}$
\begin{eqnarray}
P [Q(x), \lambda| \{x_{\rm i}\}]
&=&\frac{P [\{x_{\rm i}\}|Q(x),\lambda]\, {\cal P}[Q(x),\lambda]}
{P(\{x_{\rm i}\})}
\nonumber
\\
&=& \frac{ {\cal P} [Q(x),\lambda] \prod_{i=1}^N Q(x_{\rm i})}
{\int[dQ(x)] d\lambda 
{\cal P} [Q(x),\lambda] 
\prod_{i=1}^N Q(x_{\rm i})}\,.
\end{eqnarray}
The least square estimator of $Q(x)$ is then, as usual,
\begin{eqnarray}
Q_{\rm est}(x|\{x_{\rm i}\})&=&
\int d \lambda [dQ(x)]  \; Q(x)\, P [Q(x),\lambda|\{x_{\rm i}\}] 
\nonumber
\\
&=&
\frac{\langle Q(x) Q(x_1) Q(x_2)\cdots
  Q(x_N)\rangle^{(Q,\lambda)}}
{\langle Q(x_1)Q(x_2) \cdots Q(x_N) \rangle ^{(Q,\lambda)}}\,,
\label{Qest_d}
\end{eqnarray}
where $\langle \cdots \rangle^{(Q,\lambda)}$ stands for averaging over
$Q$ and $\lambda$ with respect to the prior; similarly, $\langle
\cdots \rangle^{(Q)}$ means averaging only over ${\cal P} [Q
|\lambda]$.

If $\lambda$ was fixed, then the averaging in Eq.~(\ref{Qest_d}) would
have one integral less---a simpler problem. However, varying it may
allow the Occam factor that arises from volumes in the $Q$-space to
find some $\lambda^*$ that creates the shortest (thus the most
probable) description of the data (cf.~Section~\ref{sm_scale_sec}).
By achieving the optimal balance between the `goodness of fit' and
closeness to the reference, this may resolve the problem of a possible
erroneous bias towards $Q^*$.

As mentioned above, the solution of this model is rather simple. We
write $n(x)$ for the data count in the bin $x$, $\sum_x n(x)=N$.
Then, leaving aside the $\lambda$ integral for a while, we have (see
Appendix~\ref{integr_app}):
\begin{eqnarray}
\langle Q(x_1) \cdots Q(x_N) \rangle ^{(Q)}&=&
\int [d Q] 
\frac{{\rm e}^
{- \lambda D_{\rm KL}(Q^*||Q)}}
{{\cal Z}_Q(\lambda)} \delta (\sum_x Q(x) -1)
\prod_x Q(x)^{n(x)} 
\label{ans_d_1}
\\
&=&
\frac{{\rm e}^{S[Q^*]}}{{\cal Z}_Q(\lambda)} 
\int [dQ]  \delta (\sum_x Q(x) -1) 
\prod_x Q(x)^{n(x)+\lambda Q^*(x)} 
\label{ans_d_2}
\\
&=&
\frac{{\rm e}^{S[Q^*]}}{{\cal Z}_Q(\lambda)}
\;\frac{\prod_{x=1}^K \Gamma \Big(n(x)+\lambda Q^*(x) +1 \Big)}
{\Gamma(\lambda +N +K)},
\label{ans_d_3}
\end{eqnarray}
where $S[Q^*]$ is just the entropy of the reference distribution.
${\cal Z}_Q(\lambda)$ is given by the same integral,
Eq.~(\ref{ans_d_1},~\ref{ans_d_3}), but with $n(x)=N=0$. Therefore, if
we now integrate over $\lambda$ using the steepest descent technique,
then the most probable value of the inverse temperature is determined
by
\begin{equation}
\sum_{j=1}^{N} \frac{1}{\lambda^*+K+j-1}
-\sum_{x=1}^K \sum_{j=1}^{n(x)} \frac{Q^*(x)}{\lambda^* Q*(x) +j} =0.
\end{equation}
Numerically (cf.~Section~\ref{it_toynum_sec}) this equation has one
nontrivial solution for large $N$, so that the Occam factor seems to
work again.  Unfortunately, however, we were unable to obtain
satisfying analytical results with exception of some trivial
asymptotic limits. If, on the other hand, we keep $\lambda$ fixed,
then [see Eq.~(\ref{Qest_d})]
\begin{equation}
Q_{\rm est} (x|\{x_{\rm i}\}; \lambda)=\frac{n(x)+\lambda Q^*(x)+1}
{N+\lambda+K}
\label{Qest_res_d}
\end{equation}
The simplicity of this result is intriguing. Our initial suggestion to
average the actual undersampled data with the well sampled smoothing
distribution turns out to have deep roots in Bayesian inference with
the prior Eq.~(\ref{prior_d})!

Note the presence of `$+1$' in Eqs.~(\ref{ans_d_3},~\ref{Qest_res_d}).
Due to this term the estimated distribution is a smoothed version of
the counts $n(x)$ even if $\lambda \to 0$. In this theory, no bin has
an estimated probability of zero, so that $D_{\rm KL}(Q^*||Q)$ always
is well defined, and observing the next sample in any bin never is a
completely unexpected event.  This summand, which has so many
desirable consequences, appears because we define the prior,
Eq.~(\ref{prior_d}), on the space of $Q$'s. Changing the variables
from probabilities to likelihoods, $-\phi(x)=\log Q(x)$, creates a
Jacobian which effectively adds one count in every bin. Maximal
likelihood estimation of parameters does not have this feature, and
this is yet another argument in favor of the Bayesian formulation.

Even though this toy model has an exact solution, it still is
instructive to perform a saddle point (large $N$) analysis in the hope
that some useful knowledge reusable in more complex settings will be
gained.  With the usual change of variables,
\begin{equation}
Q(x)={\rm e}^{-\phi(x)},\,\,\,\,\, Q^*(x)={\rm e}^{-\phi^*(x)},
\end{equation}
leaving the $\lambda$ integral aside again, and replacing the
$\delta$-function by its Fourier representation, we get the following
expression for the correlation functions:
\begin{eqnarray}
\langle Q(x_1) \cdots Q(x_N)\rangle^{(Q)} &=&
\int \frac{[d\phi(x)]}{{\cal Z}_Q(\lambda)} \frac{d\mu}{2\pi}
{\rm e}^{-H [\phi(x),i\mu,\lambda,\phi^*(x)] - \sum_x [n(x)+1]
  \phi(x) },
\\
H [\phi(x),i\mu,\lambda,\phi^*(x)] &=&
\lambda \sum_x Q^*(x) [\phi(x)-\phi^*(x)]
+i\mu (\sum_x {\rm e}^{-\phi(x)} -1).
\end{eqnarray}
Calculating up to one--loop corrections using the steepest descent
techniques, this Hamiltonian results in
\begin{eqnarray}
\langle Q(x_1) \cdots Q(x_N)\rangle^{(Q)} &=&
{\rm e}^{-H_{\rm eff} [\phi_{\rm cl}(x),
  \lambda,\phi^*(x)] -\sum_x [n(x)+1]
  \phi_{\rm cl}(x)} ,
\label{corrthruexp_d}
\\
H_{\rm eff} [\phi_{\rm cl}(x), \lambda,\phi^*(x)]&=&
H [\phi_{\rm cl}(x),i\mu=N+\lambda+K, \lambda,\phi^*(x)]
\nonumber
\\
&+&
\frac{K}{2}\log \left(1+\frac{N}{\lambda}\right) +
\frac{1}{2}\sum_x (\phi^*(x)-\phi_{\rm cl}(x)),
\label{seff_d}
\\
Q_{\rm cl}(x) &\equiv& {\rm e}^{-\phi^{\rm cl}(x)} = 
\frac{n(x)+\lambda Q^*(x)+1}{N+\lambda+K}
\label{Qcl_d}
\end{eqnarray}
Again, just like in Eqs.~(\ref{occam_num},~\ref{action_bcs}), the
fluctuation determinant [the last two terms in Eq.~(\ref{seff_d})] has
a different $\lambda$ dependence than the data and the prior terms.
This suggests, even more clearly than the exact result
Eq.~(\ref{ans_d_3}), that competition between them will select a
nontrivial most probable value of $\lambda^*$.

Note that $Q_{\rm est}$, Eq.~(\ref{Qest_res_d}), is equal to $Q_{\rm
  cl}$.  A priori one should not expect them to be similar even for
large $N$. Indeed, the matrix of second derivatives in the saddle
point calculation has eigenvalues $\sim n(x)$, and these are small for
bins with small counts. So, in principle, one could expect large
discrepancies between the exact and the classical solutions.  The fact
that they {\em are} the same inspires a hope that the saddle point
analysis may remain useful for other, more complex, problems.

Finally, we want to illustrate yet another important aspect of this
toy model. What is the predictive information for this system? In
general, it is difficult to calculate, so we consider two very simple
limits: $\lambda \ll N, \, K \ll N$, and $ 1 \ll N \ll \lambda, K \ll
\lambda$. The first case closely parallels calculations of Section
\ref{learn_distr_sec} and yields
\begin{equation}
S_1(N) \approx \frac{K}{2} \log N.
\end{equation}
In the second case, somewhat lengthy calculations lead to
\begin{equation}
S(N) = N S [Q^*] +S_1 (N),\,\,\,\,\,\,\,
\lim_{N\to\infty, N/\lambda \to 0} S_1 \approx {\rm 0},
\end{equation}
where $S [Q^*]$ is again the entropy of the reference distribution.
These results are expected: for small $\lambda$ and large $N$ this
problem is just learning a $K$--parameter distribution, so
Eq.~(\ref{S1annealed}) should hold. On the other hand, for extremely
large $\lambda$, the estimate converges to the reference distribution
regardless of the data, so the problem is effectively zero
dimensional.

If we do estimates at some large, but fixed, $\lambda$, then we should
see a crossover from the zero to the $K$ parameter regime.
\footnote{If we average over $\lambda$, and $\lambda^*(N)/N$ starts
  large and drifts to below $1$ as $N$ increases, then we will still
  observe the crossover.  This behavior of $\lambda^*$ is possible
  since at low $N$ most of the weight should go to the smoothing term,
  while at large $N$ the actual counts are to be trusted.} For generic
distributions the crossover will be smooth, since each bin starts to
add its $(1/2) \log N$ to the predictive information independently
when the estimate for that bin switches from the reference value to
the count, i.~e., when $n(x)>\lambda Q^*(x)$
[cf.~Eq.~(\ref{Qest_res_d})].  A smooth crossover from zero to the
logarithm is possible only through a faster than logarithmic growth
for some range of $N$. Indeed, from the discussion in
Section~\ref{nonparam_gen_sec}, we know that continuous addition of
extra degrees of freedom is a sign of the power--law growth of
predictive information.  It is remarkable that a discrete system as
simple as this toy model can exhibit such a rich behavior that, so
far, has been associated with only very complex nonparametric models;
we will see a numerical demonstration of this in the next section.

\section{Toy model: numerical analysis}
\label{it_toynum_sec}

If we want to be able to observe all of the features described in the
previous section, then $K$, the number of bins, should be large enough
to allow for a prominent prior--dominated (scarce data) region, but it
also should be small enough so that we can generate enough samples and
observe a cross--over to the data--driven regime in all bins. The
choice of $K=75,100,125$ reasonably satisfies both conditions and will
be used in all of our simulations.

The next important question is the generation of random distributions
from the prior, Eq.~(\ref{prior_d}). Due to the $\delta$-function
normalization constraint this is a complicated task.  However, in the
limit of large $\lambda$, $D_{\rm KL} [Q^*||Q]$ typically is small and
can be approximated by the $\chi^2$ distance
\begin{equation}
\lim_{D_{\rm KL} \to 0} D_{\rm KL} [Q^*(x)||Q(x)] = {1 \over 2}
\sum_{x=1}^K \frac{[Q(x)-Q^*(x)]^2}{Q^*(x)}.
\label{chi2_d}
\end{equation}
Then the prior, Eq.~(\ref{prior_d}), becomes a multi--variable normal
density, and generation of random distributions is easy \footnote{The
  KL divergence tends to the $\chi^2$ measure from below.  Thus
  replacing $D_{\rm KL}$ with $\chi^2$ produces a slightly narrower
  prior; this difference becomes smaller as $\lambda$ grows.  In
  principle, this can produce dramatic changes in statistics, but for
  our choices of independent parameters this turns out to be almost
  irrelevant.}.

\begin{figure}[t]
  \centerline{\epsfxsize=0.7\hsize\epsffile{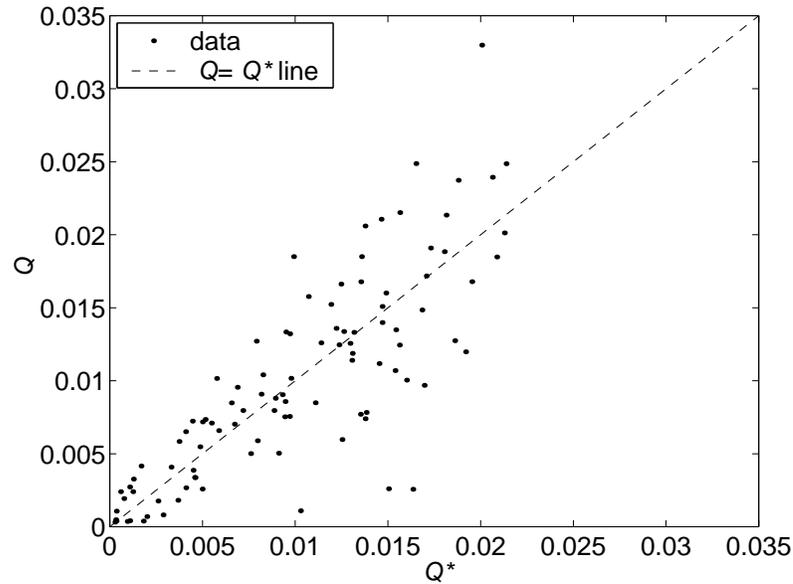}}
\caption[The reference and a random distributions]
{Comparison between the reference distribution and a typical random
  one generated with $\lambda=500$, and $K=100$. Each point
  corresponds to the values $\Big(Q^*(x), Q(x)\Big)$ in one of the
  bins $x$.}
\label{qrefdiff}
\end{figure}

The choice of $\lambda$ is motivated by the same arguments as the
choice of $K$. The asymptotic regime is reached for $\lambda K > N$,
while simulations are time limited by $N \sim 10^5$. Therefore, we
must use $\lambda$ of up to about $1000$.  On the other hand we want
to see as much of the prior--enforced learning as possible, so we
choose to work close to this limit: $\lambda=300,500,1000$. One may be
concerned that these large values will produce random distributions
that are almost identical to the reference, which would make some of
our results less interesting.  Fig.~(\ref{qrefdiff}) shows that these
fears are misplaced.  The reference and the random distributions are
similar (as we want them to be), but not excessively so.

\subsection{Learning with the correct prior}

\begin{figure}[t]
  \centerline{\epsfxsize=0.7\hsize\epsffile{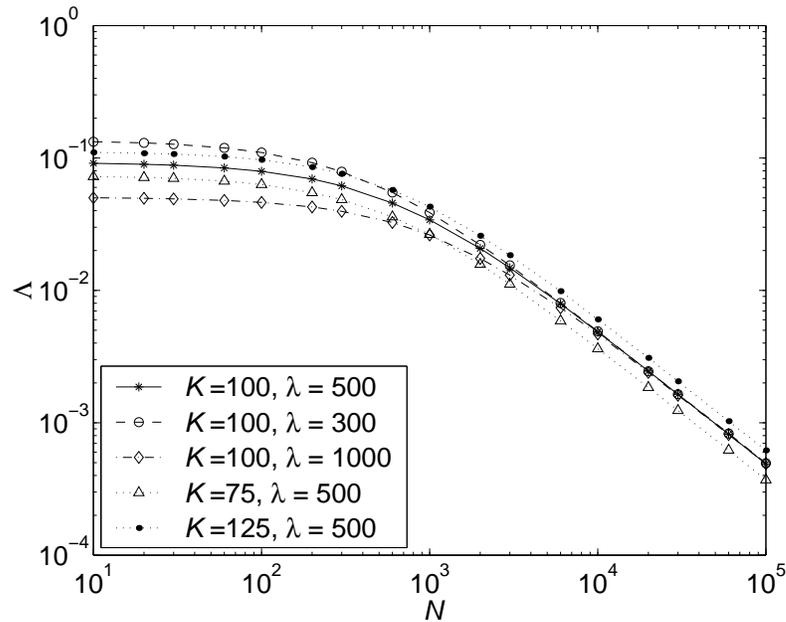}}
\caption[$\Lambda$ as a function of $\lambda$, $K$, and $N$.]
{$\Lambda$ as a function of $\lambda$, $K$, and $N$ for
  $\lambda=\lambda_a$.}
\label{it_correct}
\end{figure}

Fig.~(\ref{it_correct}) shows the $N$ dependence of the universal
learning curve $\Lambda$ [averaged KL divergence,
Eqs.~(\ref{unilcurve},~\ref{lcurve_num})], for various combinations of
$\lambda$ and $K$. All of the behavior predicted in
Section~\ref{it_toytheory_sec} is observed clearly.  The learning
curves start out flat (predictive information and the effective number
of parameters are zero) and soon enter a continuous series of
transitions that add more and more degrees of freedom.  Finally the
behavior enforced by Eq.~(\ref{s1distr}),
\begin{equation}
\Lambda(N) = \frac{K}{2N},
\end{equation}
is reached. Notice that, in agreement with the claim that the
number of active parameters depends on the comparison between the
counts and the value of $\lambda$, the asymptotic regime is reached at
earlier $N$ for smaller values of the parameter. The agreement with
this asymptotic behavior is so remarkably good that we chose not to
quote any fits: all of them are within the expected errors. 

In the pre--asymptotic regime, learning curves with larger $\lambda$
start lower since there the random distributions are very close to the
reference and are estimated much better by it.  Similarly, curves with
smaller $K$ also start lower because now there are fewer degrees of
freedom and, therefore, fewer ways to get a larger KL divergence.

\subsection{Learning with `wrong' priors}

\begin{figure}[t]
  \centerline{\epsfxsize=0.7\hsize\epsffile{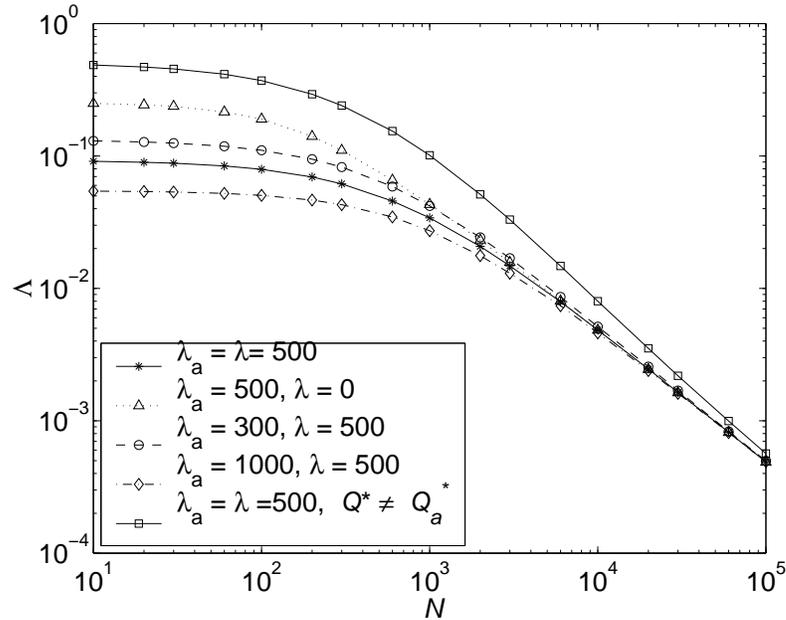}}
\caption[$\Lambda$ as a function of $\lambda$, $\lambda_a$, $Q^*_a$
and $N$.]  {$\Lambda$ as a function of $\lambda$, $\lambda_a$, $Q^*_a$
  and $N$. For all curves $K=100$.}
\label{it_incorrect}
\end{figure}

As we did for nonparametric distributions in
Section~\ref{wrong_pr_sec}, we now want to investigate the performance
of the learning algorithm on atypical data sequences. That is, as
before, we will differentiate two sets of parameters: $\lambda$ and
$Q^*$, which encode the expectations of the learning machine, and
$\lambda_a$ and $Q^*_a$, which together with Eq.~(\ref{prior_d})
describe the ensemble of atypical target distributions.  Simulations
related to this question are summarized in Fig.~(\ref{it_incorrect}).
Here we have shown learning with different combinations of $\lambda
\neq \lambda_a$, and the case $\lambda=\lambda_a=500, Q^*=Q^*_a$ is
plotted as a reference. Comparing the curves with the corresponding
ones from the previous Figure, we clearly see that, even though
learning with an `incorrect' prior is possible, there are
discrepancies: convergence to the asymptotic limit is different, and
the curves start slightly higher then their `correct' counterparts.

Another interesting example shown is when $\lambda$ is `correct', but
the reference distribution itself is totally wrong.  We see that this
type of mistake is much more costly: even for $N$ as large as $10^5$
the influence of the wrong reference distribution is still strong
enough to compromise fast learning.

Note the curve with $\lambda_a=500$, $\lambda=0$. This is a case of
`trivial' learning, when the only regularization is due to the
Jacobian of the $Q \to \phi$ transformation [the `$+1$' term in
Eq.~(\ref{Qest_res_d})]. If not for it, the estimate would be
extremely overfitted and would have zeros, and the $\Lambda$ would
explode.  On the other hand, with the '$+1$' correction, $\Lambda$
starts out from a constant value, which is the KL divergence between
the target and the uniform distribution.

\subsection{Selecting $\lambda$ with the help of the data}

\begin{figure}[t]
  \centerline{\epsfxsize=0.7\hsize\epsffile{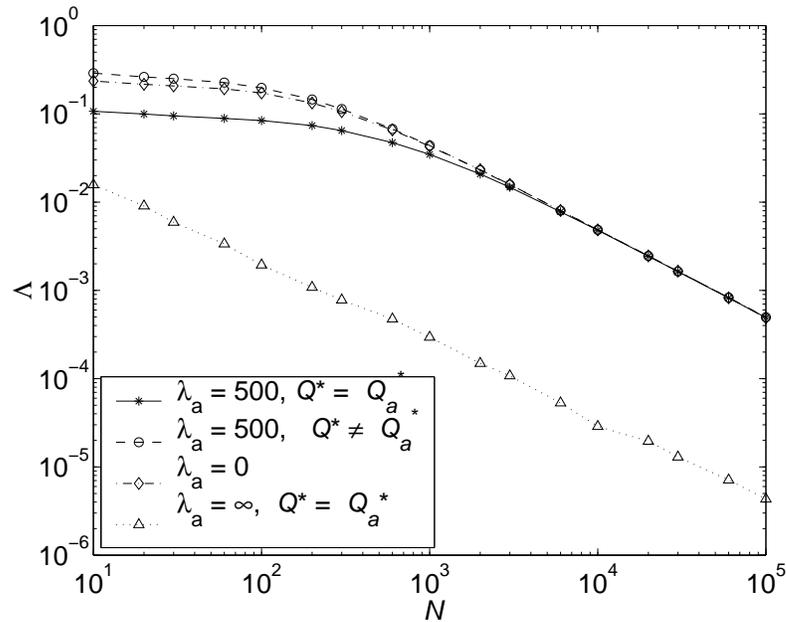}}
\caption{Learning with an adaptive $\lambda^*$.}
\label{it_lam_star}
\end{figure}

Having shown how the learning machine performs on expected and
unexpected data samples, we are now in a position to ask if the Occam
factors can select the right regularization parameter $\lambda$ as in
the case of finite dimensional models (MacKay 1992, Balasubramanian
1997) and nonparametric models (Bialek, Cal\-lan, and Strong 1996, and
Section~\ref{sm_scale_sec} of the present work). This question has an
affirmative answer, and Fig.~(\ref{it_lam_star}) shows $\lambda^*$,
the value of $\lambda$ that maximizes the correlation function
Eq.~(\ref{ans_d_3}) and minimizes the exponent in
Eq.~(\ref{corrthruexp_d}).
\begin{figure}[t]
  \centerline{\epsfxsize=0.7\hsize\epsffile{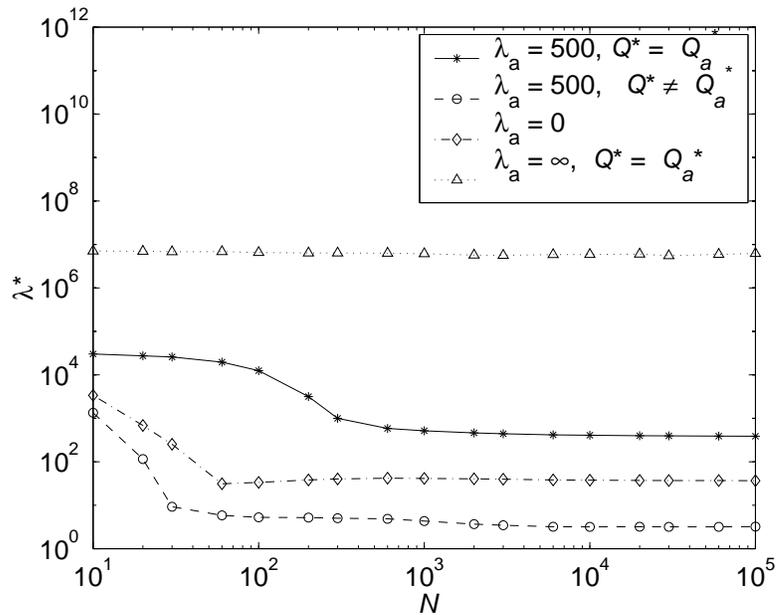}}
\caption{$\lambda^*$ for various ensembles of target distributions.}
\label{it_D_star}
\end{figure}
We show the results averaged over many runs, even though this form of
presentation is questionable because $\lambda^*$ fluctuates a lot in
different trials. These fluctuations explain the kinks on the three
lower curves. For $N$ to the left of the kinks, there are many
realizations of the data, for which there is no best value for
$\lambda$, and the correlation functions are maximized at $\lambda^*
\to \infty$. The kinks appear due to the numerically imposed finite
cutoff on possible values of the parameter.  Apart than this, the rest
of the learning curves' behavior is as hoped. For ensembles generated
at $\lambda_a=500$ and studied at $Q^*=Q^*_a$, $\lambda^*$ turns out
to be very close to 500.  If $Q^*\neq Q^*_a$, $\lambda^*$ drifts to
smaller values, letting the data, not the reference, control the
estimate. The same happens for $\lambda_a=0$ for any $Q^*$ since at
this value of the parameter the target is, again, far from the
reference.

Finally, we examine the case where the ensemble of the distributions
is qualitatively closer to the reference than the prior,
Eq.~(\ref{prior_d}), allows (this corresponds to $\eta\neq\eta_a$ for
nonparametric learning). This can be achieved by having a higher power
of the KL divergence in the exponent of the prior, but such a choice
is very difficult for numerical simulations.  So we take an easier,
but less illustrative example of $\lambda_a=\infty$; that is, the
target distribution is exactly equal to the reference.  Here, not
surprisingly, $\lambda^*$ quickly becomes very large.  Again, it often
exceeds the numerical cutoff, so the average line shown should not be
taken to literally.

Finishing up this section, in Fig.~(\ref{it_D_star}) we show the
learning curves calculated at $\lambda=\lambda^*$ for all the cases
considered above.  Comparing to the corresponding learning curves in
Fig.~(\ref{it_incorrect}), we deduce that learning with an adaptive
$\lambda$ is much faster than with a fixed wrong one, and the example
with $\lambda_a=\infty$ is particularly demonstrative \footnote{This
  remarkable performance for $\lambda_a=\infty$ is achieved with the
  upper cutoff on $\lambda^*$, and it is possible that $\Lambda$ can
  fall off even faster without the cutoff.}. The only learning curve
which starts off slightly worse than its fixed $\lambda$ analogue is
that for $\lambda_a=500, \, Q^*=Q^*_a$. Even though it improves very
quickly, this once again proves the common knowledge that for
small sample sizes nothing beats learning with the `correct' prior.

\section{Further work}
\label{further_d_sec}

The toy example we have investigated resolves the questions it was
meant to answer.  Yet it is just a toy example, and most of its value
lies in the extension to more difficult problems. 

The most straightforward, but very interesting development is one that
has been already mentioned in passing. Reversing the direction of the
KL divergence in Eq.~(\ref{prior_d}) and choosing a uniform reference
distribution $Q^*$, we obtain a prior that favors distributions with
larger entropies, i.~e., the flattest and the most regular
distributions.  Finding the least variable distribution compatible
with the data is certainly in the spirit of our work and deserves an
investigation.

This idea may be made more sophisticated when the independent variable
is a vector, $\vec{x}=\{x,y\}$. If the distribution $Q({\vec x})$ is
expected to be smooth, but the cardinalities of $x$ and $y$ ($K_x$ and
$K_y$ respectively) are large, then only the marginal distributions
$Q(x)=\sum_y Q(x,y)$ and $Q(y)=\sum_x Q(x,y)$ are sampled well for
$K_x$ and $K_y \ll N \sim K_x K_y$.  To smooth the data we might want
to weight the a priori probability of $Q(x,y)$ by the entropy
$S[Q(x,y)]$. However, this choice, though valid, is not the best.
Indeed, the entropy can be small because the marginals are very
narrow. But in the limit we are interested in, the marginals are well
defined by the data and do not require separate smoothing. So it is
not the entropy of a distribution, but its value with respect to the
entropy of the marginals that should enter a regularizing prior. This
is the mutual information $I(x,y)$ between $x$ and $y$, and it is,
once again, a meaningful information--theoretic quantity.

A more ambitious but very appealing direction is to combine these
methods with the relevant information extraction ideas of Tishby et
al.~(1999) and Bialek and Tishby (in preparation), discussed briefly
in Section~\ref{pred_discus_sec}. Recall, that these authors proposed
to compress (that is, to smooth) the variable $x$ into $\hat{x}$ so
that the mutual information $I(\hat{x},y)$ remains as close to
$I(x,y)$, as possible. There is a one parameter family of solutions to
the problem, and this parameter measures the relative importance of
compression (smoothing) and preservation of the information (fit to
the data). In many practical applications, choosing the right value of
this parameter is a problem. We can view the result of Tishby et
al.~as a classical solution to a problem with some (yet to be defined)
prior.  One can realistically expect that the value of the parameter
will, once again, be set by the Occam factor.

If this theory succeeds, we can use these results to further advance
the theory of learning a nonparametric variable.  Derivatives of
densities do not have any special meaning in the framework of
information theory. Using them, as in Eq.~(\ref{prior_bcs}) or its
reparameterization invariant version (Periwal 1997, 1998), is thus not
a preferred regularization.  Building priors that include terms with
derivatives of many different orders and fixing coefficients of the
terms by requiring that the estimate does not depend on the UV details
of the data (rounding or truncation) \footnote{This Wilsonian
  renormalization group approach was suggested by V.~Periwal.}  may
work and have a meaning in QFT, but this is an approach alien to
information theory.  Similarly, preferring distributions that are
close to their filtered version and averaging over the filters
afterwards\footnote{This idea is by W.~Bialek.}  has the same problem.
What we mean by smoothness in any formulation, including nonparametric
continuous ones, is that the independent, possibly continuous,
variable $x$ can be successfully coded in some $\hat{x}$ of finite and
small cardinality such that this compacted version explains the data
almost as well as the original one. For example, in the finite
parameter case, $\hat{x}={\bgm\alpha}$.  Developing a theory for
smoothing through compression would be a great achievement in itself,
and within this formalism we get the added advantage that the right
balance between the goodness of fit and the compression will again be
determined by the Occam factor.

This is obviously only a start to an extensive program of
generalizations, and we hope make a significant progress along these
lines in the near future.



\chapter{Conclusion: what have we achieved?}

Let us summarize our achievements and compare them to the promises we
made and the desires we expressed in the beginning of this work.  As
promised, we built a uniform and universally valid approach to
learning by using information theory and treating learning as an
ability to predict. For this we defined a new quantity, the predictive
information, and the study of it revealed that it not only measures
the information relevant to prediction of a time series, but also
defines uniquely the complexity of the process that generated the
series. Statistical mechanics gave an insight on how learning is
always annealing in the model space, and then we could illuminate
numerous connections to statistical learning and coding theories and
catch some omissions in those.  Summarizing, we indeed delivered on
the promise of a coherent re--treatment of the old knowledge.

Then we went further and showed that conventional finite parameter
models are not the only possible scenario. We investigated
nonparametric and (undersampled) discrete learning to show that the
capacity control mechanisms work much better then one might have
thought. We showed that the 'information theory only' approach to
learning can be made self consistent and does not need any
supplemental help to survive. All of these efforts constitute an
attempt to build up some new flesh around the core of ideas that are
the focal point of our attention. This flesh may seem too thin, and it
certainly is---a lot more has to be done before one can finally say:
``The End!''  However, one has to start somewhere, and we did.


\appendix

\chapter{Appendix}
\section{Summary of nonparametric learning}

\label{bcs_app}

The main problem of statistics---inferring distributions from a finite
data set--- has a wide variety of possible practical applications.
Usually, based on some a priori considerations, an observer has some
finite-dimensional model for the distribution being studied.  Then the
problem reduces to an estimation of a finite number of parameters from
a large data set, and this is relatively well-studied (see, for
example, Vapnik 1998, Balasubramanian 1997, and
Sections~\ref{testcase}--\ref{fluct_prior_sec} of the current work).
Unfortunately, reducing the problem to a finite number of parameters
heavily biases the outcome of statistical inference: the true
distribution may not even be in the chosen family. Thus lately it has
become popular to look for nonparametric solutions to the problem of
learning distributions (recent reviews are Dey et al.~1998 and Lemm
1999). As discussed in Section~\ref{fluct_prior_sec}, nonparametric
estimations necessarily are prior dependent, i.~e., Bayesian.
Therefore, the result of the inference is a probability distribution
of probability distributions, which becomes more concentrated as the
number of samples increases. Even though the result depends on the
prior, the prior may be very mildly restrictive (say only some
smoothness constraints are assumed), and then the bias is less than in
any finite parameter setting.

Recently Bialek, Callan, and Strong presented an elegant formulation
that casts nonparametric Bayesian learning in the familiar setting of
statistical mechanics or, equivalently, Euclidean Quantum Field Theory
(QFT) (Bialek, Callan, and Strong 1996). This approach and some
alternative formulations were further developed by Periwal (1997,
1998), Holy (1997), and Aida (1998).  In the present work, we have
utilized heavily the techniques and results of Bialek et al.~and
expanded or corrected some of their conclusions. To make our
presentation more self contained, we here present a brief overview of
the theory augmented with some comments of our own.

Following the original reference, if $N$ i.~i.~d.~samples $\{x_i\},
i=1\dots N,$ are observed, then the probability that a particular
density $Q(x)$ gave rise to these data is given by the application of
the Bayes formula
\begin{equation}
P [Q(x)| \{x_i\}]
=\frac{P [\{x_i\}|Q(x)]\, {\cal P}[Q(x)]}{P(\{x_i\})}
= \frac{ {\cal P} [Q(x)] \prod_{i=1}^N Q(x_i)}{\int[dQ(x)] {\cal P} [Q(x)] 
\prod_{i=1}^N Q(x_i)}\,,
\end{equation}
where ${\cal P}[Q(x)]$ encodes our a priori expectations of $Q$.
Specifying this prior on a space of functions defines a QFT, and the
optimal least squares estimator is then given by a ratio of correlation
functions
\begin{eqnarray}
Q_{\rm est}(x|\{x_i\})&=&
\int [dQ(x)]\, Q(x)\, P [Q(x|\{x_i\})] 
\\
&=&
\frac{\langle Q(x)Q(x_1)Q(x_2)\dots
  Q(x_N)\rangle^{(Q)}}
{\langle Q(x_1)Q(x_2)\dots Q(x_N) \rangle^{(Q)}}\,,
\label{solution_bcs}
\end{eqnarray}
where $\langle\dots\rangle^{(Q)}$ means averaging with respect to the
prior. Since $Q(x) \ge 0$, it is convenient to define an unconstrained
field $\phi(x)$
\begin{equation}
Q(x)\equiv \frac{1}{l_0} \, {\rm e}^{-\phi(x)},
\label{qthruphi_bcs}
\end{equation}
Unlike another possible choice, $Q(x)= [\phi(x)]^2$, (Holy 1997) this
definition puts $Q$ and $\phi$ in one--to--one correspondence.

The next step is to select a prior that regularizes the infinite
number of degrees of freedom and allows learning. We require that when
we estimate the distribution $Q(x)$ the answer must be everywhere
finite. Also we want the prior ${\cal P}[\phi]$ to make sense as a
continuous theory, so that the statistics of $\phi(x)$ on large scales
are not affected, for example, by discretization or round--off errors
in $x$ on small scales. This implies that we should look for a
renormalization group invariant prior (the first steps along this
direction were performed in Aida 1998) \footnote{As noted by
  V.~Periwal in private communication, one may hope to construct a
  complete theory of nonparametric learning in many dimensions by
  choosing a renormalization group compliant prior. That is, the
  prior's (many) parameters have to be defined to change with the
  renormalization group flow in such a way that the resulting
  correlation functions do not depend on the cutoff scale, which is in
  its turn due to round--off, discretization, or filtering.}.
Simpler, but almost equally satisfying, is any ultraviolet (UV)
convergent prior.  For $x$ in one dimension, a minimal choice that is
the easiest for theoretical (and, accidentally, numerical) analysis is
\begin{equation}
\label{prior_bcs}
{\cal P}[\phi(x)]= \frac{1}{\cal Z}
\exp \left[ -\frac{l}{2}\int dx
\left(\frac{\partial \phi}{\partial x}
\right)^2 \right]\, \delta \left[{1 \over {l_0}}\int dx\, {\rm
e}^{- \phi(x)} -1 \right],
\end{equation}
where ${\cal Z}$ is the normalization constant, and the
$\delta$-function enforces normalization of $Q$. The coefficient $l$
defines a scale below which variations in $\phi$ are considered to be
too rapid, thus we refer to $l$ as the {\em smoothness scale}. By
making this scale local, one may also achieve reparameterization
invariance of learned results (Periwal 1997, 1998).

The resulting field theory was solved by Bialek et al.~up to one--loop
corrections in the limit of large $N$ using standard semiclassical
techniques:
\begin{eqnarray}
\langle Q(x_1)\cdots Q(x_N) \rangle^{(Q)} &\approx& \frac{1}
{l_0^N}\exp \Big( -H_{\rm eff} [\phi_{\rm cl} (x) ; \{x_i\};l]
-\sum_{j=1}^N \phi_{\rm cl}(x_j) \Big),
\label{dist_bcs}
\\
H_{\rm eff}[\phi_{\rm cl} (x) ; \{x_i\};l]  &=&
{l \over 2}
\int dx(\partial_x \phi_{\rm cl})^2
+ \frac{1}{2} \left(\frac{N}{ll_0}\right)^{1/2} 
\int dx\, {\rm e}^{-\phi_{\rm cl}(x)/2},
\label{action_bcs}
\end{eqnarray}
\begin{equation}
l \partial^2_x \phi_{\rm cl}(x) + \frac{N}{l_0}
\exp [-\phi_{\rm cl}(x)]
= \sum_{j=1}^N \delta(x-x_j)\,,
\label{stationary_bcs}
\end{equation}
where $\phi_{\rm cl}$ is the `classical' solution to the field theory.
In the effective Hamiltonian [Eq.~(\ref{action_bcs})], the first term
is due to the value of the prior at $\phi_{\rm cl}$, while the second
one is the infinite dimensional determinant arising from one--loop
integration over fluctuations around the classical solution.
Calculating this determinant is the most technically involved step in
the solution, and this can be done using a standard van Vleck
technique (see, for example, Chapter 7 of Coleman 1988). This term is
a direct analog of Occam factors that appear in finitely
parameterizable models (MacKay 1992, Balasubramanian 1997) and allow
one to build a complexity penalizing razor.

Using the WKB method, the authors have shown that the solutions [both
the classical approximation $Q_{\rm cl}=(1/l_0) \exp (-\phi_{\rm
  cl})$, and the optimal least squares estimator $Q_{\rm est}$,
Eq.~(\ref{solution_bcs})] are non--singular even at finite $N$ and are
essentially self consistent averagings of fluctuations (samples) over
regions of a (local) size
\begin{equation}
\xi \sim \left[l /NQ_{\rm cl} (x)\right]^{1/2}.  
\label{lsize_bcs}
\end{equation}
It was assumed implicitly that the target distribution $P(x)$ being
learned varies negligibly over this length scale, and then the WKB
method can be used again to show that the fluctuations in the
estimate, $\psi(x)\equiv \phi(x) - [-\log P(x)]$ behave at large $N$
as
\begin{eqnarray}
\left< \psi(x)\right> &=&\frac{l}{NP(x)}\,\partial^2_x \log P(x)+
\cdots ,
\\
\left<\left[ \delta \psi(x) \right]^2 \right> &=&\frac{1}{4}\, 
\frac{1}{\sqrt{NP(x)l}}+\cdots .
\label{psi2_app}
\end{eqnarray}
If $P(x)$ is not smooth enough, then in both of these equations one
must replace $P$ by its version smoothed over the local smoothness
scale $\xi$ (for more on this see Section~\ref{sm_scale_sec}). Note
also that the variance of the fluctuations is not uniformly small;
this is a direct result of parameterization dependence of the prior,
Eq.~(\ref{prior_bcs}) (Periwal 1998).

One of the most interesting conjectures of the Bialek et al.~(1996)
paper is that the Occam factor (the fluctuation determinant) is enough
to construct a complexity razor just as in the finite parameter
case.  Indeed, one may impose an a priori distribution on $l$ and
average over it after the correlation function, Eq.~(\ref{dist_bcs}),
is found. The kinetic and the fluctuation determinant terms of the
effective Hamiltonian, Eq.~(\ref{action_bcs}), have opposite
dependences on $l$, so at large $N$ the average should be dominated by
some $l^*$, for which $H_{\rm eff}$ is minimal. The data itself should
select the best smoothness scale consistent with the finite parameter
Minimal Description Length paradigm of Rissanen (1983, 1989) and the
Occam's complexity razor of MacKay (1992) and Balasubramanian (1997).
With the same implicit assumption of a very smooth $P(x)$ the authors
have concluded that
\begin{equation}
l^* \sim N^{1/3}.
\label{lstarbcs}
\end{equation}
If $P(x)$ is not smooth enough, a different dependence of $l^*$ on $N$
should be expected (cf.~Section~\ref{sm_scale_sec}).

This approach has a few shortcomings, most of them arising from
reparameterization noninvariance and omission of clear identification
of the above--men\-tioned smooth target assumption. Some of these problems
are discussed in our present work, and some have been analyzed by
Periwal (1997, 1998).

\section{Correlation function evaluation}
\label{integr_app}

In the step from Eq.~(\ref{ans_d_2}) to Eq.~(\ref{ans_d_3}) we have
performed the integral of the following form
\begin{equation}
{\cal I} = \int_0^1 dt_1 \,dt_2 \cdots dt_K \,\prod_{j=1}^K t_j^{z_j} \delta
\left(1 - \sum_{k=1}^K t_k\right),
\label{integral_ap}
\end{equation}
where $t_j$ was $Q(x)$, $z_j$ was $n(x) +\lambda Q^*(x)$, and the
limit of integration for each variable is $1$ because probabilities
are normalized to one. This integral may be calculated in a
straightforward manner by integrating out each variable in turn. This
creates a product of $\Beta$--functions, which can be simply reduced
to the final result, Eq.~(\ref{gamma_result}). However, this ease of
integration is a consequence of the simplicity of our toy model.  Some
other models currently under investigation (see Section
\ref{further_d_sec}) involve similar integrals, but they are
sufficiently different to prohibit easy exact calculations.  Keeping
these possible applications in mind, we want to show a trickier
integration method of Eq.~(\ref{integral_ap}) that may be more useful
in other problems.

\begin{figure}[t]
  \centerline{\epsfxsize=0.6\hsize\epsffile{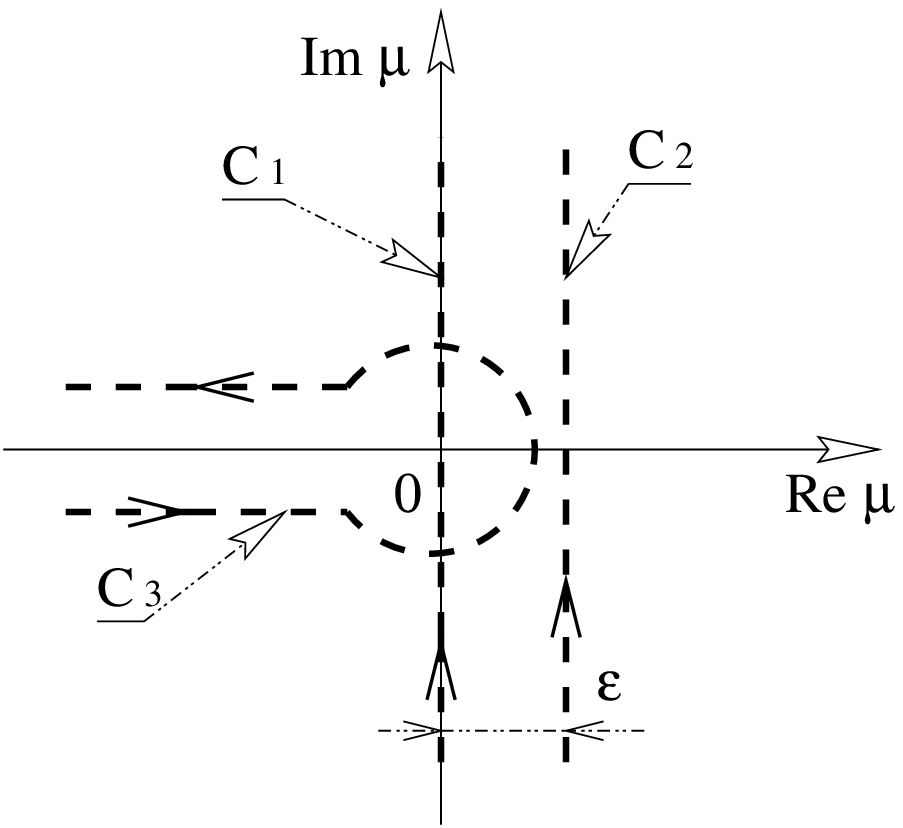}}
\caption{Integration contours.}
\label{conts}
\end{figure}

First note that due to the $\delta$--function, which enforces the
normalization, only $t_j$'s less than or equal to $1$ matter. So we
may equally well replace the upper limits of all integrals in
Eq.~(\ref{integral_ap}) by $+\infty$. Then we can replace the delta
function by its Fourier representation, shift the contour of
integration by a small $\varepsilon$ to the right [the contours and
the directions of the integrations are shown in Fig.~(\ref{conts})],
and exchange the order of the integrals
\begin{eqnarray}
{\cal I} &=& \int_0^{\infty} dt_1 \,dt_2 \cdots dt_K \,
\prod_{j=1}^K t_j^{z_j} 
\int_{C_1} \frac{d\mu}{2\pi i}\, {\rm e}^
{\mu \left(1 - \sum_{k=1}^K t_k\right)}
\\
&=& \lim_{\varepsilon \to 0}
\int_{C_2} \frac{d\mu}{2\pi i} \, {\rm e}^{\mu}
\prod_{j=1}^K
\left[
\int_0^{\infty} dt_j \,t_j^{z_j} {\rm e}^{-\mu t_j}
\right].
\label{shifted_cont}
\end{eqnarray}
Now, since we shifted the contour, ${\rm Re} \mu >0$. Therefore each of the
internal integrals in Eq.~(\ref{shifted_cont}) is a $\Gamma$
function:
\begin{equation}
{\cal I} = \prod_{j=1}^{K} \Gamma(z_j +1) 
\times \lim_{\varepsilon \to 0}
\int_{C_2} \frac{d\mu}{2\pi i} \,
\frac{{\rm e}^{\mu}}{\prod_j \mu^{z_j+1}}
\end{equation}

In more complex cases we probably would have stopped here, trading $K$
real integrals of Eq.~(\ref{integral_ap}) for one contour integration
in the complex plane---this is why this method may be of use. However,
our toy model is {\em very} easy, so we can proceed further and do the
$\mu$ integration. Remembering that $\varepsilon$ is small, we can now
use the Jordan's lemma and bend the integration contour $C_2$ into
$C_3$. Then using a well known formula for the integral representation
of the inverse $\Gamma$--function [Gradshtein and Ryzhik 1965,
Eq.~(8.315)] we get
\begin{equation}
{\cal I}=
\int_0^1 dt_1 \,dt_2 \cdots dt_K \,\prod_{j=1}^K t^{z_j} \delta
\left(1 - \sum_{k=1}^K t_k\right)=
\frac{\prod_{j=1}^{K} \Gamma(z_j +1)}
{\Gamma(\sum_{j=1}^K z_j + K)}
\label{gamma_result}
\end{equation}
Note that, in the case of Eq.~(\ref{ans_d_2}), the sum of $z_j$'s is
just $N$.


\prefacesection{Bibliography}
\sectionmark{Bibliography}
\chaptermark{Bibliography}

\begin{description}
    \item Abarbanel, H.~D.~I., Brown, R., Sidorowich, J.~J., \&
  Tsimring L.~S. (1993). The analysis of observed chaotic data in
  physical systems, {\em Revs. Mod. Phys.} {\bf 65}, 1331--1392.
    \item Adami, C., \& Cerf N.~J. (2000). Physical complexity of
  symbolic sequences, {\it Physica D} {\bf 137}, 62--69. See also
  adap-org/9605002.\footnote{Where available, we give references to
    the Los Alamos e--print archive. Papers may be retrieved from the
    web site {\em http://xxx.lanl.gov/abs/*/*}, where {\em */*} is the
    reference; thus Adami and Cerf (2000) is found at {\em
      http://xxx.lanl.gov/abs/adap-org/9605002}.  For preprints this
    is a primary reference; for published papers there may be
    differences between the published version and the e--print. }
    \item Aida, T.~(1999).  Field theoretical analysis of on--line
  learning of probability distributions, {\it Phys. Rev. Lett.} {\bf
    83}, 3554--3557.  See also cond-mat/9911474.
    \item Akaike, H. (1974a). Information theory and an extension of
  the maximum likelihood principle, in {\em Second international
    symposium of information theory}, B.~Pet\-rov and F.~Csaki,
  eds.  Budapest: (Akademia Kiedo, Budapest).
    \item Akaike, H.~(1974b). A new look at the statistical model
  identification, {\em IEEE Trans. Automatic Control.} {\bf 19},
  716--723.
    \item Atick, J.~J.~(1992).  Could information theory provide an
  ecological theory of sensory processing?  In {\em Princeton Lectures
    on Biophysics}, W.~Bialek, ed., pp.~223--289 (World Scientific,
  Singapore).
    \item Attneave, F.~(1954).  Some informational aspects of visual
  perception, {\em Psych.~Rev.} {\bf 61}, 183--193.
    \item Balasubramanian, V.~(1997).  Statistical inference, Occam's
  razor, and statistical mechanics on the space of probability
  distributions, {\em Neural Comp.} {\bf 9}, 349--368.  See also
  cond-mat/9601030.
    \item Barlow, H.~B.~(1959).  Sensory mechanisms, the reduction of
  redundancy and intelligence, in {\em Proceedings of the Symposium on
    the Mechanization of Thought Processes, vol.~2}, D.~V.~Blake and
  A.~M.~Uttley, eds., pp.~537--574 (H.~M.~Stationery Office, London).
    \item Barlow, H.~B.~(1961).  Possible principles underlying the
  transformation of sensory messages, in {\em Sensory Communication},
  W.~Rosenblith, ed., pp.~217--234 (MIT Press, Cambridge).
    \item Barlow, H.~B.~(1983).  Intelligence, guesswork, language,
  {\em Nature} {\bf 304}, 207--209.
    \item Bennett, C.~H.~(1990).  How to define complexity in physics,
  and why, in {\it Complexity, Entropy and the Physics of
    Information}, W.~H.~Zurek, ed., pp.~137--148 (Addison--Wesley,
  Redwood City).
    \item Bialek, W.~(1995).  Predictive information and the
  complexity of time series, NEC Research Institute technical note.
    \item Bialek, W., Callan, C.~G., \& Strong, S.~P.~(1996). Field
  theories for learning probability distributions, {\it
    Phys.~Rev.~Lett.}  {\bf 77}, 4693--4697.  See also\\
  cond-mat/9607180.
    \item Bialek, W., Fairhall, A., Miller, J., \& Nemenman, I., in
  preparation.
    \item Bialek, W., Nemenman, I., \& Tishby N.~(2000).
  Predictability, complexity and learning, preprint. Available at
  physics/0007070.
    \item Bialek, W., \& Nemenman, I.~(2000). Learning continuous
  distributions: simulations with field theoretic priors, in
  preparation.
    \item Bialek, W., \& Tishby, N.~(1999). Predictive information,
  preprint.  Available at cond-mat/9902341.
    \item Bialek, W., \& Tishby, N.~(in preparation). Extracting
  relevant information.
    \item Brenner, N., Bialek, W., \& de Ruyter van Steveninck, R.
  (2000). Adaptive rescaling optimizes information transmission, {\em
    Neuron} {\bf 26}, 695--702.
    \item Bruder, S.~D.~(1998).  Ph.D.~Dissertation, Princeton
  University.
    \item Capaldi, E.~A., ~Robinson, G.~E., \& Fahrbach S.~E.~(1999).
  Neuroethology of spatial learning: The birds and the bees, {\em
    Annu.~Rev.~Psychol.} {\bf 50}, 651--682.
    \item Carpenter, R.~H.~S., \& Williams, M.~L.~L.~(1995). Neural
  computation of log likelihood in control of saccadic eye movements,
  {\em Nature} {\bf 377}, 59--62.
    \item Chaitin, G.~J.~(1975). A theory of program size formally
  identical to information theory, {\em J.~Assoc.~Comp.~Mach.}  {\bf
    22}, 329--340.
    \item Clarke, B.~S., \& Barron A.~R.~(1990).
  Information--theoretic asymptotics of Bayes methods, {\em IEEE
    Trans.~Inf.~Thy.} {\bf 36}, 453--471.
    \item Clarke, B.~S., \& Barron A.~R.~(1994). Jeffreys' prior is
  asymptotically least favorable under entropy risk, {\em
    J.~Stat.~Planning and Inference} {\bf 41}, 37--60.
    \item Coleman, S.\ (1988). {\em Aspects of symmetry}, (Cambridge
  University Press, Cam\-bridge).
    \item Cover, T.~M., \& King, R.~C.~(1978).  A convergent gambling
  estimate of the entropy of English, {\em IEEE Trans.~Inf.~Thy.} {\bf
    24}, 413--421.
    \item Cover, T.~M., \& Thomas, J.~A.~(1991).  {\em Elements of
    Information Theory} (Wiley, New York).
    \item Crutchfield, J.~P., \& Feldman, D.~P.~(1997). Statistical
  complexity of simple 1--d spin systems, {\em Phys.~Rev.~E} {\bf 55},
  1239--1243R. See also cond-mat/9702191.
    \item Crutchfield, J.~P., Feldman, D.~P., \& Shalizi C.~R.~(1999).
  Comment on ``Simple measure for complexity,'' preprint.  Available
  at chao-dyn/9907001.
    \item Crutchfield, J.~P., \& Shalizi, C.~R.~(1998). Thermodynamic
  depth of causal states: objective complexity via minimal
  representation,  {\em Phys.~Rev.~E} {\bf 59}, 275--283.  See also
  cond-mat/9808147.
    \item Crutchfield, J.~P., \& Young, K.~(1989). Inferring
  statistical complexity, {\em Phys.~Rev. Lett.} {\bf 63}, 105--108.
    \item Ebeling, W., \& P{\"o}schel, T.~(1994). Entropy and
  long-range correlations in literary English, {\em Europhys.~Lett.}
  {\bf 26}, 241--246.
    \item Feldman, D.~P., \& Crutchfield, J.~P.~(1998). Measures of
  statistical complexity: why? {\em Phys.~Lett.~A} {\bf 238},
  244--252.  See also cond-mat/9708186.
    \item Gell--Mann, M., \& Lloyd, S.~(1996). Information measures,
  effective complexity, and total information, {\it Complexity} {\bf
    2}, 44--52.
    \item de Gennes, P.--G.~(1979). {\em Scaling Concepts in Polymer
    Physics} (Cornell University Press, Ithaca and London).
    \item Gradshtein, I.~S., Ryzhik I.~M., (1965).  {\em Tables of
    integrals, series, and products}, 4th ed. (Academic Press, New
  York).
    \item Grassberger, P.~(1986).  Toward a quantitative theory of
  self--generated complexity, {\it Int.~J.~Theor.~Phys.} {\bf 25},
  907--938.
    \item Haussler, D., Kearns, M., Seung, S., \& Tishby, N.~(1996).
  Rigorous learning curve bounds from statistical mechanics, {\em
    Machine Learning} {\bf 25}, 195--236.
    \item Hawken, M.~J., \& Parker, A.~J.~(1991). Spatial receptive
  field organization in monkey V1 and its relationship to the cone
  mosaic, in {\it Computational models of visual processing}, M.~S.
  Landy and J.~A.~ Movshon, eds., pp.~83--93 (MIT Press, Cambridge)
    \item van Helden, J., Andre, B., \& Collado-Vides, J.\ (1998).
  Extracting regulatory sites from the upstream region of yeast genes
  by computational analysis of oligo\-nucleotide frequencies, {\em J.\ 
    Mol.\ Biol.}, {\bf 281}, 827--842.
    \item Herschkowitz, D., \& Nadal, J.--P.~(1999). Unsupervised and
  supervised learning: mutual information between parameters and
  observations, {\it Phys.~Rev.~E}, {\bf 59}, 3344--3360.
    \item Hilberg, W.~(1990). The well--known lower bound of
  information in written language: Is it a misinterpretation of
  Shannon experiments? (in German) {\it Frequenz} {\bf 44}, 243--248.
    \item Holy, T.~E.~(1997).  Analysis of data from continuous
  probability distributions, {\em Phys.~Rev.~Lett.}  {\bf 79},
  3545--3548.  See also physics/9706015.
    \item Kemeney, J.~G.~(1953).  The use of simplicity in induction,
  {\em Philos.~Rev.} {\bf 62}, 391--315.
    \item Knudsen, E.~I., du Lac S., \& Esterly S.~D.~(1987).
  Computational maps in brain, Ann.~Rev.~Neurosci. {\bf 10}, 41--65.
    \item Kolmogoroff, A.~(1939).~ Sur l'interpolation et
  extrapolations des suites stationnaires, {\em C.~R.~Acad.~Sci.
    Paris} {\bf 208}, 2043--2045.
    \item Kolmogorov, A.~N.~(1941).  Interpolation and extrapolation
  of stationary random sequences (in Russian), in {\em Izv.~Akad.
    Nauk.~SSSR Ser.~Mat.} {\bf 5}, 3--14; translation in {\em Selected
    Works of A.~N.~Kolmogorov, vol.~II.}  A.~N.~Shiryagev, ed., pp.
  272--280 (Kluwer Academic Publishers, Dordrecht, The Netherlands).
    \item Kolmogorov, A.~N.~(1965).  Three approaches to the
  quantitative definition of information, {\em Prob.~Inf.~Trans.}
  {\bf 1}, 4--7.
    \item Li, M., \& Vit{\'a}nyi, P.~(1993). {\em An Introduction to
    Kolmogorov Complexity and its Applications} (Springer-Verlag, New
  York).
    \item Lloyd, S., \& Pagels, H.~(1988). Complexity as thermodynamic
  depth, {\em Ann.~Phys.} {\bf 188}, 186--213.
    \item Logothetis, N.~K., \& Sheinberg, D.~L.~(1996).  Visual
  object recognition, {\em Annu.~Rev. Neurosci.} {\bf 19}, 577--621.
    \item Lopes, L.~L., \& Oden, G.~C.~(1987). Distinguishing between
  random and nonrandom events, {\em J.~Exp.~Psych.: Learning, Memory,
    and Cognition} {\bf 13}, 392--400.
    \item Lopez--Ruiz, R., Mancini, H.~L., \& Calbet, X.~(1995). A
  statistical measure of complexity, {\em Phys.~Lett.~A} {\bf 209},
  321--326.
    \item MacKay, D.~J.~C.~(1992).  Bayesian interpolation, {\em
    Neural Comp.} {\bf 4}, 415--447.
    \item Manning C., \& Sch\"{u}tze H.~(1999). {\it Foundations of
    statistical natural language processing} (MIT Press, Cambridge,
  MA).
    \item Opper, M.~(1994). Learning and generalization in a two-layer
  neural network: the role of the Vapnik--Chervonenkis dimension, {\it
    Phys.~Rev.~Lett.} {\bf 72}, 2113--2116.
    \item Opper, M., \& Haussler, D.~(1995). Bounds for predictive
  errors in the statistical mechanics of supervised learning, {\it
    Phys.~Rev.~Lett.}  {\bf 75}, 3772--3775.
    \item Pereira F., Tishby N., \& Lee L.~(1993). Distributional
  clustering of English words, in {\it Proceedings of the 31st Annual
    Meeting of the Association for Computational Linguistics},
  pp.~183--190.
    \item Periwal, V.~(1997).  Reparametrization invariant statistical
  inference and gravity, {\em Phys.~Rev.~Lett.} {\bf 78}, 4671--4674.
  See also hep-th/9703135.
    \item Periwal, V.~(1998).  Geometrical statistical inference, {\em
    Nucl.~Phys.~B} {\bf 554[FS]}, 719--730.  See also
  adap-org/9801001.
    \item P\"oschel, T., Ebeling, W., \& Ros\'e, H.~(1995). Guessing
  probability distributions from small samples, {\it J.~Stat.~Phys.}
  {\bf 80}, 1443--1452.
    \item Press, W., Flannery, B., Teukolsky S., Vetterling, W.~(1988)
  {\em Numerical recipes in {\em C}} (Cambridge University Press ,
  Cambridge).
    \item Rieke, F., Warland, D., de Ruyter van Steveninck, R., \&
  Bialek, W.~(1997).  {\em Spikes: Exploring the Neural Code} (MIT
  Press, Cambridge).
    \item Rissanen, J.~(1978).  Modeling by shortest data
  description, {\em Automatica}, {\bf 14}, 465--471.
    \item Rissanen, J.~(1983). A universal prior for integers and
  estimation by minimum description length, {\em Ann.~Stat.} {\bf 11},
  416--431.
    \item Rissanen, J.~(1984). Universal coding, information,
  prediction, and estimation, {\em IEEE Trans.~Inf.~Thy.}  {\bf 30},
  629--636.
    \item Rissanen, J.~(1986). Stochastic complexity and modeling,
  {\em Ann.~Statist.} {\bf 14}, 1080--1100.
    \item Rissanen, J.~(1987). Stochastic complexity, {\em J.~Roy.
    Stat.~Soc.~B}, {\bf 49}, 223--239, 253--265.
    \item Rissanen, J.~(1989). {\em Stochastic Complexity and
    Statistical Inquiry} (World Scientific, Singapore).
    \item Rissanen, J.~(1996).  Fisher information and stochastic
  complexity, {\em IEEE Trans.~Inf. Thy.} {\bf 42}, 40--47.
    \item Saffran, J.~R., Aslin, R.~N., \& Newport, E.~L.~(1996).
  Statistical learning by 8--month--old infants, {\em Science} {\bf
    274}, 1926--1928.
    \item Saffran, J.~R., Johnson, E.~K., Aslin, R.~H., \& Newport, E.
  L.~(1999).  Statistical learning of tone sequences by human infants
  and adults, {\em Cognition} {\bf 70}, 27--52.
    \item Seung, H.~S., Sompolinsky, H., \& Tishby, N.~(1992).
  Statistical mechanics of learning from examples,  {\em Phys.~Rev.~A}
  {\bf 45}, 6056--6091.
    \item Shalizi, C.~R., \& Crutchfield, J.~P.~(1999). Computational
  mechanics: pattern and prediction, structure and simplicity,
  preprint.  Available at \\cond-mat/9907176.
    \item Shalizi, C.~R., \& Crutchfield, J.~P.~(2000). Information
  bottleneck, causal states, and statistical relevance bases: how to
  to represent relevant information in memoryless transduction.
  Available at nlin.AO/0006025.
    \item Shiner, J., Davison, M., \& Landsberger, P.~(1999). Simple
  measure for complexity, {\em Phys.~Rev.~E} {\bf 59}, 1459--1464.
    \item Shannon, C.~E.~(1948).  A mathematical theory of
  communication, {\em Bell Sys.~Tech.~J.} {\bf 27}, 379--423,
  623--656.  Reprinted in C.~E.~Shannon and W.~Weaver, {\em The
    Mathematical Theory of Communication} (University of Illinois
  Press, Urbana, 1949).
    \item Shannon, C.~E.~(1951).  Prediction and entropy of printed
  English, {\em Bell Sys.~Tech.~J.} {\bf 30}, 50--64. Reprinted in
  N.~J.~A.~Sloane and A.~D.~Wyner, eds., {\em Claude Elwood Shannon:
    Collected papers} (IEEE Press, New York, 1993).
    \item Smirnakis, S., Berry III, M.~J., Warland, D.~K., Bialek, W.,
  \& Meister, M.~(1997).  Adaptation of retinal processing to image
  contrast and spatial scale, {\em Nature} {\bf 386}, 69--73.
    \item Sole, R.~V., \& Luque, B.~(1999).  Statistical measures of
  complexity for strongly interacting systems, preprint.  Available at
  adap-org/9909002.
    \item Solomonoff, R.~J.~(1964).  A formal theory of inductive
  inference, {\em Inform.~and Control} {\bf 7}, 1--22, 224--254.
    \item Tishby, N., Pereira, F., \& Bialek, W.~(1999).  The
  information bottleneck method, in {\em Proceedings of the 37th
    Annual Allerton Conference on Communication, Control and
    Computing}, B.~Hajek and R.~S.~Sreenivas, eds., pp.~368--377
  (University of Illinois). See also physics/0004057
    \item Vapnik, V.~(1998). \emph{Statistical Learning Theory} (John
  Wiley \& Sons, New York).
    \item Vit{\'a}nyi, P., \& Li, M.~(2000). Minimum description
  length induction, Bayesianism, and Kolmogorov Complexity, {\em IEEE
    Trans.~Inf.~Thy.}  {\bf 46}, 446--464. See also cs.LG/9901014.
    \item Weigend, A.~S., \& Gershenfeld, N.~A., eds. (1994). {\em
    Time series prediction: Forecasting the future and understanding
    the past} (Addison--Wesley, Reading MA).
    \item Wiener, N.~(1949).  {\em Extrapolation, Interpolation and
    Smoothing of Time Series} (Wiley, New York).
\end{description}


\end{document}